\documentclass[twocolumn, times]{aastex631}
\usepackage{graphicx}
\usepackage{amsmath}
\usepackage{amssymb}
\usepackage{natbib}
\usepackage{hyperref}
\usepackage[all]{hypcap}
\usepackage{xcolor}
\usepackage{soul}
\usepackage{mathtools}
\usepackage{bm}
\usepackage{multirow}
\let\tablenum\relax
\usepackage{siunitx}
\DeclareSIUnit \Jy {Jy}

\hypersetup{
    colorlinks=true,
    linkcolor=blue,
    filecolor=magenta,
    urlcolor=blue,
}

\newcommand{\vect}[1]{\boldsymbol{#1}}

\newcommand{\Speak}{S_{\rm peak}}

\defcitealias{Mahajan2023}{MK23}
\newcommand{\cta}{\citetalias{Mahajan2023}}

\begin{document}

\title{Intrinsic Emission of PSR B1937+21 at 327 MHz}

\author[0000-0002-6317-3190]{Nikhil Mahajan}
\affil{Department of Astronomy and Astrophysics, University of Toronto, 50 St. George Street, Toronto, ON M5S 3H4, Canada}

\author[0000-0002-5830-8505]{Marten H. van Kerkwijk}
\affil{Department of Astronomy and Astrophysics, University of Toronto, 50 St. George Street, Toronto, ON M5S 3H4, Canada}

\correspondingauthor{Nikhil Mahajan}
\email{mahajan@astro.utoronto.ca}

\begin{abstract}
  At 327 MHz, the observed emission of PSR B1937+21 is greatly affected by scattering in the interstellar medium, on a timescale of order the pulse period.
  We use the bright impulsive giant pulses emitted by the pulsar to measure the impulse response of the interstellar medium and then recover the intrinsic emission of the pulsar by deconvolution -- revealing fine structure on timescales not normally observable.
  We find that the intrinsic widths of the main pulse and interpulse in the pulse profile are similar to those measured at higher frequencies.
  We detect 60,270 giant pulses which typically appear as narrow, $\sim\!100$ ns bursts consisting of one to few nanoshots with widths $\lesssim\!10$ ns.
  However, about $10\%$ of the giant pulses exhibit multiple bursts which seem to be causally related to each other.
  We also report the first detection of giant micropulses in PSR B1937+21, primarily associated with the regular main pulse emission.
  These are distinct from giant pulses not only in the phases at which they occur, but also in their larger widths, of order a microsecond, and steeper energy distribution.
  These measurements place useful observational constraints on emission mechanisms for giant pulses as well as the regular radio emission of millisecond pulsars.
\end{abstract}


\section{Introduction}
\label{sec:introduction}

The nature of pulsar radio emission remains an open question \citep{ Melrose2021, Philippov2022}: we still lack sufficient understanding of pulsar magnetospheres as well as the physical mechanisms that generate both regular pulse emission and bright transients such as giant pulses.
Since some of the proposed mechanisms have strong frequency dependence, it helps to gather observational constraints on the intrinsic emission over as large a range in radio frequency as possible.

Radio signals from pulsars are distorted by the effects of propagation through the interstellar medium (ISM) such as dispersion, birefringence, and scintillation due to multi-path scattering \citep{Rickett1990}.
As a result, the observed radio signal from a pulsar differs significantly from the true intrinsic emission.
While dispersion and birefringence are easily mitigated by inverse filtering \citep{Hankins1975}, there is no general technique for inverting the effects of multi-path scattering.

Since the scattering timescale scales roughly as $\nu^{-4}$, pulsar emission is significantly more scattered at low radio frequencies.
Scattering-induced broadening can wash away finer structure in pulse profiles, especially in millisecond pulsars where the scattering timescales can be of order the pulse period.
For instance, at frequencies below \qty{120}{MHz}, the pulse profile of \object{PSR B1937+21} (a \qty{1.55}{ms} pulsar) is almost entirely washed out by scattering, and the main pulse or interpulse components are no longer distinguishable \citep{Kondratiev2016}.

One can describe the scattering seen in the observed signal $y(t)$ from a pulsar as the result of a convolution of the intrinsic emission $x(t)$ with the impulse response function of the ISM $h(t)$, i.e., one observes $y(t) = (h \ast x)(t) + \epsilon$ where $\ast$ denotes convolution and $\epsilon$ is a noise term.
In principle, if either $h$ or $x$ are known, the other can be recovered by deconvolution.

Previous attempts to infer the impulse response from observed data fall into two categories: phase retrieval using the dynamic spectrum \citep{Walker2008, Baker2022}, and using the cyclic nature of a pulsar signal to conduct cyclic spectroscopy \citep{Demorest2011, Walker2013}.
Phase retrieval techniques usually assume that the impulse response is very sparse in some basis, and are ineffective otherwise \citep{Oslowski2022}.
Cyclic spectroscopy, on the other hand, assumes that the pulsar signal is a cyclostationary signal.
\cite{Walker2013} use cyclic spectroscopy to infer the impulse response and the intrinsic pulse profile of PSR B1937+21 at 430 MHz.
However, many pulsars including PSR B1937+21 exhibit phenomena such as giant pulses, nulling, or mode changing which violate the cyclostationarity requirement.
Thus far, neither of these techniques have been used to coherently recover the actual intrinsic emission of the pulsar, i.e., in voltages.

In our previous work, \citeauthor{Mahajan2023} (\citeyear{Mahajan2023}, hereafter \cta), we used the fact that \object{PSR B1937+21} emits bright impulsive giant pulses which can be used as independent, but noisy, measurement of the impulse response function (IRF).
Using this technique, we were able to successfully model the time-varying IRF of the ISM along the line-of-sight.
In this paper, we use this time-varying IRF to recover the intrinsic emission of PSR B1937+21 at \qty{327}{MHz}.

PSR B1937+21 is of particular interest, as it is a bright and fast millisecond pulsar, which is very well-studied across the electromagnetic spectrum.
In radio, it has a very stable regular pulse emission \citep{Jenet2001}, but also, as previously mentioned, emits giant pulses \citep{Cognard1996}, highly-energetic narrow bursts with timescales as short as a few nanoseconds \citep{Soglasnov2004}.
Giant pulses have also been observed in a handful of others pulsars, which seem to share the property of having a high magnetic field strength at the light cylinder, $B_\mathrm{LC}$, such as the Crab Pulsar \citep{Hankins2003}, PSR B0540-69 \citep{Johnston2003}, and PSR B1957+20 \citep{Knight2006}.
As we will show later, PSR B1937+21 also appears to emit bright, but less energetic, bursts called ``giant micropulses''.
Giant micropulses were first observed in the Vela pulsar \citep{Johnston2001} and PSR B1706-44 \citep{Johnston2002a}.

In the next section, we describe the observational and signal processing techniques used to recover the intrinsic emission of the pulsar. We discuss the recovered intrinsic pulse profile in Section~\ref{sec:regular_pulse}.
We describe our search for transient pulses in the intrinsic emission signal in Section~\ref{sec:transient}, and the properties of the detected giant micropulses and giant pulses in Section~\ref{sec:gmp} and~\ref{sec:gp}, respectively.
Finally, we discuss our findings in Section~\ref{sec:discussion} and our conclusions in Section~\ref{sec:conclusions}.

\section{Observations and Data Reduction}
\label{sec:observations}

We observed PSR B1937+21 for \qty{6540}{s} on MJD 58245 and \qty{1590}{s} on MJD 58298 using the \qty{327}{MHz} Gregorian receiver on the Arecibo Telescope.
As described in \cta, we recorded dual-polarization raw baseband (voltage) data using the Puerto Rico Ultimate Pulsar Processing Instrument (PUPPI) backend.
For our analysis, we use 19 contiguous bands of width \qty{3.125}{MHz} each for a total bandwidth of \qty{59.375}{MHz} centered on \qty{327}{MHz} (from \qty{297.3125} to \qty{356.6875}{MHz}).

We use the \textsc{Baseband} \citep{baseband} and \textsc{Pulsarbat} \citep{pulsarbat} software packages to read and process the raw baseband data.
We pre-process the data by flattening the passband to correct for the effects of a polyphase filter bank which is used in the instrument backend.
Since we measure the passband directly from the data, this has the added effect of attenuating bright narrowband radio-frequency interference (RFI).

At the beginning of each observation, a correlated signal (generated by a pulsed noise diode) is injected into both polarization cables, which we used to measure the relative phase between the two polarizations.
This relative phase is largely dominated by a cable delay, characterized by a linear phase gradient as a function of frequency, as we also found from the pulsar signal itself in \cta.
We correct for this time shift by shifting the signals relative to each other by the measured delay, and compensate for the remaining, non-time-shift relative phases (all below \qty{0.1}{rad}) by rotating the signals by a constant phase in each of the 19 bands.

We then dedisperse the data coherently, using dispersion measures of \qty{71.0201}{pc.cm^{-3}} on MJD $58245$, and \qty{71.0169}{pc.cm^{-3}} on MJD $58298$ (taken from \cta).
We also correct for Faraday rotation using a constant rotation measure (RM) of \qty{9.35}{rad.m^{-2}} for both observations by applying the corresponding transfer function to the baseband signal, which corrects for the time delays between the left- and right-circularly polarized signals emitted by the pulsar.
The RM was determined by fitting a Faraday rotation curve to the frequency-dependent polarization angle in the folded pulse profile, and is consistent with previous RM measurements \citep{Yan2011, Dai2015, Wahl2022} given the uncertainty introduced by ionospheric contributions.

In these pre-processed, dedispersed data, we conduct a giant pulse search to model an IRF -- which, with our correction of Faraday rotation in the baseband signal can now safely be assumed to be independent of polarization (Section~\ref{ssec:irf}).
Next, we use the IRF to recover the intrinsic emission signal via a deconvolution technique (Section~\ref{ssec:deconv}).
Finally, we calibrate the polarization (Section~\ref{ssec:pol_cal}) and normalize the data to estimate fluxes (Section~\ref{ssec:flux}).
Note that all these steps are conducted on baseband data, and the output is thus also calibrated baseband signal.

\subsection{Impulse Response Function}
\label{ssec:irf}

Using the processed baseband data, we follow the techniques described in \cta\, to search for giant pulses and model the time-varying impulse response of the ISM, $H(\nu, t)$.
Since our goal in this work is to recover the intrinsic signal via deconvolution, we are more sensitive to noise.
Hence, while in \cta, where the goal was to try to understand the structure of the wavefield, it made sense to include also fainter but more noisy contributions to $H(\nu, t)$, here we should exclude those.
So, in this work, we use a higher stopping criterion of $\gamma = 6$ (compared to $\gamma = 5$ used in \cta).
Furthermore, we mask out regions in the $\tau-\dot{\tau}$ space which are dominated by noise.
The combined effect of these changes is that the modeled IRF is less noisy.

Another small change we made was to the criteria for giant pulse detection: we slightly relaxed the maximum difference in arrival time between the various streams, to \qty{240}{ns} (compared to \qty{200}{ns} in \cta), as we noticed that we were unnecessarily ignoring pulses which, while having more temporal structure over the full band, are still usable as approximate impulses in individual frequency bands.
With this relaxed constraint, we detect 14,414 giant pulses (compared to 13,025 in \cta).


The wavefields are solved independently for each frequency band, so the modeled IRFs have arbitrary phases relative to each other, i.e. the dot product between an observed giant pulse and the modeled IRF has significantly different phases in each frequency band.
The phase differences between adjacent bands, however, are largely consistent between giant pulses, as expected given that giant pulses should still be impulsive for neighbouring bands.
In these phase differences, we see small variations over time, which can be modeled well by a slowly time-varying curve\footnote{The changes have the right amplitude and frequency dependence to be due to small changes in ionospheric DM. We did not explore fitting for this explicitly.}.
By correcting for these phase differences, we effectively get a wideband IRF.

\subsection{Deconvolution}
\label{ssec:deconv}

With an IRF, we can now recover the intrinsic signal via deconvolution. Following \cta, we use a regularized inverse filter,
\begin{equation}
G(\nu) = \frac{H^*(\nu)}{|H|^2 + \mu}
\end{equation}
where $H(\nu)$ is the impulse response in the frequency domain, and $\mu > 0$ is a regularization factor.
If $y(t)$ is the observed signal, then $\hat{x}(t) = (g \ast y)(t)$ approximates the intrinsic signal, $x(t)$.
Here, $g(\tau)$ is the time-domain representation of $G(\nu)$.

When the IRF is perfectly known and $\mu$ is the inverse of the signal-to-noise ratio, $G(\nu)$ is the Wiener filter, and $\hat{x}(t)$ is the best least-squares approximation of $x(t)$.
In our case, however, the modeled IRF is noisy, and we do not know the signal-to-noise statistics of the intrinsic signal.
Thus, we use a constant $\mu$ determined by trial-and-error depending on the use case (see Sections~\ref{sec:regular_pulse} and \ref{sec:transient}).

In general, a high $\mu$ applies stronger regularization can lead to low-level deconvolution artifacts (of the order the length of the IRF), whereas a low $\mu$ boosts the noise terms lowering the signal-to-noise ratio in the deconvolution output.
Thus, a low $\mu$ is better suited for cases where we average across a large section of data, such as a folded pulse profile, where the high signal-to-noise ratio can compensate for the increase in noise terms post-deconvolution.
A high $\mu$ is better used when analyzing transient bursts in short signal segments, as we are more noise-sensitive and the deconvolution artifacts are much wider than a short transient, and can be treated as additive Gaussian noise.

\subsection{Polarization Calibration}
\label{ssec:pol_cal}

In this work, we use the PSR/IEEE convention for Stokes parameters \citep{Straten2010}.
We calibrate the polarization by modelling how the Stokes parameters change with parallactic angle due to feed rotation \citep{Johnston2002, Straten2004}.
As PSR B1937+21 has a high degree of linear polarization, and our observations span a large range in parallactic angle (\ang{156} on MJD 58245 and \ang{80} on MJD 58298), we can use the observed pulsar emission to self-calibrate.

Specifically, we use that the recorded signal (or voltages) can be described as $\vect{v_{\mathrm{out}}} = \vect{J} \vect{v_{\mathrm{in}}}$ where $\vect{J}$ is a $2 \times 2$ complex-valued Jones matrix which describes how the input signal, $\vect{v_{\mathrm{in}}}$, is transformed.
The Jones matrix usually includes three components: the responses of the feed and receiver backend, and a rotation matrix for the rotation of the feed relative to the sky.

Since we did not observe a reliable calibrator source with known polarization state, we have to make assumptions about the average polarization state of PSR B1937+21's emission in order to sufficiently constrain the feed response.
After correcting for the relative phases and gains using noise diode observations, we found that the polarization angle of the pulsar's emission is well-described by a Faraday rotation curve and the average circular polarization is close to zero.
Other observations of the PSR B1937+21 which measure polarization also show a very small circular polarization fraction \citep{Wang2023, Dai2015, NanoGrav, Stairs1999}.

Thus, we decided to solve for the feed response by assuming that both Stokes' $U$ and $V$ are zero.
Consequently, the polarization angles presented in this paper have an arbitrary (but constant) rotation relative to the sky, and the circular polarization values are systematically inaccurate by a few percent in the polarized fraction.
Despite these inaccuracies, conducting a polarization calibration in this manner is still a significant improvement over not doing it.

\subsection{Flux Density Estimation}
\label{ssec:flux}

Since we did not observe a known unpolarized flux calibrator, we cannot conduct a robust absolute flux calibration of our data.
Thus, we use flux measurements from the literature and assume that the mean flux density of the pulsar is \qty{400}{mJy} at \qty{327}{MHz} \citep{Foster1991}.
Though the pulsar is known to have a spectral index of around $-2.6$, we assume a constant flux density across all frequency bands (thus, assuming a zero spectral index), both for simplicity and to ensure the spectra of giant pulses would display well.
A benefit of normalizing the data in both observations relative to the mean flux density is that flux densities of various emission phenomena can be fairly compared between the observations.

We note that even if we had observed a flux calibrator, it would have been difficult to use the fluxes, since interstellar scintillation causes the overall magnification of the pulsar's emission to vary significantly over timescales of order a month \citep{Ramachandran2006}.
As described in \cta, we cannot measure this overall magnification, and instead assume that the total integrated intensity of the IRF is $1$.

\begin{figure*}
  \centering
  \includegraphics[width=0.92\textwidth,trim=0 0 0 0,clip]{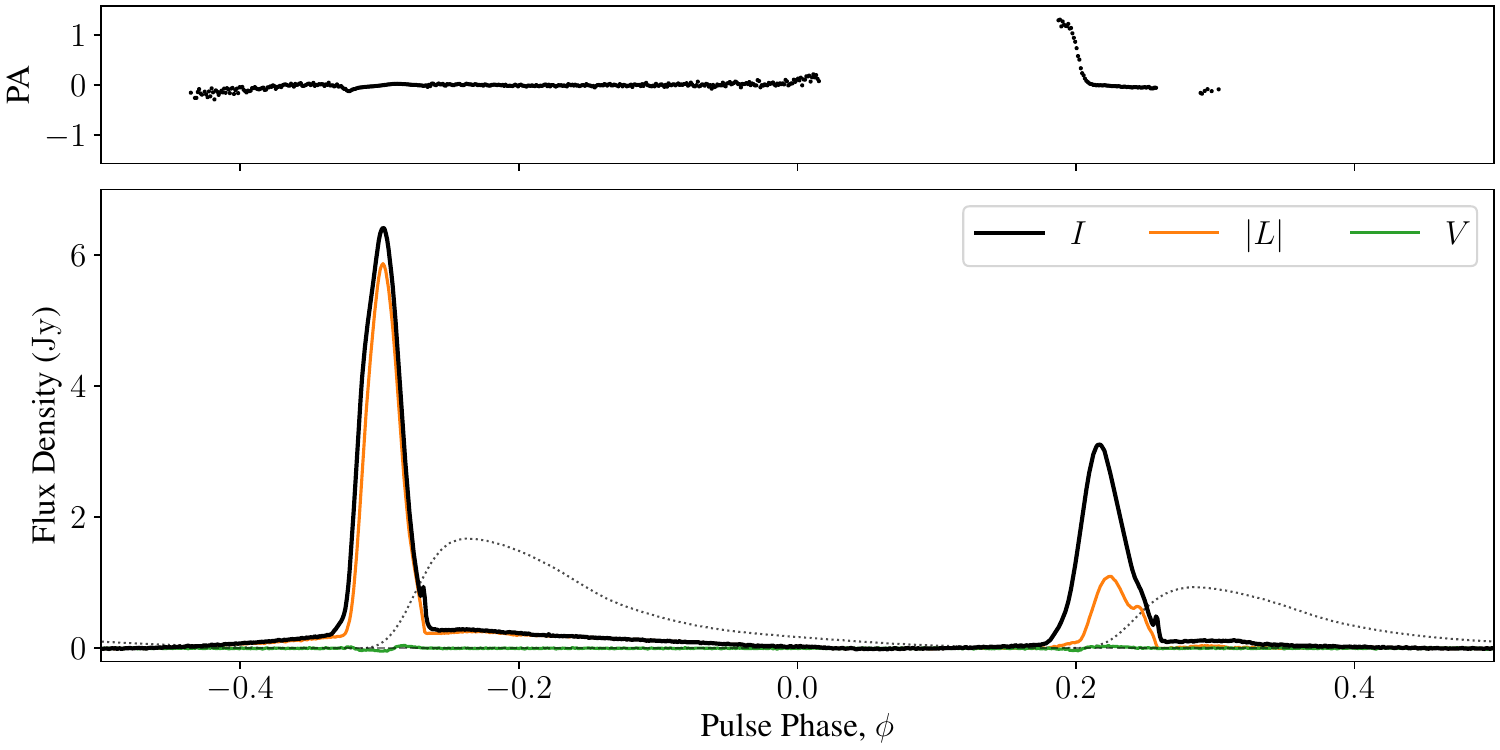}
  \caption{
    The intrinsic pulse profile, found by descattering the baseband data from MJD 58245, folded using $1000$ bins in pulse phase, with polarization angle, PA, shown wherever $|L| / \sigma_I > 7$.
    The pulse profile obtained without descattering is shown as a dotted line for comparison.
    Note that the polarization calibration assumed that Stokes U and V were zero averaged over the pulse profile (see Sect.~\ref{ssec:pol_cal} for details).
    Hence, the absolute polarization angle is not calibrated, and Stokes V is not an accurate measurement of the pulsar's circular polarization profile (though it should not be too far off, since other measurements find that the true average circular polarization is close to zero).
    \vspace{1ex}
  }\label{fig:profile}
\end{figure*}

\section{Intrinsic Pulse Profile}
\label{sec:regular_pulse}

To construct the intrinsic pulse profile, we use a regularization factor of $\mu = 0.01$ during deconvolution\footnote{We normalize the IRF to have a total energy of $1$, i.e. the mean value of $|H(\nu)|^2$ is $1$. Thus, the numerical value of $\mu$ is relative to this total energy.}.
We then fold the resulting signal into a full-Stokes pulse profile using pulsar timing models provided by the NANOGrav collaboration \citep{NanoGrav}.
We have not attempted to calibrate the absolute pulse phase to match that of other studies.
The intrinsic pulse profile for the data from MJD 58245 is shown in Figure~\ref{fig:profile}.
The intrinsic profile on MJD 58298 agrees with this profile (as shown in \cta).

There are two primary pulse components, the main pulse (MP) which peaks at a pulse phase of $\phi = -0.297$ and the interpulse (IP) which peaks at $\phi = +0.217$.
For both MP and IP, there are small narrow bumps on the trailing shoulder which are caused by giant pulse emission.
The linear polarization fractions for the MP and IP are $\sim\!81\%$ and $\sim\!30\%$, respectively.
We show Stokes $V$ for reference, but this may deviate from the true circular polarization profile of PSR B1937+21 given the assumption of zero total circular polarization used in our polarization calibration (Section~\ref{ssec:pol_cal}).

The MP arrives \qty{756.5}{\us} after the IP (\ang{174.8} in pulse phase) in agreement with measurements made by \cite{Foster1991}.
The pulse widths at half maximum, $W_{50}$, are \qty{51.2}{\us} (\ang{11.8}) and \qty{54.5}{\us} (\ang{12.6}) for the MP and IP, respectively.
These widths are consistent with those seen in the intrinsic profile at \qty{430}{MHz} recovered by \cite{Walker2013}, and also with widths measured from observed profiles at higher frequency (and thus less affected by scattering) by \citet{Kramer1999}.
The leading edge of the IP exhibits a $\sim\!\ang{80}$ jump in polarization angle, which has also been observed at \qty{610}{MHz} by \cite{Stairs1999}.

Around both components we see faint, wide bumps.
We expect that some portion of these bumps could be an artifact of our noisy deconvolution method:
since our modeled IRF does not capture the true IRF completely, a small portion of the observed signal is not correctly ``deconvolved'' but is instead scattered further, convolved with something like the cross-correlation of the modeled IRF and the un-modeled component of the true IRF.
Any such descattering mismatch, however, should affect the MP and IP in the same way, while in our results the low-level bumps around the MP and IP do not look alike (even when adjusting for the different peak intensities of the components).
Thus, we believe that at least some part of these bumps is real.
This is supported by the results in \cite{Walker2013} who recover the intrinsic pulse profile at 430 MHz using cyclic spectroscopy and also find wider low-level features around both the MP and the IP.

\section{Search for bright bursts}
\label{sec:transient}

We already know that PSR B1937+21 emits bright narrow bursts in the form of giant pulses.
In fact, we use them to model the IRF.
However, the giant pulses found previously have a selection bias due to the requirement that they be very impulsive and bright.
Thus, we conduct a new search for bright bursts in the intrinsic emission signal to conduct a proper analysis of the transient bursts exhibited by the pulsar.

We use a regularization factor of $\mu = 1$ for the deconvolution step for this search.
The stronger regularization prevents small values of $|H(\nu)|^2$ from amplifying noise in the recovered signal, but also implies a slight reduction in the fraction of energy that will be recovered for an intrinsically narrow burst.
For every pulse period in our data, we extract baseband signal snippets aligned by pulse phase.
We have 4,198,508 and 1,020,697 pulsar rotations on MJD 58245 and 58298, respectively.

For each snippet, we take the signals from the 19 contiguous frequency bands and ``stitch them together'' in the frequency domain to form a \qty{59.375}{MHz} baseband signal for each polarization, with a time resolution of $\approx\!\qty{16.84}{ns}$.
We then take the squared modulus to compute intensities, subtracting the underlying regular pulse emission to ensure the mean intensity in the absence of a burst is zero.
The regular pulse emission is well-described as amplitude-modulated noise and thus follows a $\chi^2$ distribution with $2$ degrees of freedom, with a scale parameter which is a function of pulse phase.
We can estimate the scale parameter in an outlier-robust manner using the median pulse profile, and multiplying by the mean-to-median ratio of the distribution which is invariant to the scale parameter.
We then add the polarizations together to get the total intensity (Stokes $I$) and apply a running uniform filter with a width of 3 samples ($\sim\!\qty{50}{ns}$).
This uniform filter is used to avoid a bias against impulses which occur between two samples (and thus have as low as half the actual flux density in a given discrete sample).
Bright bursts are detected when the flux density for a sample of this signal is above a threshold chosen such that the expected number of false detections across our data is $0.5$, based on the measured noise statistics.
This threshold is pulse-phase dependent as the underlying regular emission contributes to the variance of the noise.
In practice, we expect even fewer false detections since we employ a minimum fluence cutoff as described later.

\begin{figure*}
  \centering
  \includegraphics[width=0.95\textwidth,trim=0 0 0 0,clip]{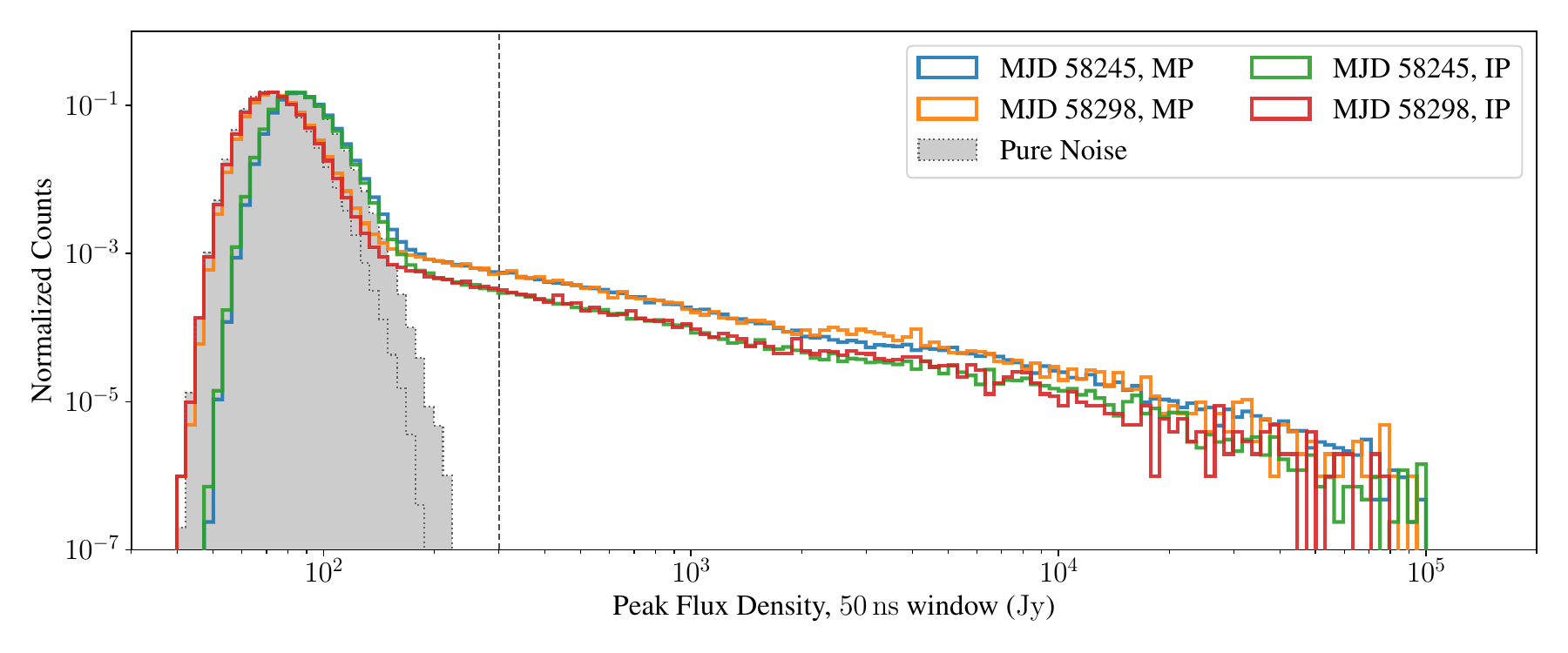}
  \caption{
    Normalized histograms of the peak flux densities, $\Speak$, detected for every individual pulse rotation, over \qty{25}{\us} snippets around the MP and IP giant pulse regions (with flux densities measured using a sliding \qty{50}{ns} window).
    The vertical dotted line at \qty{300}{Jy} is the $\Speak$ cutoff used when detecting a burst.
    Shown in grey are the expected histograms if the data consisted of only noise, which is modeled as an appropriately scaled $\chi^2$ distribution.
    \vspace{10pt}
  }\label{fig:gp_cutoff}
\end{figure*}

In Figure~\ref{fig:gp_cutoff}, we show the histograms for the peak flux density, $\Speak$, in \qty{25}{\us} regions centered where MP and IP giant pulses are emitted, constructed using all pulsar rotations for each observation.
We also show the distribution expected if the data contained only noise, which is modeled as a $\chi^2$ distribution with the scale parameter appropriate for the given pulse phase.
Evidently, there are a significant number of bright giant pulses, which cause the observed distribution to deviate at higher peak flux densities.
The observed pulsar emission is brighter during the observation on MJD 58298, likely due to scintillation-induced magnification.
Since we normalize the data to match flux densities, this means the noise appears to be at a lower flux level in MJD 58298.
The fact that the MP and IP distributions of the two observations match in the high-flux tails (for both MP and IP) shows that the giant pulse flux distributions do not vary relative to the averaged pulsar flux.

When detecting bursts, we often find clumps of samples above the flux density threshold, sometimes with small gaps in between them.
In order to close the gaps and ensure we do not miss portions of a burst, we include all samples within \qty{200}{ns} of a detection as part of the detected burst.
We measure the fluence, $E$, of a burst by integrating the flux density; for its pulse phase, we use the centroid of the detected burst.

In this work, we only consider bursts with $\Speak \ge \qty{300}{Jy}$ (measured in a window of size \qty{50}{ns}) and a fluence of $E \ge \qty{35}{Jy.\us}$.
A typical fluence measurement has an error of around $3-4$ \unit{Jy.\us}, implying that, at the cut-off, we measure the fluence at $\sim\!10\sigma$.
For comparison, the MP component in the regular pulse emission has a peak flux density of $\sim\!\qty{6}{Jy}$ and an average fluence of $\sim\!\qty{350}{Jy.\us}$.
Thus, our weakest detected bursts are at least 50 times as bright as the MP, while having at least $1/10$th the fluence.

\begin{figure*}
  \centering
  \includegraphics[width=0.95\textwidth,trim=0 0 0 0,clip]{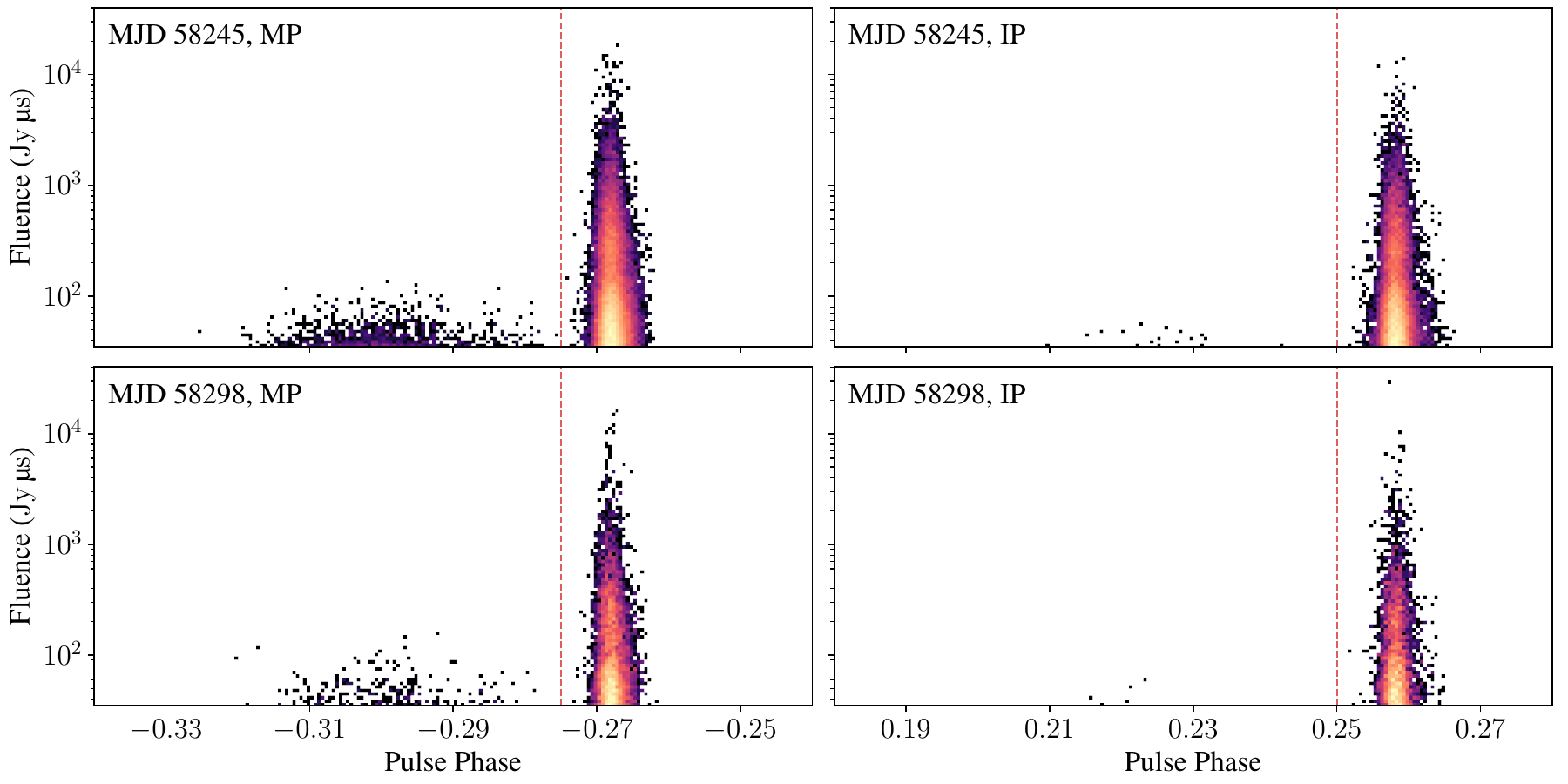}
  \caption{
    The pulse phase -- fluence distribution for bursts with $E > \qty{35}{Jy.\us}$ (with the colormap logarithmic in normalized counts).
    For both the MP (left) and IP (right) pulse phase regions, we see two distinct populations of bursts (separated by a red dashed lines): giant micropulses (GMP) and giant pulses (GP).
    The GMPs occur rarely, in the same phase window as the regular emission, while the GPs occur at much higher rates and fluences, in a narrow pulse phase window trailing the regular emission.
    \vspace{10pt}
  }\label{fig:gp_phase}
\end{figure*}

The pulse phase -- fluence distribution of all detected bursts is shown in Figure~\ref{fig:gp_phase}.
We find that there are two distinct populations of bursts in both the MP and IP, which we will refer to as giant pulses (GP) and giant micropulses (GMP).
GPs, on the trailing side (later in pulse phase) for both MP and IP, are characterized by the narrow pulse phase range in which they occur and a fluence distribution which extends to large values.
GMPs, on the leading side, are characterized by a wider pulse phase range, lower fluences and are coincident with the pulse components in the regular pulse emission.

\begin{figure}
  \centering
  \includegraphics[width=0.47\textwidth,trim=0 0 0 0,clip]{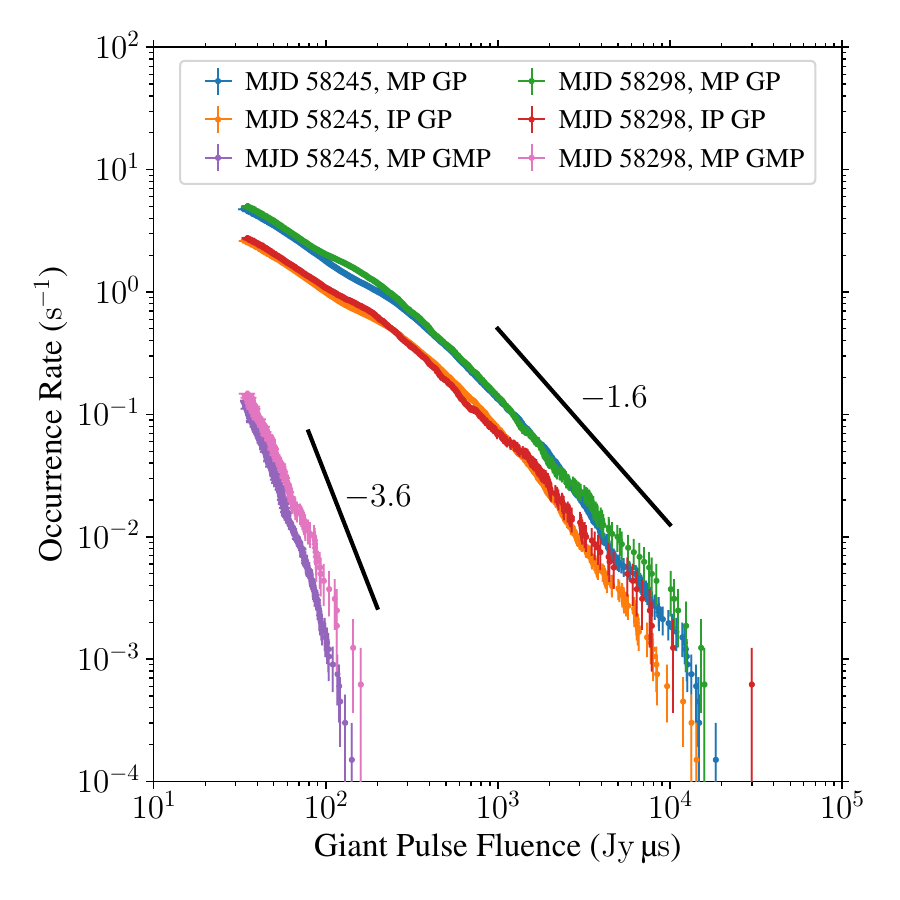}
  \caption{
    Reverse cumulative fluence distributions for detected bursts, for both observations (with IP GMPs omitted because of their low numbers).
    For reference, power laws with indices of $-1.6$ and $-3.6$ are shown.
    One sees that the MP and IP GP distributions are similar, with a turnover at a few \qty{100}{Jy.\us}.
    The distribution of the MP GMPs is far steeper, providing further evidence that GMPs are a different type of burst.
  }\label{fig:gp_power}
\end{figure}

For the purposes of measuring statistics, we divide detected bursts into GPs and GMPs based on their pulse phase, with the partition line denoted by a red dashed line in Figure~\ref{fig:gp_phase} at $\phi = -0.275$ in the MP region, and $\phi = +0.25$ in the IP region.
We summarize our statistics in Table~\ref{tab:stats}, and present reverse cumulative fluence distributions in Figure~\ref{fig:gp_power}.

\begin{deluxetable}{lrrrrr}
\tablecaption{Detection statistics for bright bursts emitted by PSR B1937+21 for both observations.\label{tab:stats}}
\setlength\tabcolsep{7.5pt}
\tablehead{
  &\multicolumn{2}{c}{MJD 58245}&
  \multicolumn{2}{c}{MJD 58298}\\
  \colhead{Component} &
  \colhead{$N$} & \colhead{$r$} &
  \colhead{$N$} & \colhead{$r$} &
  \colhead{$N_{\rm tot}$}
  }
\startdata
MP GP  & 30913 & 4.73   & 7922 & 4.98 & \multirow{2}*{$\Big\}~$60270}\\
IP GP  & 17078 & 2.61   & 4357 & 2.74 \\
MP GMP &   847 & 0.13   &  235 & 0.15 & \multirow{2}*{$\Big\}~$\phn1104}\\
IP GMP &    17 & \nodata &   5 & \nodata \\
\enddata
\tablecomments{For each observation, we list the total number of detected pulses $N$ as well as the rate $r$ (in $\unit{s^{-1}}$, using the exposure times of 6540 and $\qty{1590}{s}$).
  We do not list occurrence rates for the IP GMPs given their low number of detections.
  The last column lists the total number $N_{\rm tot}$ of GP and GMP, i.e., combining MP and IP, as well as observation.
}
\end{deluxetable}
\vspace{-1em}
\section{Giant Micropulses}
\label{sec:gmp}

\begin{figure*}
  \centering
  \includegraphics[width=0.88\textwidth,trim=0 0 0 0,clip]{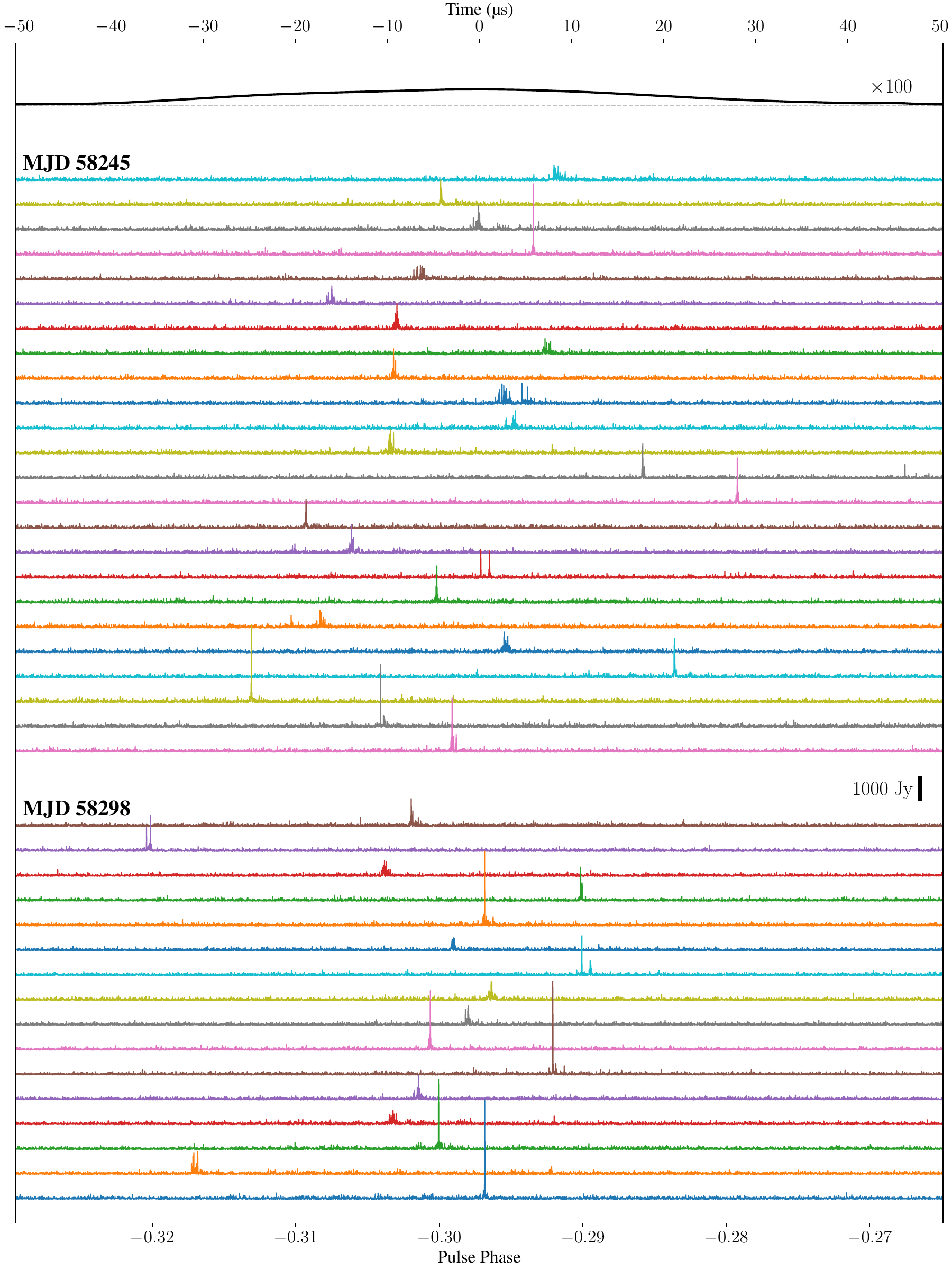}
  \caption{
    The total intensity signals of all ``giant micropulses'' (GMP) in the main pulse region that have fluence above $\qty{85}{Jy.\us}$.
    The signals are offset from each other by \qty{1}{kJy}, for which a vertical black bar is provided as a reference.
    At the top, the main pulse component from the averaged intrinsic pulse profile is shown for comparison (multiplied with a factor 100).
    The peak flux density of GMPs greatly exceeds that of the average pulse profile but their fluence is modest: only $\sim\!25\%$ of the average profile even for the energetic GMPs shown here.
  }\label{fig:gp_gmp}
\end{figure*}

From our search for bright bursts in the intrinsic emission, we have found a total of 1104 giant micropulses, mostly in the MP region of pulse phase.
The highest fluences we observe for GMPs are \qty{160}{Jy.\us} and \qty{60}{Jy.\us} in the MP and IP regions, respectively, which can be compared to the typical fluence for the regular emission components of $\sim\!350$ and $\sim\! \qty{200}{Jy.\us}$, respectively.
The fluence distribution for MP GMPs seems to follow a power-law, such that $P(E > E_0) \propto {E_0}^\alpha$ with $\alpha$ being approximately in the range from $-3.6$ to $-4.0$.

In Figure~\ref{fig:gp_gmp}, we show the total intensity signals of all MP GMPs with fluence above $\qty{85}{Jy.\us}$ from both observations.
It is clear these are not statistical flukes.
The bursts seem to occur in clumps, of roughly $\sim\!\qty{1}{\us}$ width, but with many exhibiting narrower sub-clumps or even single-sample peaks, indicating structure on timescales of $10 - 20$ \unit{ns}.
The brightest MP GMP we observe has a peak flux density of \qty{4}{kJy}.
For the MP GMPs, the median pulse phase of occurrence is $\phi = -0.300$ and the interquartile range (IQR; the pulse phase range of the middle $50\%$ of the MP GMPs) is \qty{15.4}{\us} (\ang{3.6}).
Thus, MP GMPs are coincident with the regular MP emission, but narrower in pulse phase range\footnote{If we assume normal distributions, the equivalent IQR for the regular pulse emission is ${\rm IQR} = {\rm FWHM} / 1.75 \approx \qty{29.3}{\us}.$}.

Previous single-pulse studies of PSR B1937+21 at \qty{430}{MHz} and \qty{1410}{MHz} did not find these GMPs and reported stable single-pulse behavior \citep{Jenet2001, Jenet2004}, with pulse-to-pulse fluctuations consistent with being caused by interstellar scintillation.
These previous non-detections cannot be due to lack of sensitivity, since the GMPs we find are quite bright.
Instead, it seems likely that scattering-induced broadening caused them to become undetectable, as their contribution to the total pulse brightness is only modest.
Additionally, commonly-used metrics to measure single-pulse stability such as the coefficient of variation\footnote{In pulsar literature, this is often referred to as the ``modulation index''.} are insensitive to rare, low-fluence bursts such as the GMPs we detect.

\section{Giant Pulses}
\label{sec:gp}

\begin{figure*}
  \centering
  \includegraphics[width=0.88\textwidth,trim=0 0 0 0,clip]{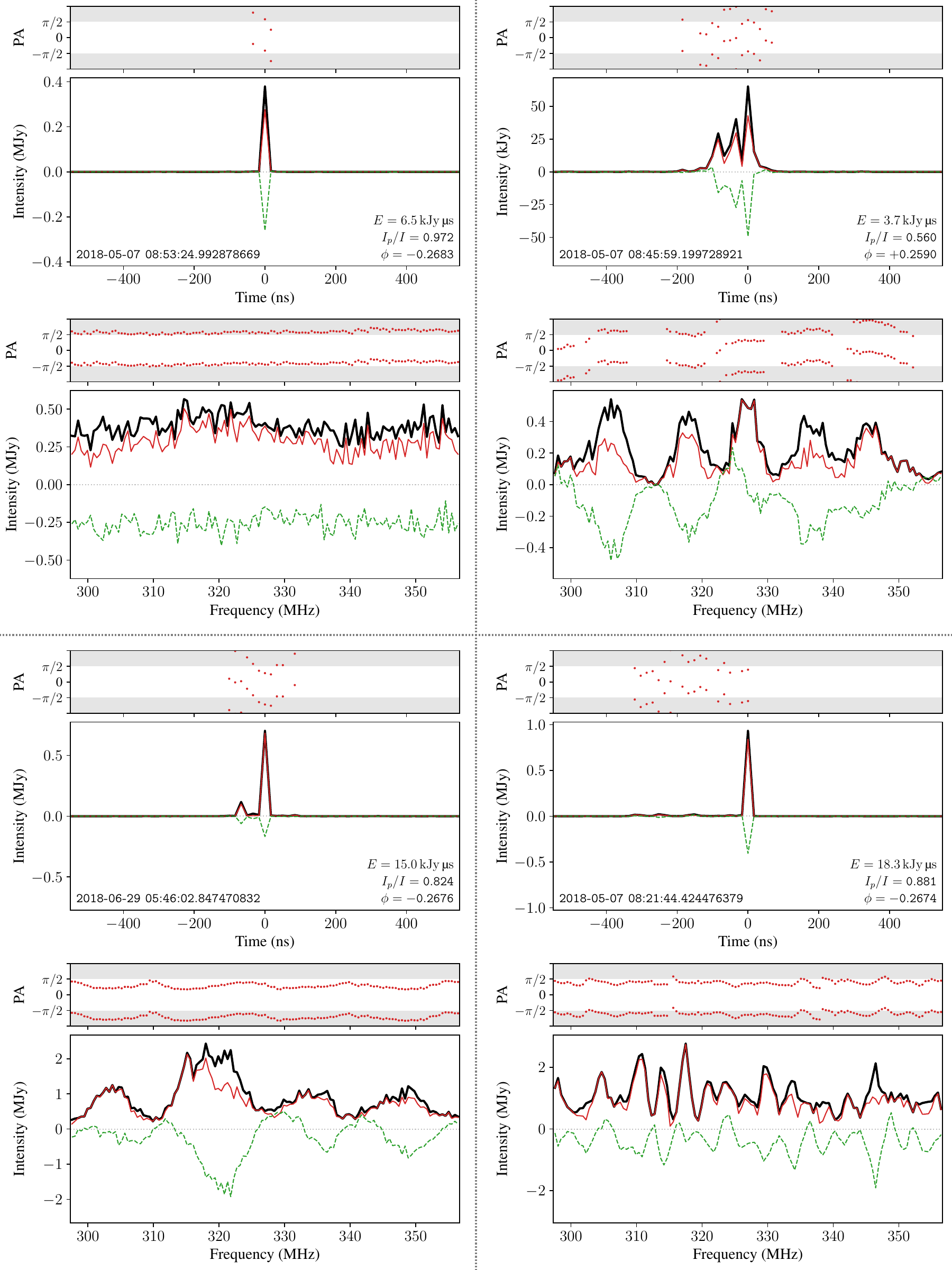}
  \caption{
    The 160 most energetic giant pulses in our dataset (4 per figure).
    For each giant pulse, the top panels show the time-domain signal, and the bottom panels show the frequency spectrum.
    Plotted are total intensity (black), linear polarization (red), and circular polarization (green).
    Also provided is the UTC timestamp corresponding to $t = 0$, the total fluence of the giant pulse, the degree of polarization and the pulse phase at which the giant pulse occurs.
    This is part of a figure set of 40 figures. The remaining figures are attached to the end of this paper as Figures~\ref{fig:figset_first} -- \ref{fig:figset_last}.
  }\label{fig:gp_brights}
\end{figure*}

In our burst search, we find over 60,000 giant pulses (GPs).
As is evident in Figure~\ref{fig:gp_cutoff}, this is a conservative count, primarily due to our choice of using a cutoff of $\Speak \ge \qty{300}{Jy}$.
Based on the distribution of peak flux density, we expect that there are $> 125,000$ GPs above the noise floor in our data and that the lower bound for the fluence of a GP, if there is one, is $\lesssim \qty{10}{Jy.\us}$, the fluence of a burst with $\Speak \sim \! \qty{200}{Jy}$ across \qty{50}{ns}.

The median pulse phases at which GPs occur are $\phi = -0.268$ and $\phi = +0.258$ for MP GPs and IP GPs, respectively. The MP GPs trail the IP GPs by \qty{738}{\us} (\ang{170.5}). The IQR for both MP GPs and IP GPs is \qty{2.3}{\us} (\ang{0.54}).
Relative to the peaks of the pulse components in the regular emission, the MP GPs trail by \qty{45.8}{\us} (\ang{10.6}) and the IP GPs trail by \qty{64.4}{\us} (\ang{14.8}).
The widths of the GP components as well as the separation between the GP components and the regular emission peaks are consistent with those found at higher frequencies by \cite{Kinkhabwala2000}.

In Figure~\ref{fig:gp_power}, we see that the tail of the fluence distribution seems to follow a power-law, as we found above for the GMPs.
For GPs with fluence above $\qty{1}{kJy.\us}$, the power-law index is roughly in the range from $-1.6$ to $-2.0$ for both observations and both components.
The distribution appears to be slightly curved, getting steeper at higher GP fluence.
Similar turnovers have been seen in previous GPs studies of PSR B1937+21 \citep{Soglasnov2004, McKee2019} and the Crab \citep{Popov2007, Karuppusamy2010, Lin2023b}.

We find that MP GPs occur $\sim\!1.8$ times more often than IP GPs, and thus the most energetic GPs tend to come from the MP region.
However, accounting for the lower occurrence rate, we see no evidence that IP GPs are intrinsically less energetic than MP GPs.
In fact, the highest fluence we measure is for an IP GP with \qty{29.6}{kJy.\us}.
The brightest GP we detect has a peak flux density of \qty{934}{kJy} across $\sim\!\qty{15}{ns}$, thus, having a implied brightness temperature of $T_b = 10^{41}\,\unit{K}$, using a distance to PSR B1937+21 of \qty{3}{kpc} \citep{Ding2023}. On the lower end, the weakest GPs we detect have $\Speak \ge \SI{300}{\Jy}$ across \SI{50}{ns} resulting in $T_b = 10^{36.5}\,\unit{K}$.

In Figure~\ref{fig:gp_brights} and the associated figure set, we show the full-Stokes time-domain signals as well as the frequency spectra for the most energetic GPs we observe.
The first pulse shown (top left panels) is a bright narrow GP of fluence \qty{6.5}{kJy.\us} which seems to be localized (in time) to a single sample, and (consequently) has a relatively flat spectrum across our entire bandwidth.
If we assume this GP is a Gaussian pulse, it would need to have a FWHM of $\lesssim\!\qty{10}{ns}$ to explain our observation: a wider pulse would spill over into neighbouring samples much more.

Generally, the GPs we detect seem to consist of clumps or bursts made up of one to a few of these ``nanoshots'', as can be seen in the other panels of Figure~\ref{fig:gp_brights}, as well as in the extended figure set.
For GPs with multiple nanoshots, we see the corresponding periodic modulations in the spectra, reflecting how the nanoshots interfere with each other.
The median width of a GP is $\sim\!\qty{100}{ns}$ (measured by the maximum extent across detections for a GP).

Most GPs exhibit just one clump of nanoshots, but some seem to exhibit multiple clumps.
We analyze these multi-burst GPs further in Section~\ref{ssec:multiburst} below, before turing to the polarization statistics of GPs in Section~\ref{ssec:gp_pol}.

\subsection{Multi-burst Giant Pulses}
\label{ssec:multiburst}

\begin{figure*}
  \centering
  \includegraphics[width=0.9\textwidth,trim=0 0 0 0,clip]{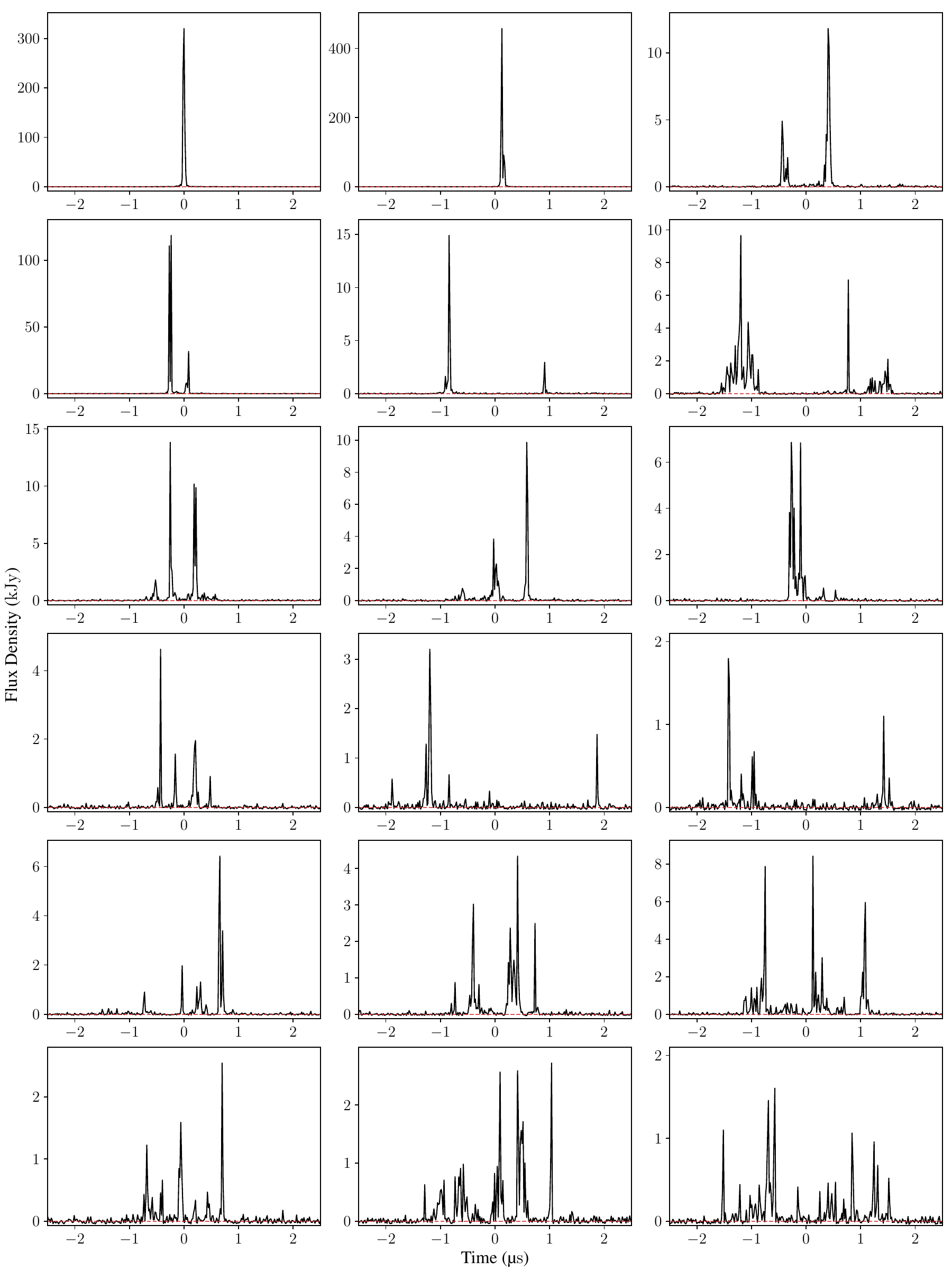}
  \caption{
    A hand-picked selection of GPs, chosen to demonstrate morphological variety, showing progressively increasing number of ``bursts''.
    The sample spacing is \qty{16.84}{ns}, and for all plots, $t = 0$ is chosen to be the mean of the earliest and latest detections in time.
  }\label{fig:gp_examples}
\end{figure*}

In Figure~\ref{fig:gp_examples}, we show the varied morphology of GPs we find.
While the typical GP looks like the examples in the first two panels, a significant number of them show distinct multiple bursts.
In order to measure the multi-burst statistics of GPs, we must consistently define what constitutes a burst.
As described in Section~\ref{sec:transient}, we define a ``detection'' as a sample of a total intensity signal (with \qty{50}{ns} uniform filtering applied) which exceeds a flux density threshold.
We define ``gaps'' as parts of the signal which are more than 5 samples away from a detection.
Thus, two adjacent bursts must necessarily have a gap of $\ge\qty{185}{ns}$ between them.
This minimum gap size was chosen to be about double the typical width of a GP of $\qty{100}{ns}$ as described earlier.
Once all the gaps have been identified, we can count the number of distinct bursts in a GP.
Table~\ref{tab:multi} summarizes the statistics of the multi-burst GPs we find.

\begin{deluxetable}{crrrr}
\tablecaption{Multi-burst statistics of giant pulses. \label{tab:multi}}

\tablehead{\colhead{No. of} & \twocolhead{\dotfill MJD 58245\dotfill} & \twocolhead{\dotfill MJD 58298\dotfill} \vspace{-0.5em} \\
\colhead{bursts} & \colhead{MP} & \colhead{IP} & \colhead{MP} & \colhead{IP}}
\startdata
$\ge 0$ & 4198508 & 4198507 & 1020697 & 1020697 \\
$\ge 1$ &   30913 &   17078 &    7922 &    4357 \\
$\ge 2$ &    2112 &    1531 &     689 &     464 \\
$\ge 3$ &     159 &     141 &      54 &      53 \\
$\ge 4$ &      20 &      17 &       7 &       8 \\
$\ge 5$ &       2 &       2 &       2 &       3 \\
$\ge 6$ &      -- &      -- &      -- &       1 \\
\enddata
\tablecomments{Two bursts are considered separate if they are more than $\qty{185}{ns}$ apart.}
\end{deluxetable}
\vspace{-1.25em}
If bursts were independent of each other and their occurrence in the same pulse rotation was due to coincidence, we would expect to see significantly fewer multi-burst GPs than we actually observe.
For example, on MJD 58245, using binomial statistics, we would expect to see $\sim\!115$ MP GPs with $\ge 2$ bursts and $< 1$ MP GPs with $\ge 3$ bursts if the bursts occurred independently of each other.
In addition to this, we find that multi-burst GPs occur much closer to each other in time than would be expected from the full pulse phase range at which GPs are observed.
For GPs where we detect two bursts, we find that the median time difference between the centroids of the bursts is $\qty{0.42}{\us}$, much smaller than the expected $\sim\!\qty{2}{\us}$ if the bursts were randomly sampled from the overall GP pulse phase distribution.
Therefore, the multiple bursts of these GPs seem to be causally related to each other, and occur in much tighter groupings compared to the pulse phase range.
We find that the pairwise time differences or fluence ratios between bursts have no preferred value or apparent pattern.

From Table~\ref{tab:multi}, we can see that for a GP with $n \ge 1$ bursts, the probability of observing the $(n + 1)$-th burst is roughly consistent for all $n$ and across different GP components.
However, this probability is significantly higher than the probability of observing the first burst (i.e., the occurrence rate of GPs).
We can roughly reproduce the observed multi-burst statistics if we set the probability of observing the first burst to be the MP and IP GP occurrence rates, $P_{\rm mp} = 0.0075$ and $P_{\rm ip} = 0.0042$, respectively, and the probability of observing each subsequent burst to be roughly $P' \sim\! 0.075$ for both components.
Thus, we see that while the occurrence rate is different between MP GPs and IP GPs, it does not appear to affect the probability of observing subsequent bursts.
The higher and consistent probability of observing a subsequent burst for both components further supports the notion that the bursts in a multi-burst GP are causally related.
A similar causal relationship between multiple bursts of a giant pulse has been recently reported in observations of the Crab \citep{Lin2023b}.

\begin{figure}
  \centering
  \includegraphics[width=0.47\textwidth,trim=0 0 0 0,clip]{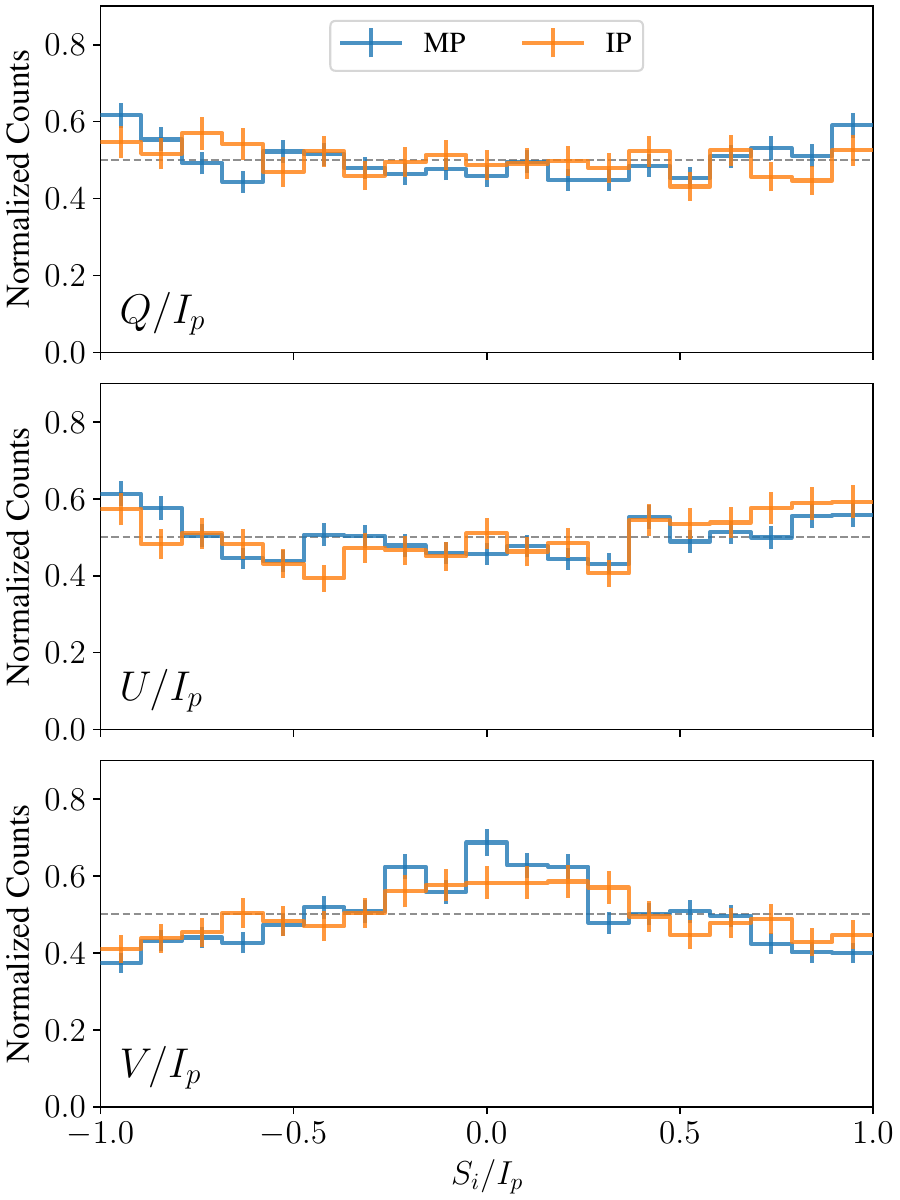}
  \caption{
    Polarization statistics for bright ($>\qty{300}{Jy.\us}$) and narrow ($<\qty{1}{\us}$) GPs for the two GP components.
    We show the distribution of Stokes $Q$, $U$, and $V$ normalized by $I_p$, the fraction of the intensity that is polarized, which are the coordinates on the Poincar\'e sphere.
    One sees that GPs are very close to randomly polarized, nearly uniformly sampling the Poincar\'e sphere.
    \vspace{1ex}
  }\label{fig:gp_polfrac}
\end{figure}

\begin{figure}
  \centering
  \includegraphics[width=0.47\textwidth,trim=0 0 0 0,clip]{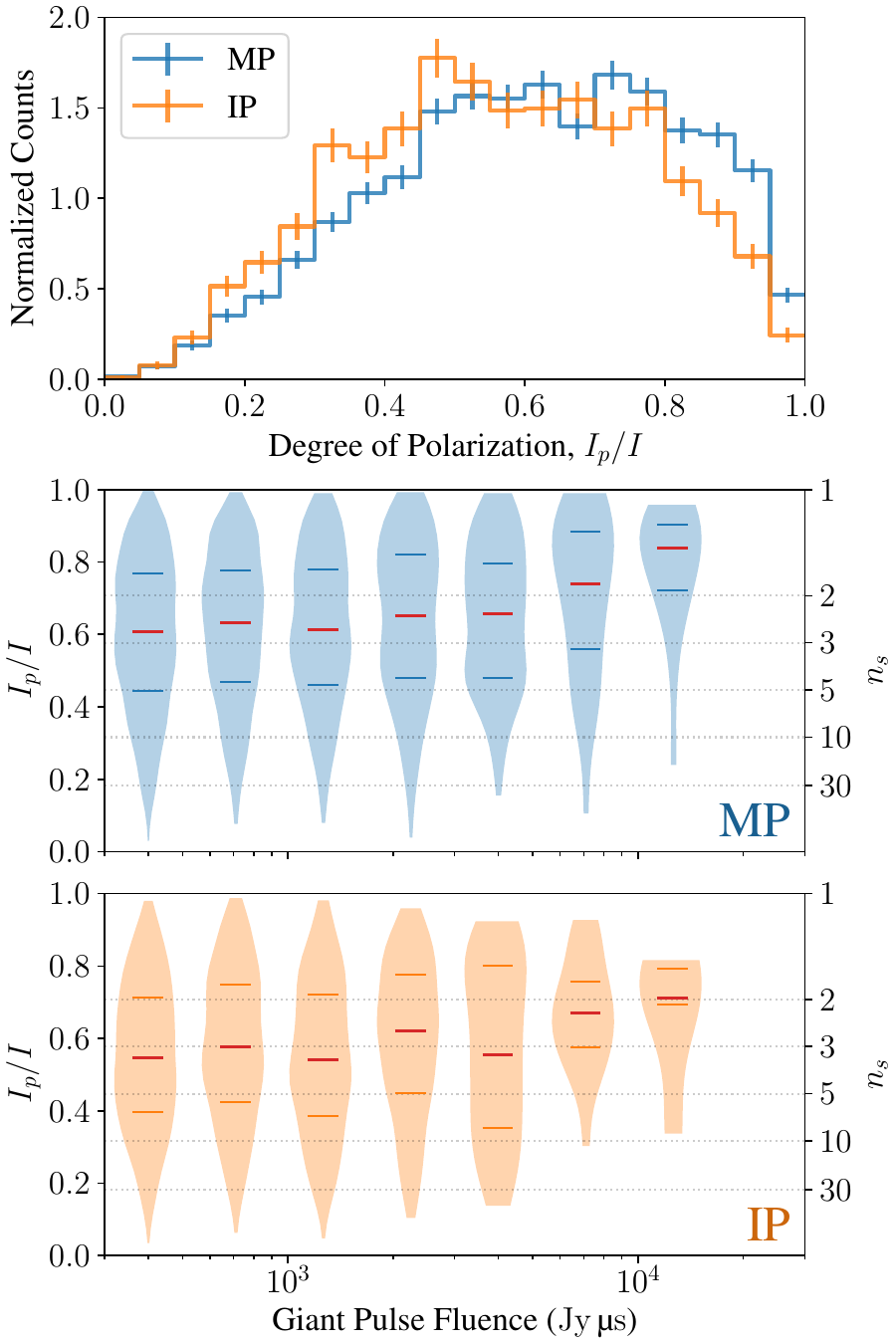}
  \caption{
    {\it Top Panel:} Degree of polarization distribution for MP and IP GPs.
    {\it Lower Panels:} Degree of polarization distributions at different GP fluence levels, shown as violin plots. The horizontal lines denote the quartiles, with the red line being the median.
    The right vertical axis also shows the implied effective number of impulses, $n_s \sim (I_p / I)^{-2}$ (see text).
    One sees that more energetic GPs appear to be have higher degree of polarization.
    \vspace{1ex}
  }\label{fig:gp_violin}
\end{figure}

\subsection{Polarization of Giant Pulses}
\label{ssec:gp_pol}

Polarization may give insight in the nature of GPs, so we determined statistics for a sample of bright and narrow GPs, with fluence $> \qty{300}{Jy.\us}$ and width $< \qty{1}{\us}$ (which ensures sufficient signal-to-noise).
We first measure the time-averaged Stokes parameters $I, Q, U, V$ across a GP, determine the polarized intensity,
\begin{equation}
    I_p = \sqrt{Q^2 + U^2 + V^2}
\end{equation}
and construct $(Q / I_p, U / I_p, V / I_p)$, i.e., points on the Poincar\'e sphere.
In Figure~\ref{fig:gp_polfrac}, we show that the polarizations of GPs almost uniformly sample the Poincar\'e sphere, suggesting that GPs are generally randomly polarized, with a possibly small preference towards low circular polarization.

As described previously, the GPs we observe appear to consist of nanoshots which are essentially impulses across our bandwidth.
If individual nanoshots are assumed to have roughly the same fluence and be randomly and fully polarized, then on average, we have $\langle (I_p / I)^{2} \rangle = 1/n_s$ where $n_s$ is the number of nanoshots \citep{Lin2023}.
In Figure~\ref{fig:gp_violin}, we show distributions of the degree of polarization, $I_p/I$, of GPs.
More energetic GPs appear to have a higher degree of polarization.
If interpreted in terms of number of nanoshots, this suggests higher fluence GPs contain fewer nanoshots, or that they are more likely to be dominated by a single bright nanoshot.
The latter is expected if nanoshots are sampled from a similar fluence distribution as we observe for GPs.
For the brighter bursts, the distribution is steeper, so the brightest nanoshot will, on average be accompanied by fainter nanoshots than will be the case for the brightest nanoshot in a fainter burst (which, at the threshold used here, is at the point where the GP fluence distribution starts to turn over).

\section{Discussion}
\label{sec:discussion}

\subsection{Giant Pulses and Giant Micropulses}

We observe two types of bright narrow bursts which we call giant pulses and giant micropulses.
In the pulsar literature, GPs were somewhat arbitrarily defined as bright pulses with more than ten times the fluence of the average pulse \citep{Cognard1996}.
When bright bursts were observed in the Vela pulsar \citep{Johnston2001}, they were called ``giant micropulses'' as they did not meet this fluence cutoff.
Now, with more sensitive instruments and improved techniques, we can measure low-fluence bursts that are clearly part of the same population as the more energetic GPs, just arising in the fainter part of the distribution, and thus, relevant to our understanding of the GP emission mechanism.

Here, we describe the key differences we see between these two populations in our data.
GPs occur over a very narrow phase range and trail the regular pulse components.
Their fluence distributions are not as steep, with power-law indices $\alpha \gtrsim -2$, and extend to extremely high fluence.
GPs typically seem to consist of one or maybe a few bursts, each having one to few nanoshots, which have widths of order \qty{10}{ns}.
GMPs, on the other hand, occur over a pulse phase range roughly coincident with the regular pulse components.
Their fluence distribution is much steeper, with power-law indices $\alpha \sim\!-4$, and, therefore, does not extend to very high fluences.
GMPs seem to be broader, with widths of order \qty{1}{\us}, although with structure at the tens of ns level.

These differences suggest that, at least in PSR B1937+21, GPs and GMPs are emitted from different parts of the magnetosphere and thus likely have different underlying emission mechanisms.
GMPs, given the overlap in phase with the regular pulse emission, may be transient bursts from the polar gap, whereas GPs are generally thought to originate at or beyond the light cylinder, given that they usually occur at the same phases as high-energy emission \citep{Cusumano2003}.

PSR B1937+21 appears to be the first pulsar where both these phenomena are observed, allowing them to be compared and distinguished.
It is possible that previous detections of, say, GPs in other pulsars are, in fact, detections of GMPs in the sense described here.
For instance, the GPs we reported in the mode-changing PSR B1957+20 at 327 MHz seem to generally have very steep fluence distributions, with $\alpha \sim\! -4.5$, except for the GPs in the main pulse during the ``high mode'', for which we found $\alpha \sim\! -1.6$ for the more energetic bursts \citep{Mahajan2018}.
A possible interpretation may be that the former are actually GMPs, and only bursts from the energetic component which emit during the high mode are GPs like the ones observed here in PSR B1937+21.
It would be interesting to apply the techniques introduced here to descatter the emission of PSR~B1957+20, and see if the emission phases at which bursts are emitted give further clues.

\subsection{Comparison with the Crab Pulsar and Implications for the Giant Pulse Emission Mechanism}

The properties we measure in this work should provide useful observational constraints for the radio emission mechanism of GPs, especially by comparing also with other GP emitters.
For the comparison, the Crab pulsar is perhaps the most useful, as it is very well studied.
It has a similarly strong magnetic field at the light cylinder, $B_{\rm LC} \sim\! 10^6 \, \unit{G}$, but a much longer spin period, $\sim\!\qty{33}{ms}$, thus potentially allowing one to tease apart properties that depend mostly on field strength and others that depend also on the spatial scale.

We find that the GPs are emitted across a very narrow pulse phase range of \ang{0.54}.
If we assume that this phase range is entirely due to relativistic beaming, then we can infer that the bulk Lorentz factor of the emission region is $\gamma \sim \! 2 / {\rm IQR} \approx 200$.
In the Crab, the GPs in the main pulse at \qty{1.7}{GHz} have an IQR of approximately \ang{1.5} \citep{Lin2023b} which, using the same argument, implies $\gamma \sim \! 75$.

We also find that GPs are mostly randomly polarized (almost uniformly distributed around the Poincar\'e sphere), with a slight preference for low circular polarization.
In the Crab Pulsar, the polarization properties of the giant pulses are similar but not identical: its GPs are preferentially linearly polarized, although the polarization angle is (nearly) random \citep{Hankins2016, Lin2023}.

We find that some GPs consist of multiple bursts, which are causally linked and seem to be grouped much more tightly than the overall pulse phase range.
This again is like that observed for GPs in the Crab Pulsar \citep{Sallmen1999, Hankins2007, Lin2023b}.
One could interpret the reduced phase range as evidence for even higher $\gamma$ factors for the emitting plasma, but it could also just reflect a maximum duration, corresponding to a maximum spatial extent over which emission can be generated.
We also find that the probability of generating a subsequent burst is roughly invariant to the number of bursts in a GP, as again seems to be the case in the Crab Pulsar \citep{Lin2023b}, suggesting an emission process where a transient event can preferentially induce another transient event with some consistent probability.

Each individual GP burst seems to consist of a couple of very closely spaced nanoshots, with the average degree of polarization suggesting an effective number of nanoshots of about three.
This is similar to what is seen for the Crab Pulsar \citep{Hankins2007}, although there the typical number of nanoshots seems to be somewhat higher, about five \citep{Lin2023}.
Furthermore, in the Crab pulsar, the nanoshots in a given microburst are spread over about \qty{1}{\us}, while for PSR~B1937+21 the bursts typically seem to last less long, about \qty{100}{ns}.
It may be meaningful that the ratio between the two is roughly the ratio of the spin periods.
Finally, in the Crab Pulsar, the nebular scattering screen resolves the nanoshots, which is only possible if the plasma emitting them travels at highly relativistic speeds, with $\gamma\simeq10^4$ \citep{Bij2021, Lin2023}, implying it emits nanoshots over $\sim\!10^3$ light cylinder radii along the line of sight.
For PSR~B1937+21, if the Lorentz factor were the same, the duration would imply a similar extent, relative to the light cylinder radius.

One of the most promising recent models for GPs is that they are emitted by merging plasmoids which form in the current sheet in the striped wind beyond the light cylinder (\citealt{Philippov2019}; see also \citealt{Lyubarsky2019}).
This model predicts GPs with roughly the brightness temperature that we observe, and also seems to match the morphology at least qualitatively: clumps of one to few nanoshots.
The simulations also seem to make testable predictions, e.g., for how the GP properties should change with frequency.
Since PSR B1937+21, with its fast spin period and therefore compact magnetosphere, is easier to simulate than the Crab Pulsar, with its much larger light cylinder, it may be worthwhile to try to target simulations specifically to PSR B1937+21.

\subsection{Better deconvolution technique}

While our deconvolution technique works well in recovering the intrinsic emission of the pulsar, it should be possible to improve on it.
One weakness of our method is that it is only least-squares optimal for well-known IRFs.
In the case of noisy measures of the IRF, which is what we have, total least-squares methods should be better suited to recover the intrinsic signal \citep{Huffel2002, Mastronardi2000}.
A problem is that these techniques usually require working with extremely large matrices, the size of which is determined by the length of the signals involved.
This is computationally intractable for wideband Nyquist-sampled baseband signals.
We believe, however, that it should be possible to use these methods effectively for recovering short signals, of order a few hundred samples, such as around a giant pulse.

Furthermore, we detect significantly more giant pulses in the intrinsic signal than we do in the original giant pulse search.
While many of the additional detections are faint, they could be included in the solution for the wavefield, presumably leading to a less noisy model.

\section{Conclusions}
\label{sec:conclusions}

We successfully recover the intrinsic emission, in the voltage domain, of PSR B1937+21 at \qty{327}{MHz}, where in the observed signal propagation effects have significantly smeared the signal.
From this intrinsic signal, we are able to measure the intrinsic pulse profile of the pulsar.
We also successfully find transient bursts, detecting over 60,000 giant pulses from both the main pulse and interpulse emission components.

We also discovered giant micropulses in PSR B1937+21, with over 1000 detected in our data.
It would be useful to conduct a search for these giant micropulses at higher radio frequencies to determine if previous non-detections were a result of selection bias (for example, due to less sensitive instruments) or these bursts do not occur at higher frequencies.

The observation of these transient bursts places significant observational constraints on the physical emission mechanisms of these components.
Further studies of other millisecond pulsars, especially giant pulse emitters like PSR B1957+20, can bring us closer to a coherent theory of radio emission in millisecond pulsars.

\begin{acknowledgements}
  We thank Rebecca Lin for helpful comments and insight regarding giant pulses in the Crab Pulsar.
  We also thank the referee for their helpful comments which definitely improved the manuscript.
  This research has made use of NASA's Astrophysics Data System Bibliographic Services.
  Computations were performed on the Niagara supercomputer at the SciNet HPC Consortium \citep{Loken2010, Ponce2019}.
  SciNet is funded by: the Canada Foundation for Innovation; the Government of Ontario; Ontario Research Fund - Research Excellence; and the University of Toronto.
  MHvK is supported by the Natural Sciences and Engineering Research Council of Canada (NSERC) via discovery and accelerator grants, and by a Killam Fellowship.
\end{acknowledgements}

\facility{Arecibo (327 MHz Gregorian). The data used in this publication was obtained as part of observing project P3229.}

\software{
    Pulsarbat \citep{pulsarbat},
    Baseband \citep{baseband},
    Numpy \citep{numpy},
    Astropy \citep{astropy:2013, astropy:2018, astropy:2022},
    Scipy \citep{scipy},
    Matplotlib \citep{matplotlib},
    Dask \citep{dask},
    TEMPO2 \citep{tempo2:1, tempo2:2}
}

\clearpage
\bibliographystyle{aasjournal}
\bibliography{thesis}

\begin{figure*}
  \centering
  \includegraphics[width=0.9\textwidth,trim=0 0 0 0,clip]{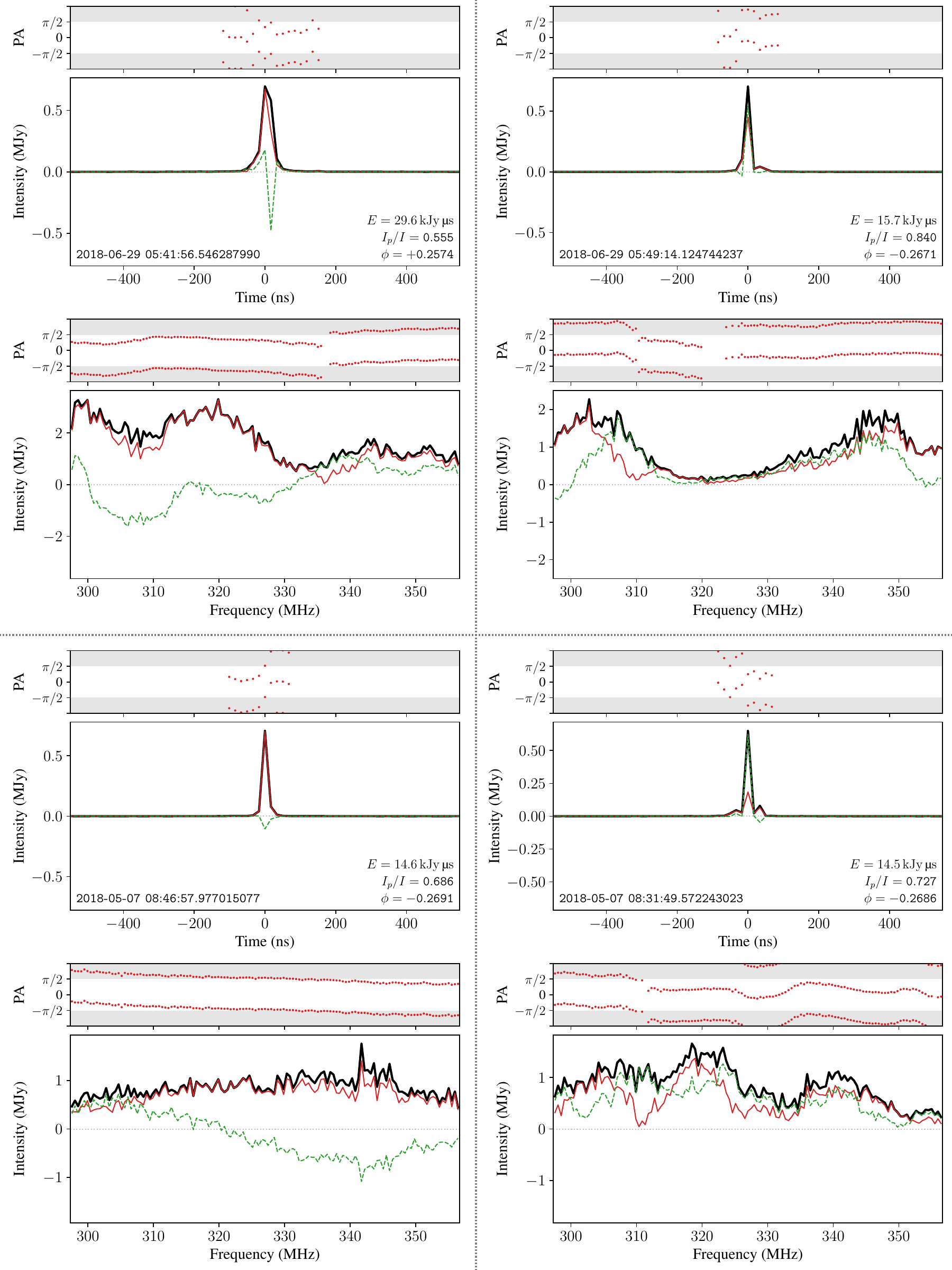}
  \caption{A selection of the most energetic giant pulses. See Figure~\ref{fig:gp_brights} for a full description.}
  \label{fig:figset_first}
\end{figure*}

\begin{figure*}
  \centering
  \includegraphics[width=0.9\textwidth,trim=0 0 0 0,clip]{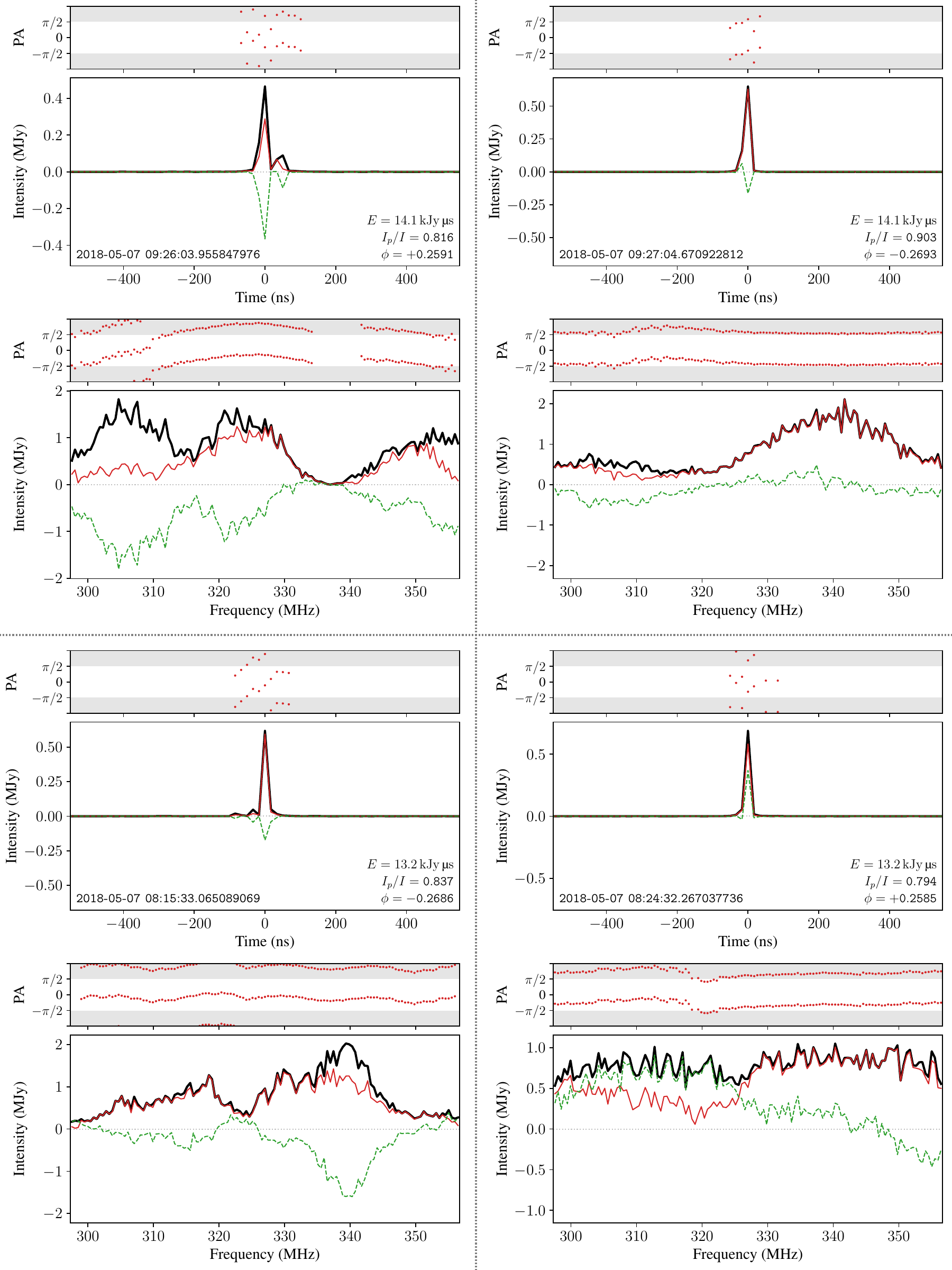}
  \caption{A selection of the most energetic giant pulses. See Figure~\ref{fig:gp_brights} for a full description.}
\end{figure*}

\begin{figure*}
  \centering
  \includegraphics[width=0.9\textwidth,trim=0 0 0 0,clip]{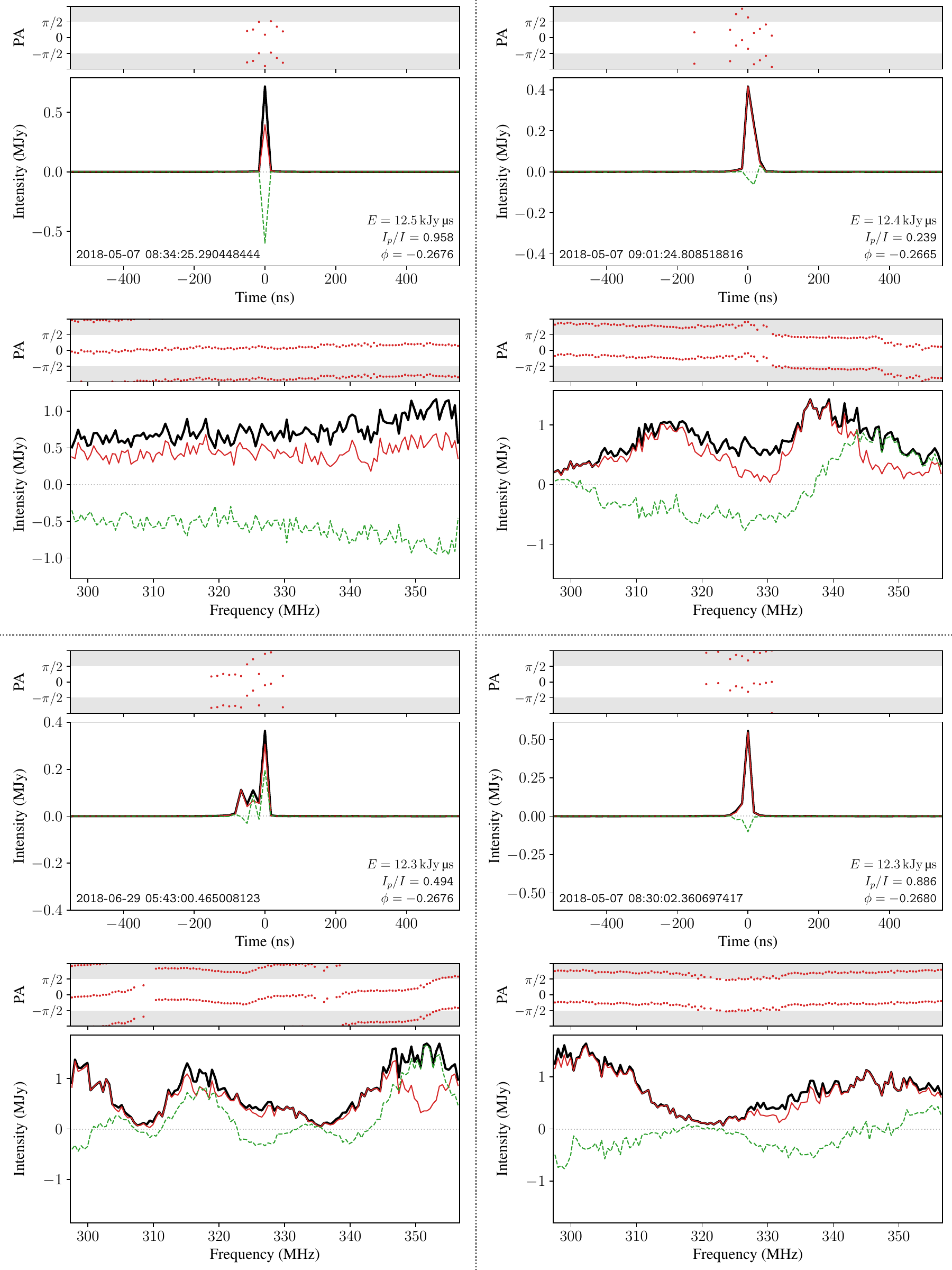}
  \caption{A selection of the most energetic giant pulses. See Figure~\ref{fig:gp_brights} for a full description.}
\end{figure*}

\begin{figure*}
  \centering
  \includegraphics[width=0.9\textwidth,trim=0 0 0 0,clip]{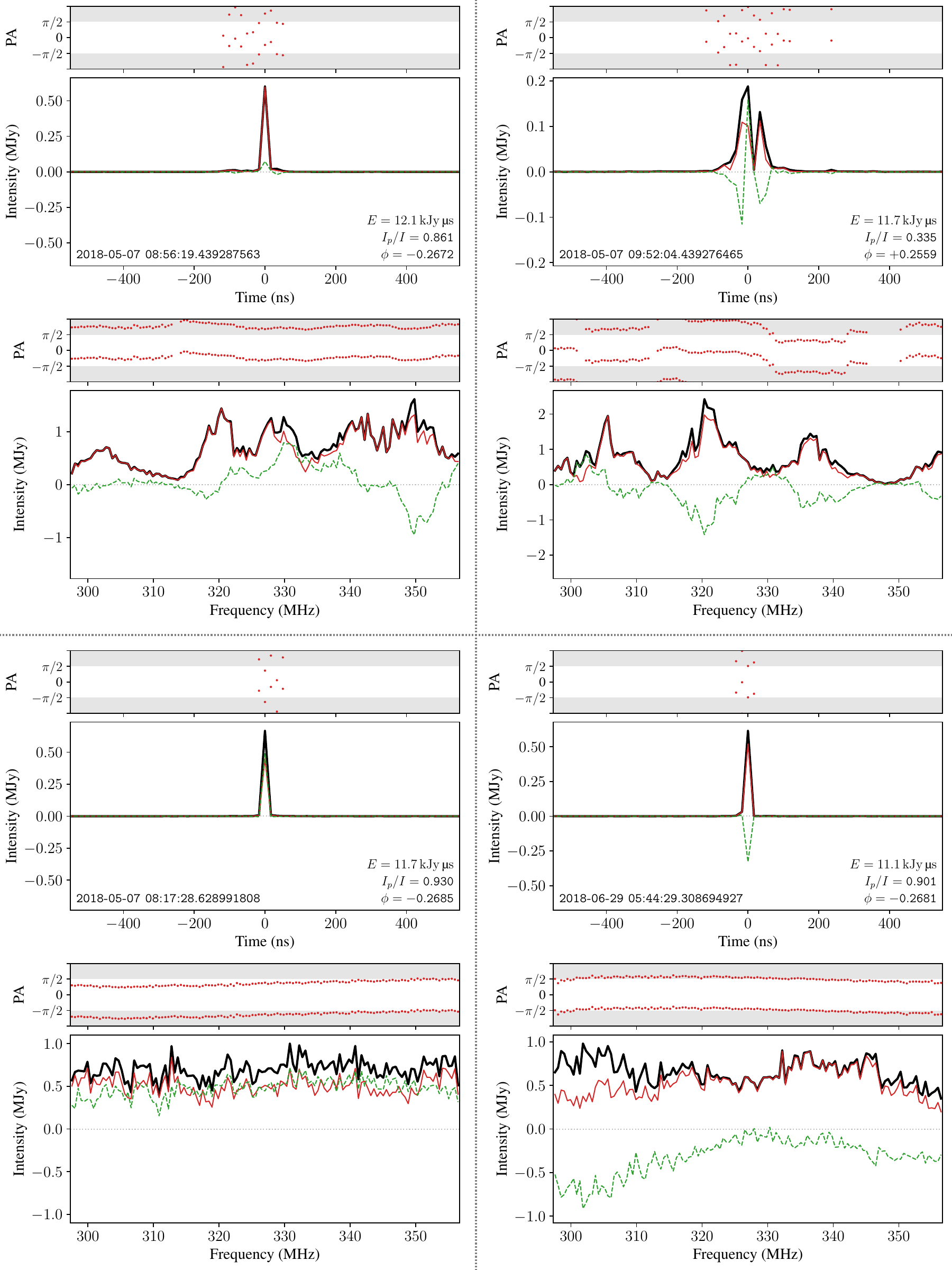}
  \caption{A selection of the most energetic giant pulses. See Figure~\ref{fig:gp_brights} for a full description.}
\end{figure*}

\begin{figure*}
  \centering
  \includegraphics[width=0.9\textwidth,trim=0 0 0 0,clip]{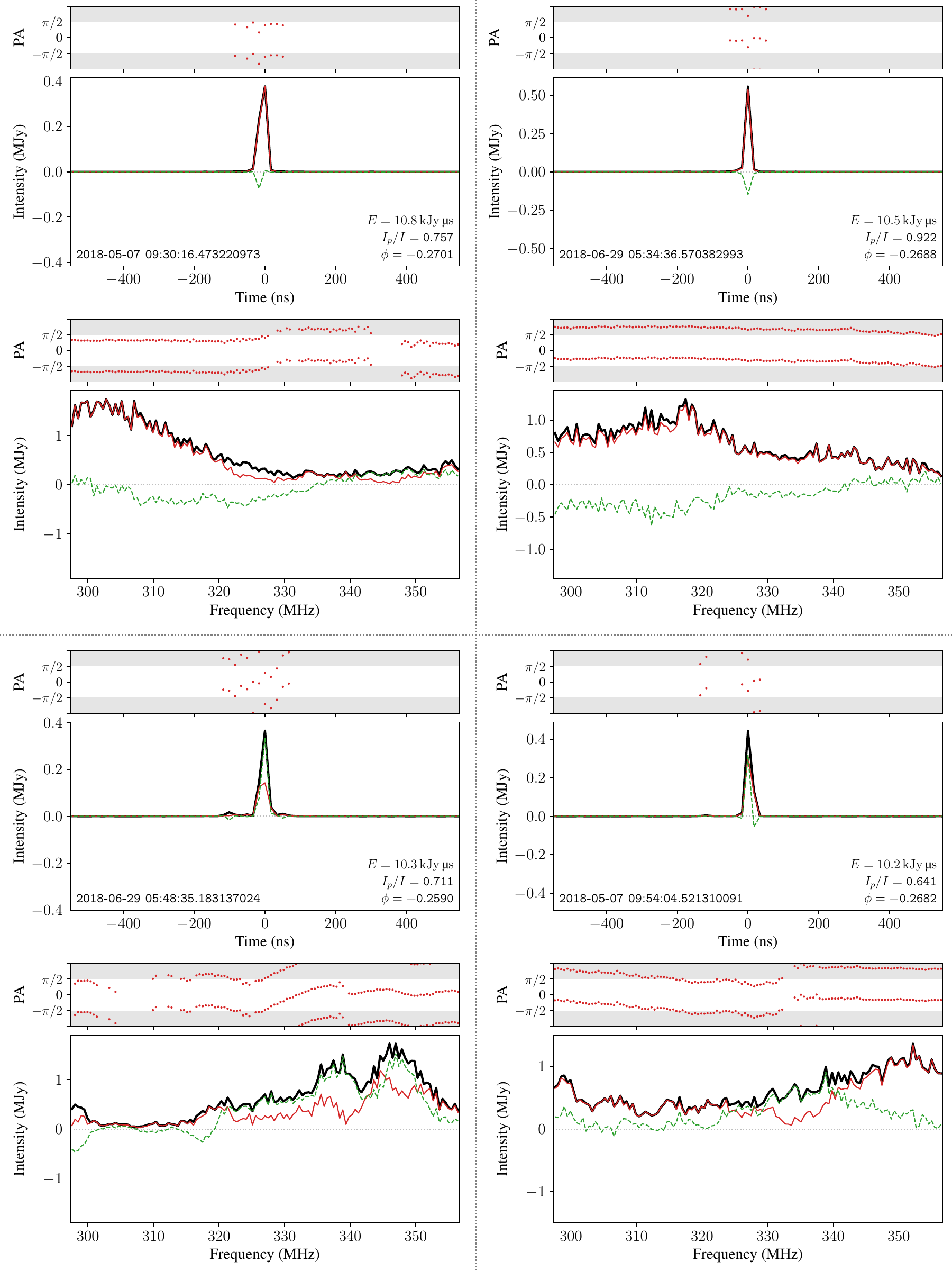}
  \caption{A selection of the most energetic giant pulses. See Figure~\ref{fig:gp_brights} for a full description.}
\end{figure*}

\begin{figure*}
  \centering
  \includegraphics[width=0.9\textwidth,trim=0 0 0 0,clip]{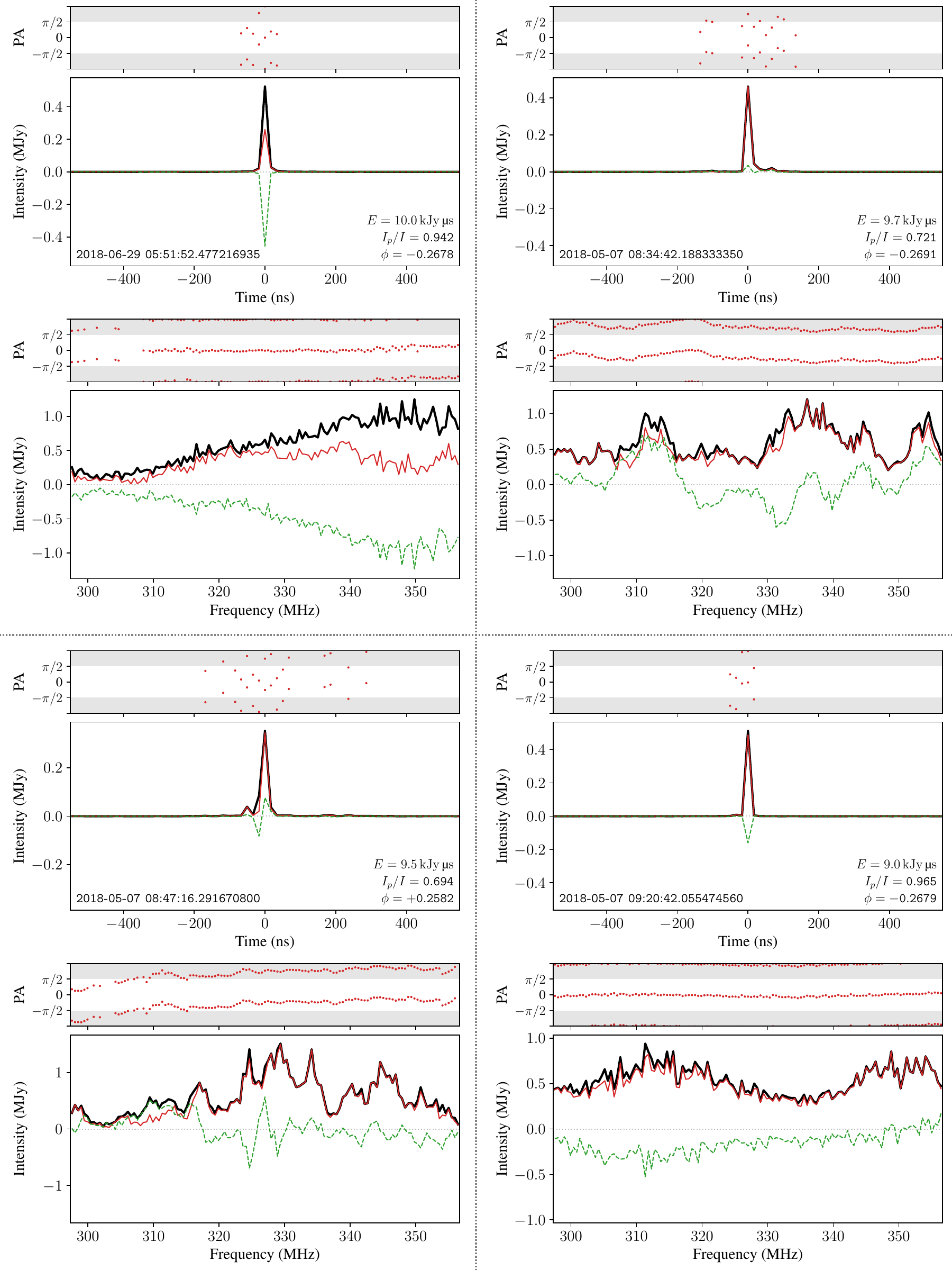}
  \caption{A selection of the most energetic giant pulses. See Figure~\ref{fig:gp_brights} for a full description.}
\end{figure*}

\begin{figure*}
  \centering
  \includegraphics[width=0.9\textwidth,trim=0 0 0 0,clip]{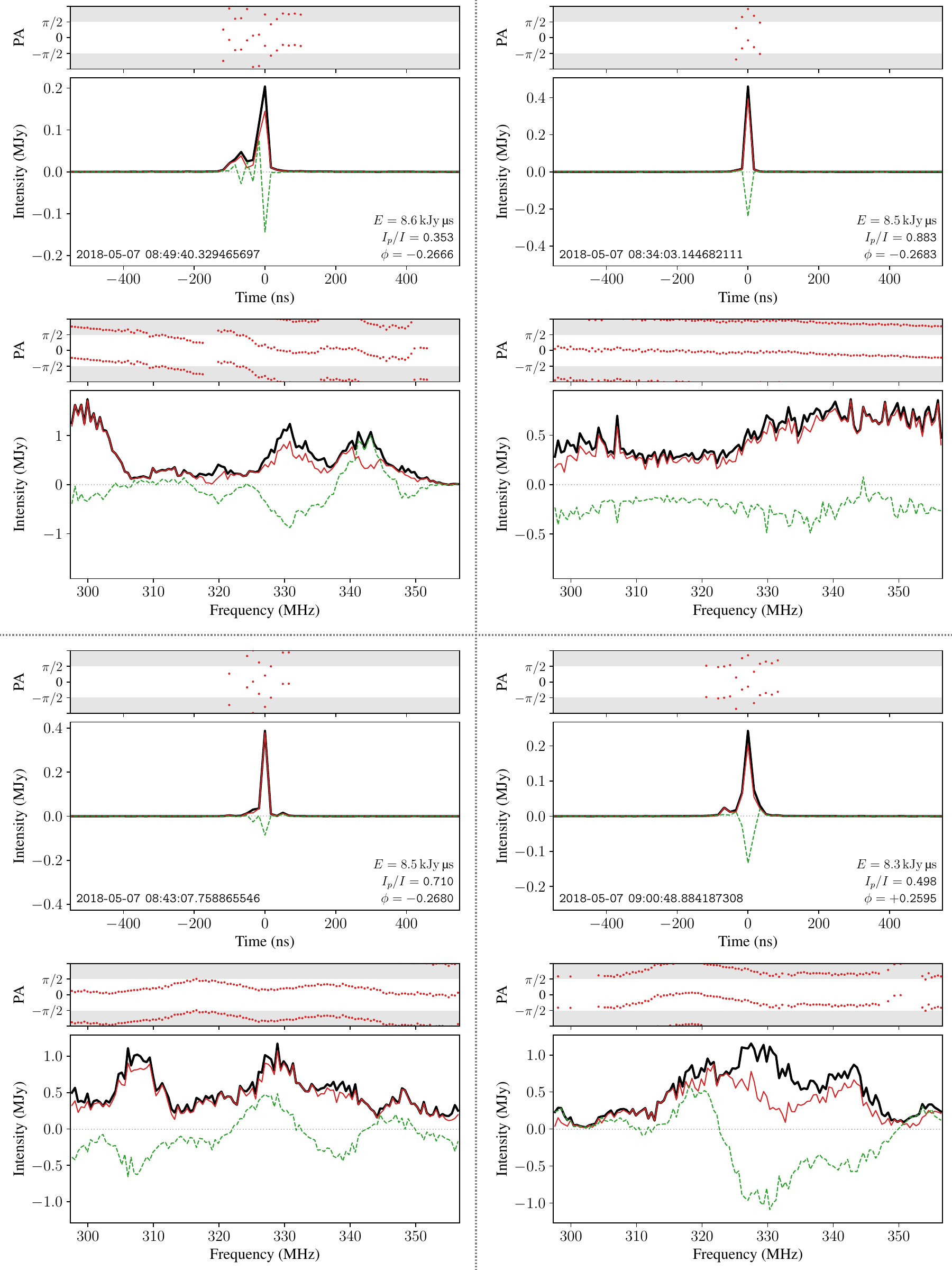}
  \caption{A selection of the most energetic giant pulses. See Figure~\ref{fig:gp_brights} for a full description.}
\end{figure*}

\begin{figure*}
  \centering
  \includegraphics[width=0.9\textwidth,trim=0 0 0 0,clip]{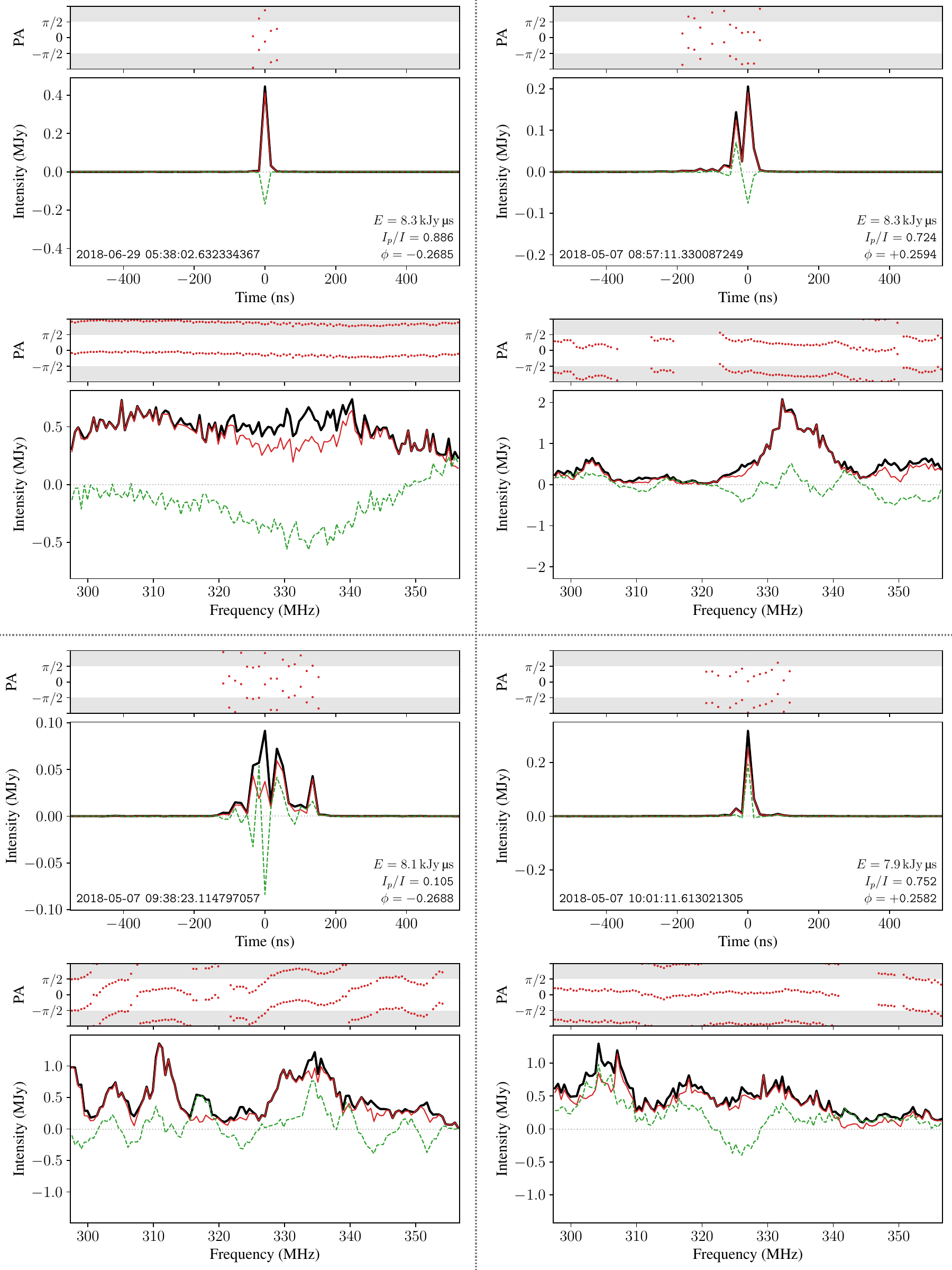}
  \caption{A selection of the most energetic giant pulses. See Figure~\ref{fig:gp_brights} for a full description.}
\end{figure*}

\begin{figure*}
  \centering
  \includegraphics[width=0.9\textwidth,trim=0 0 0 0,clip]{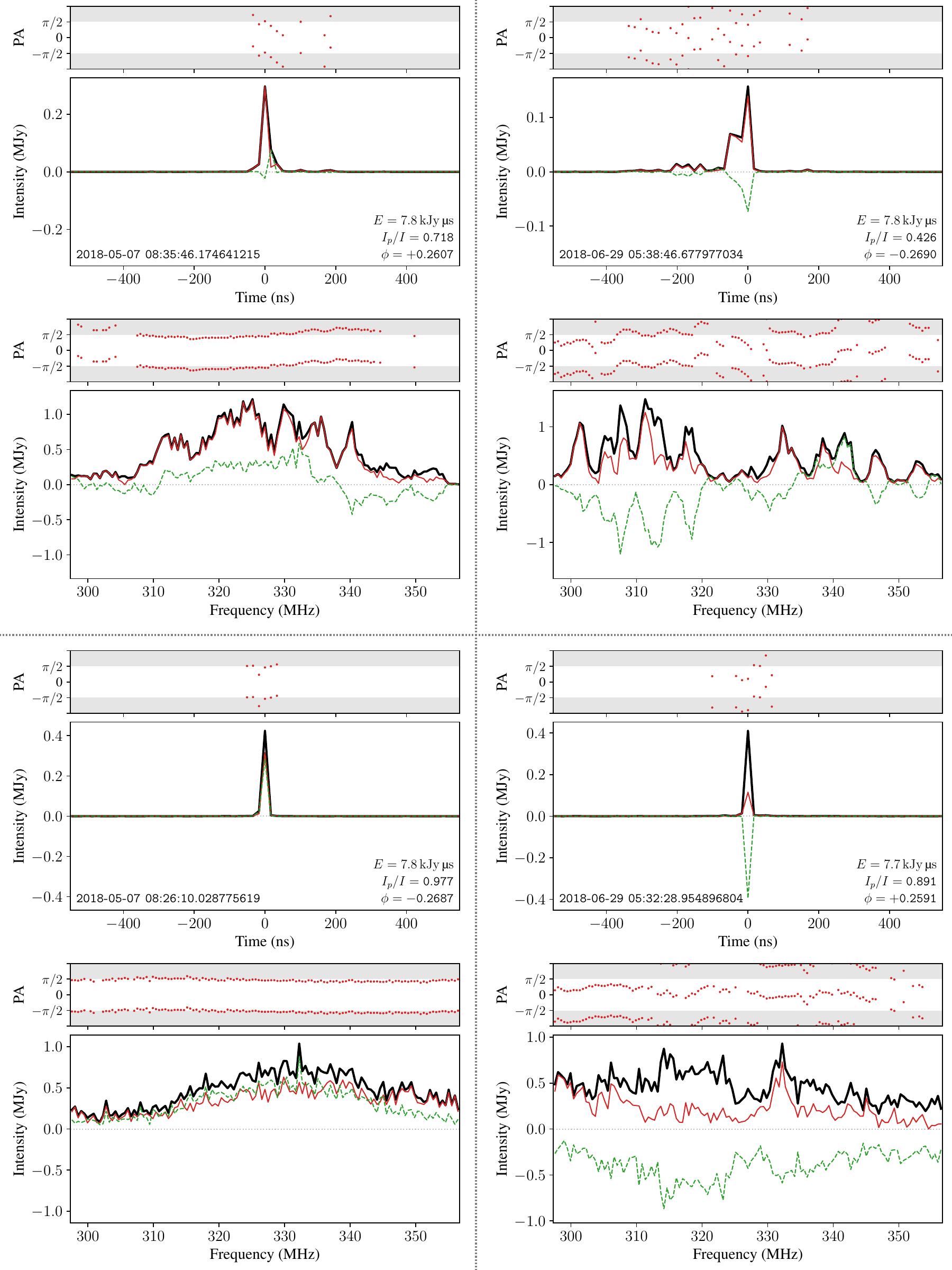}
  \caption{A selection of the most energetic giant pulses. See Figure~\ref{fig:gp_brights} for a full description.}
\end{figure*}

\begin{figure*}
  \centering
  \includegraphics[width=0.9\textwidth,trim=0 0 0 0,clip]{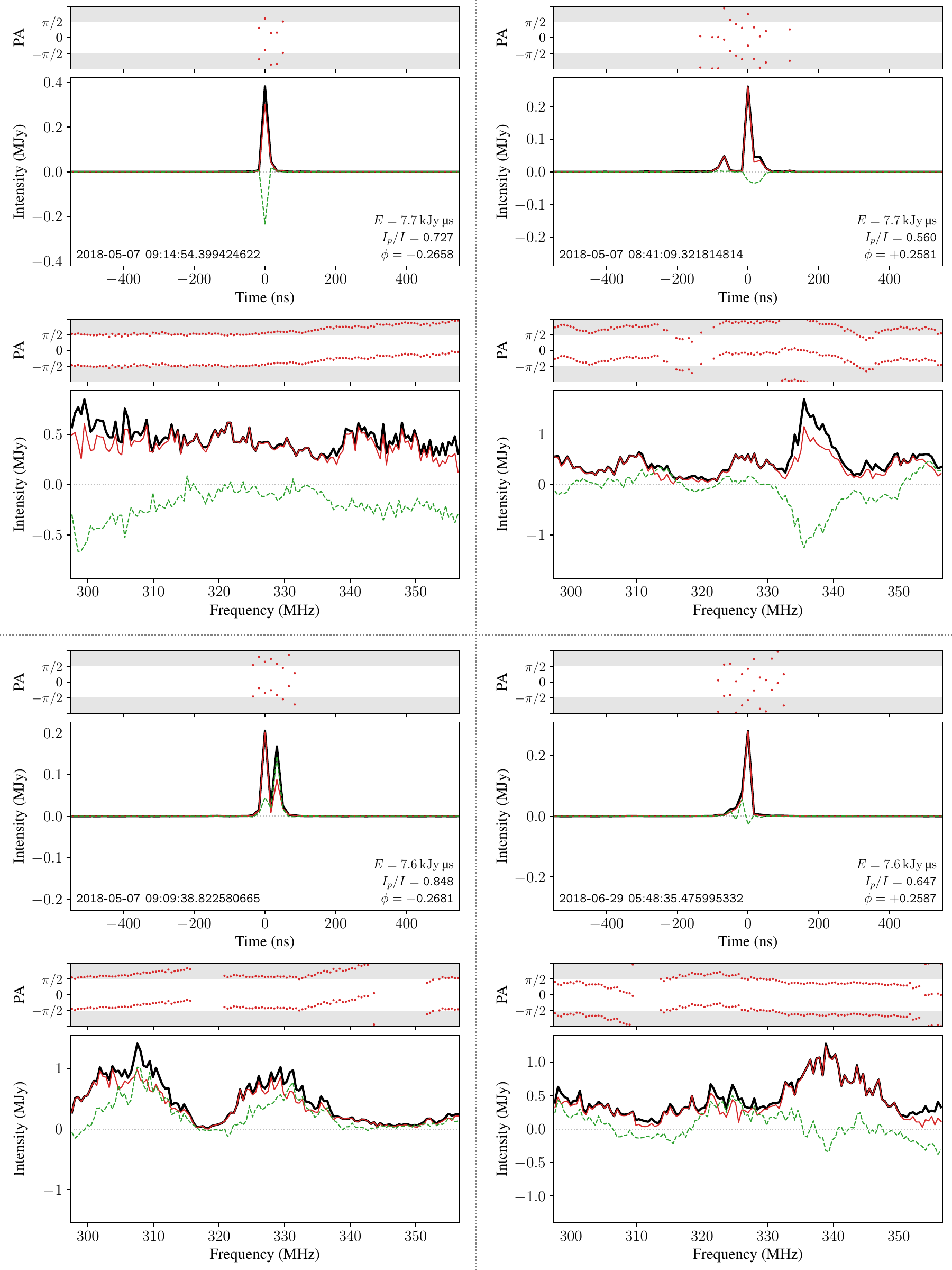}
  \caption{A selection of the most energetic giant pulses. See Figure~\ref{fig:gp_brights} for a full description.}
\end{figure*}

\begin{figure*}
  \centering
  \includegraphics[width=0.9\textwidth,trim=0 0 0 0,clip]{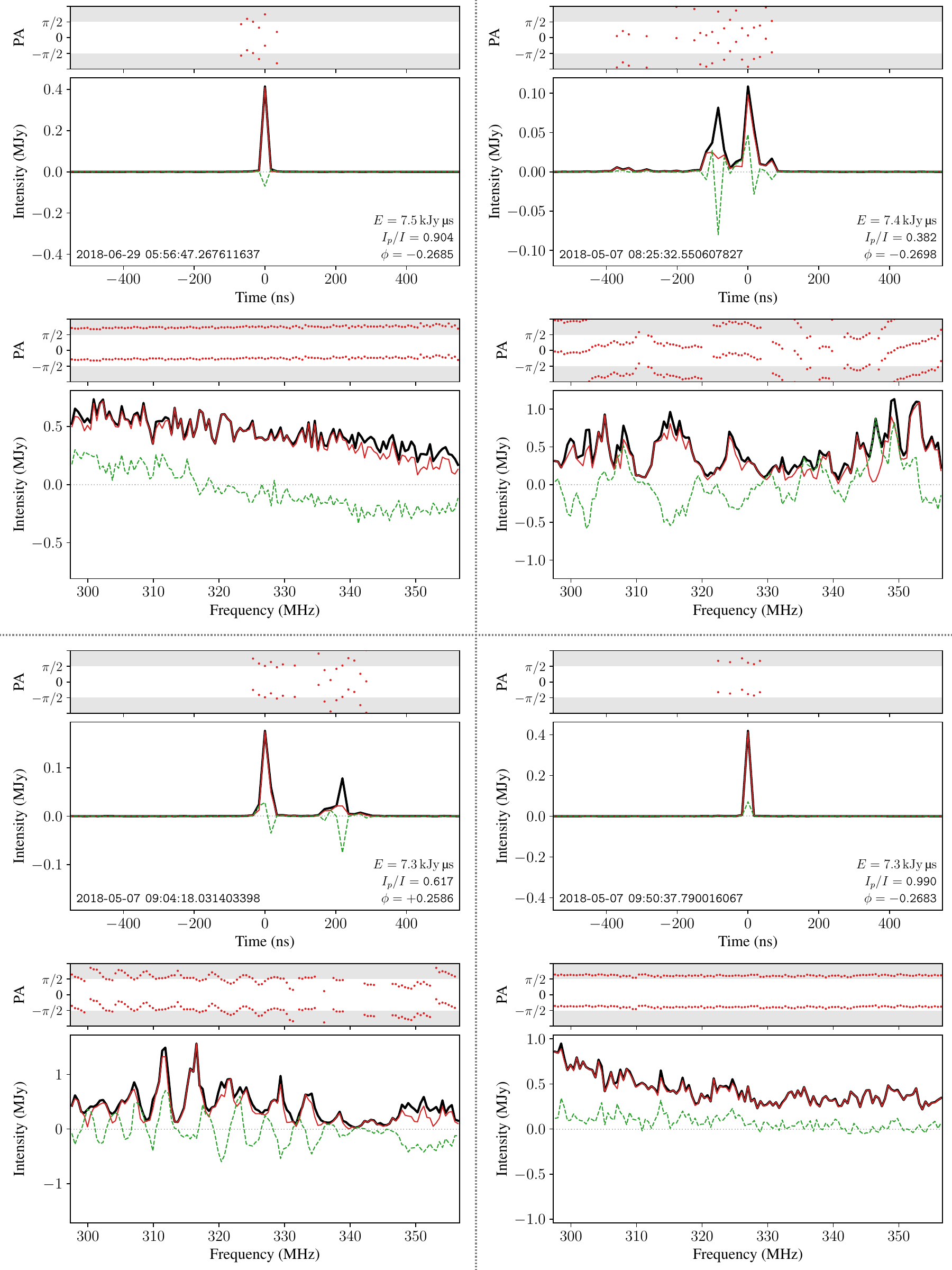}
  \caption{A selection of the most energetic giant pulses. See Figure~\ref{fig:gp_brights} for a full description.}
\end{figure*}

\begin{figure*}
  \centering
  \includegraphics[width=0.9\textwidth,trim=0 0 0 0,clip]{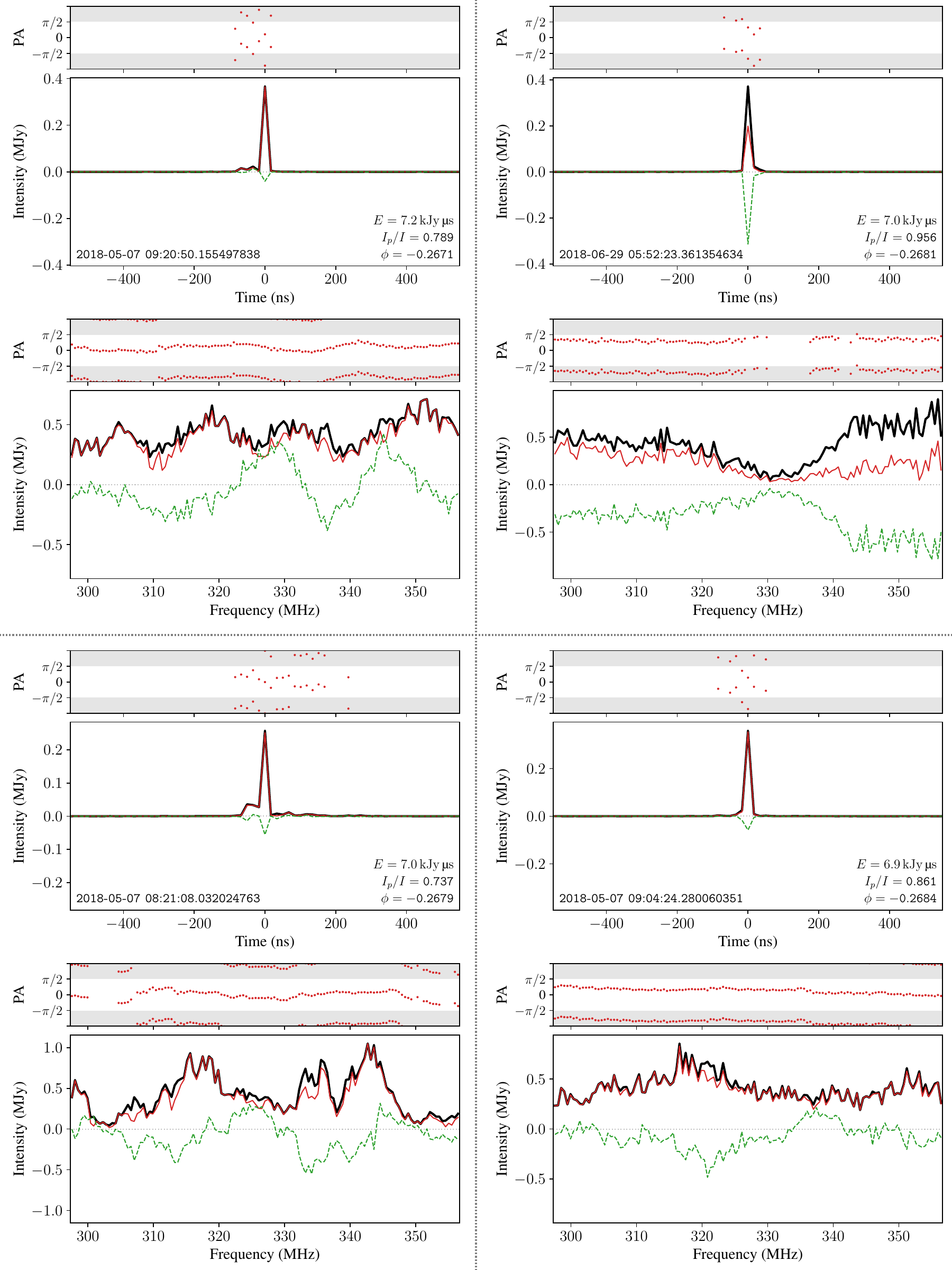}
  \caption{A selection of the most energetic giant pulses. See Figure~\ref{fig:gp_brights} for a full description.}
\end{figure*}

\begin{figure*}
  \centering
  \includegraphics[width=0.9\textwidth,trim=0 0 0 0,clip]{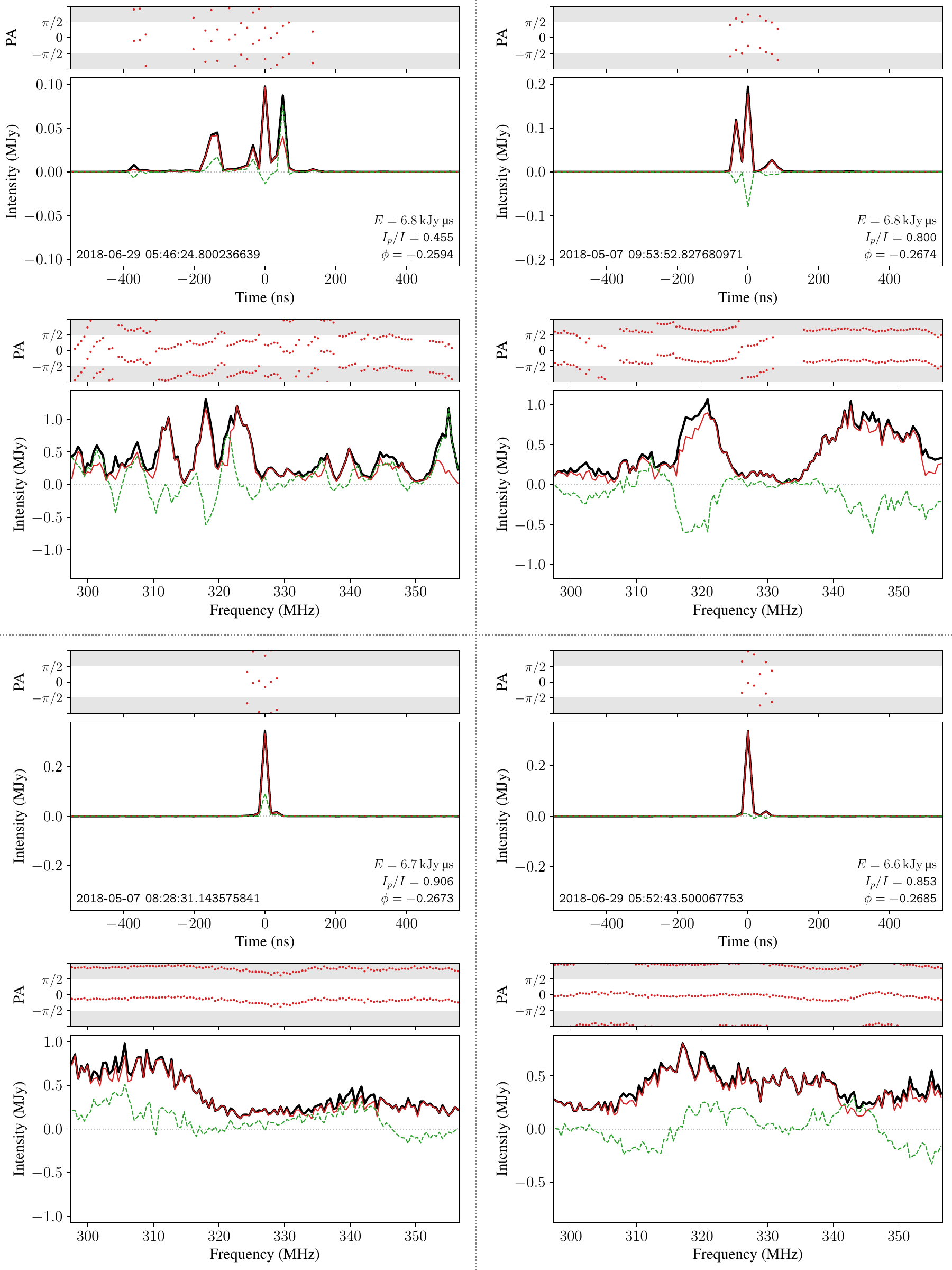}
  \caption{A selection of the most energetic giant pulses. See Figure~\ref{fig:gp_brights} for a full description.}
\end{figure*}

\begin{figure*}
  \centering
  \includegraphics[width=0.9\textwidth,trim=0 0 0 0,clip]{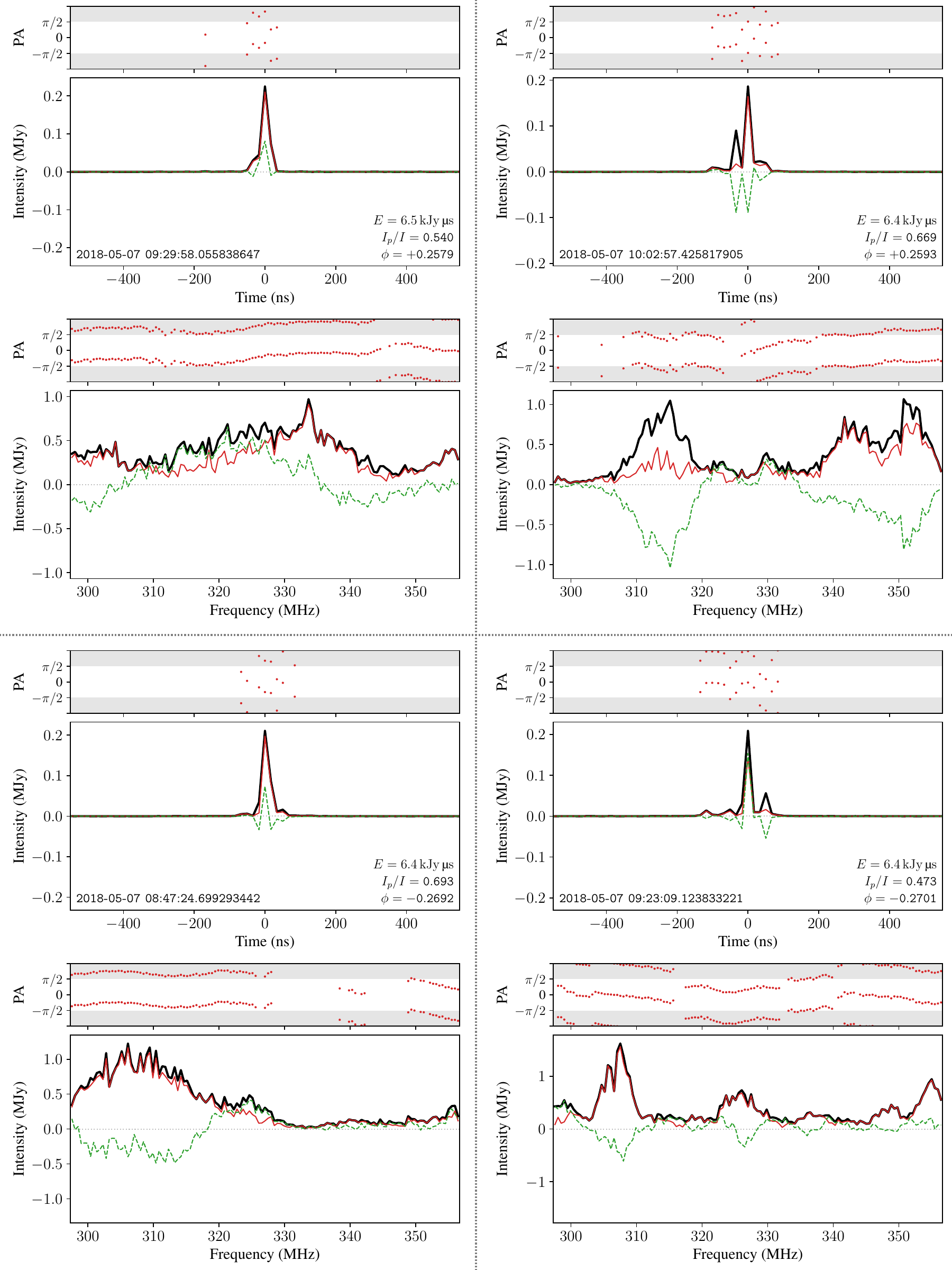}
  \caption{A selection of the most energetic giant pulses. See Figure~\ref{fig:gp_brights} for a full description.}
\end{figure*}

\begin{figure*}
  \centering
  \includegraphics[width=0.9\textwidth,trim=0 0 0 0,clip]{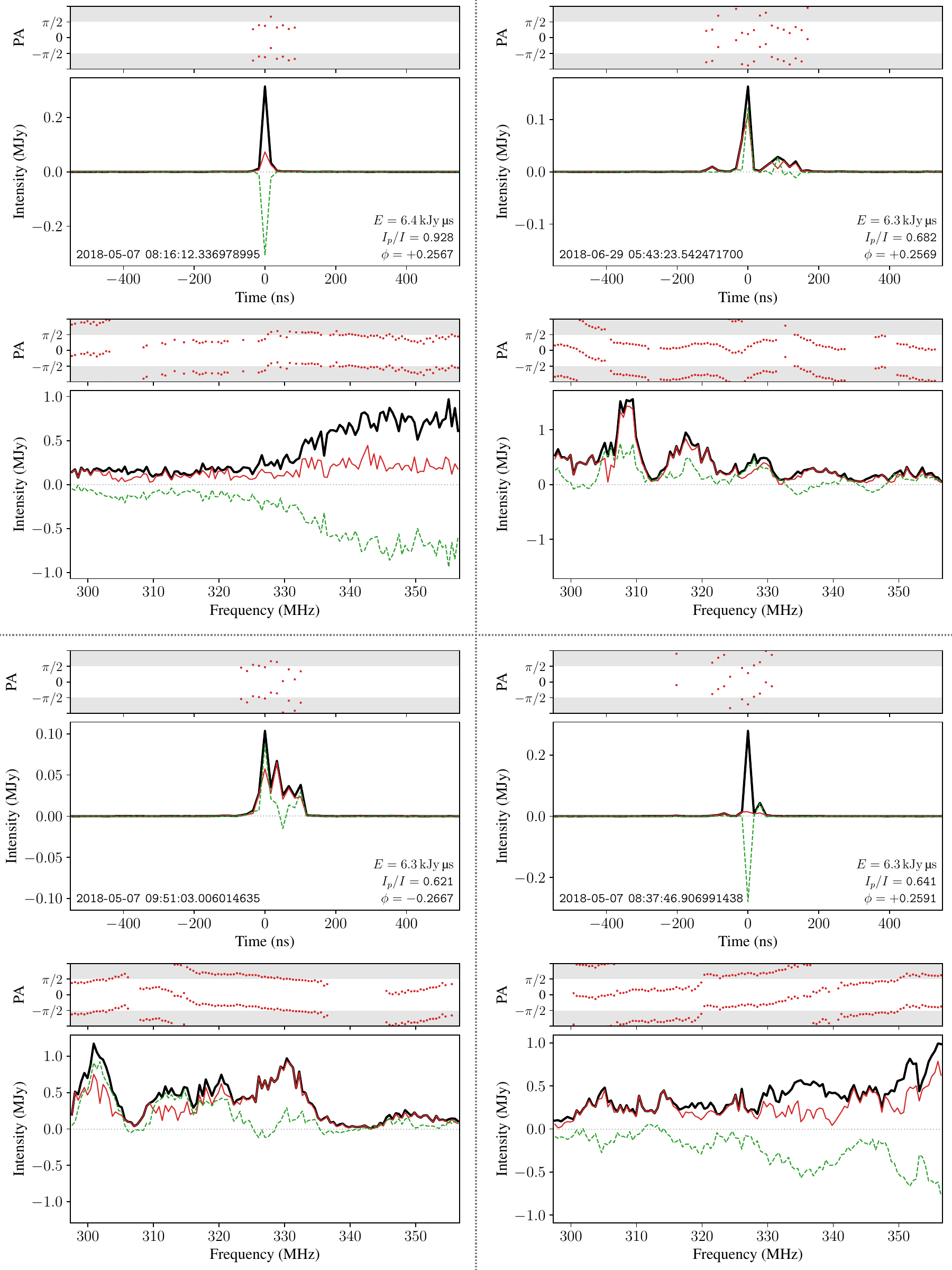}
  \caption{A selection of the most energetic giant pulses. See Figure~\ref{fig:gp_brights} for a full description.}
\end{figure*}

\begin{figure*}
  \centering
  \includegraphics[width=0.9\textwidth,trim=0 0 0 0,clip]{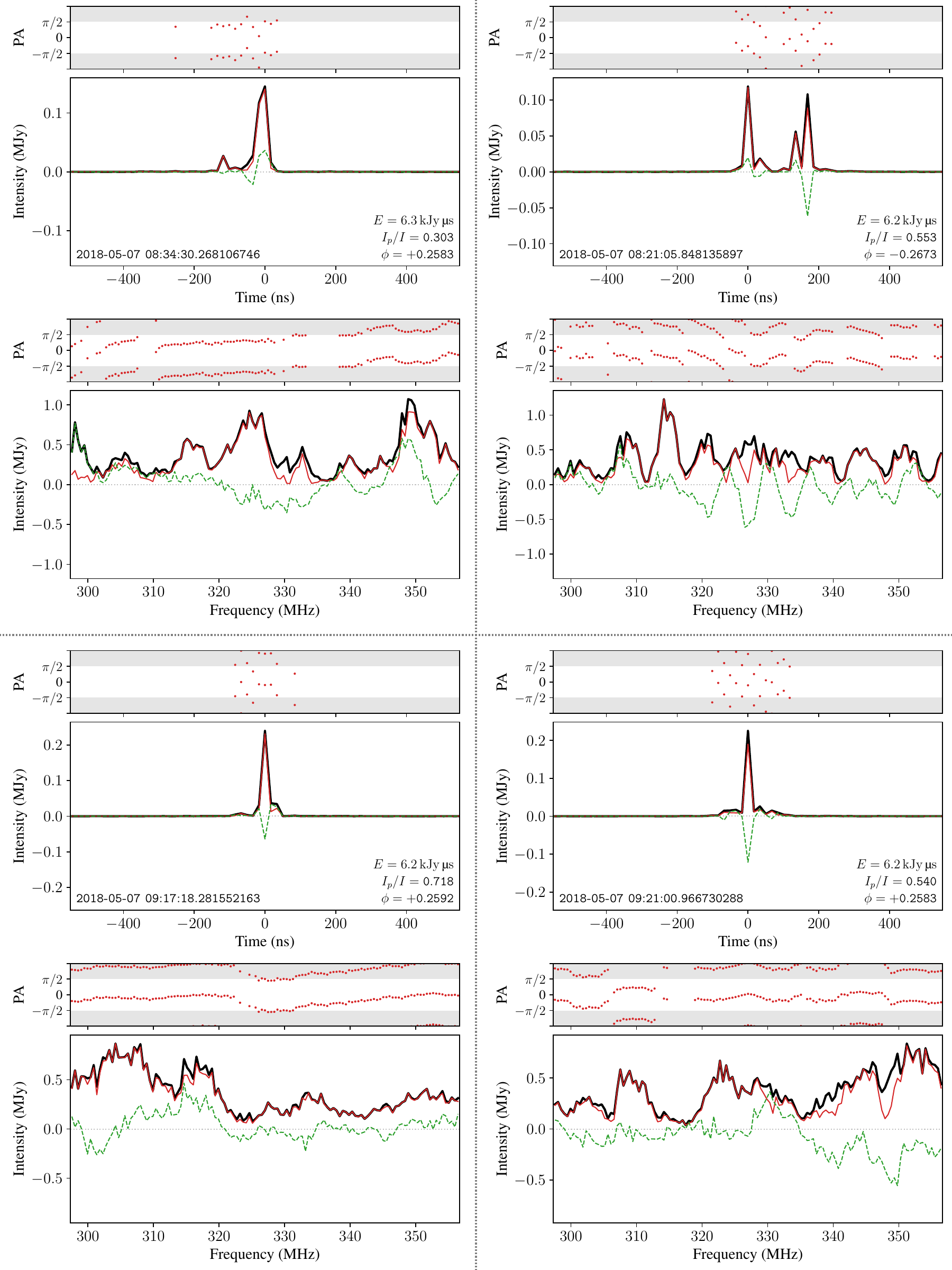}
  \caption{A selection of the most energetic giant pulses. See Figure~\ref{fig:gp_brights} for a full description.}
\end{figure*}

\begin{figure*}
  \centering
  \includegraphics[width=0.9\textwidth,trim=0 0 0 0,clip]{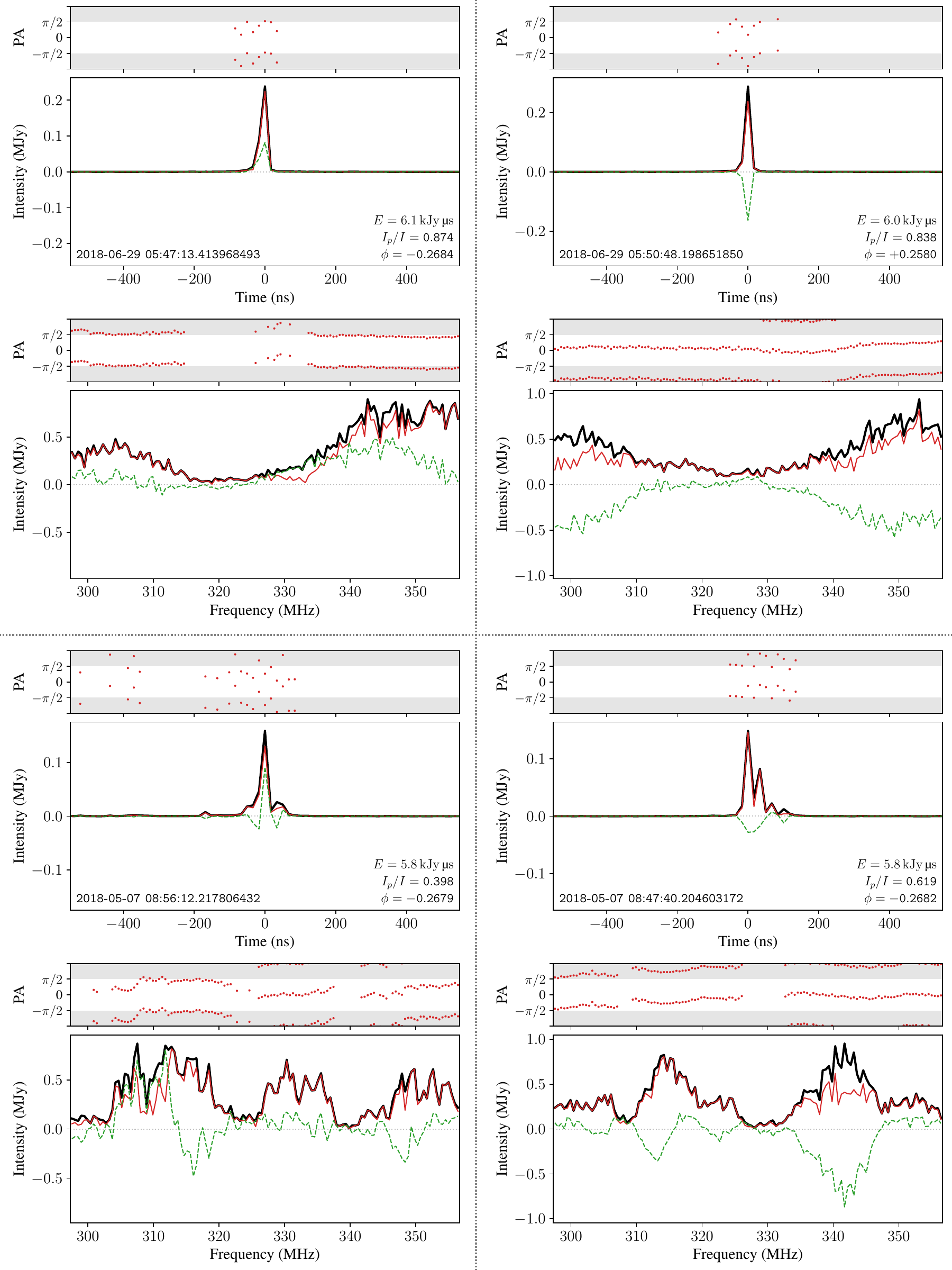}
  \caption{A selection of the most energetic giant pulses. See Figure~\ref{fig:gp_brights} for a full description.}
\end{figure*}

\begin{figure*}
  \centering
  \includegraphics[width=0.9\textwidth,trim=0 0 0 0,clip]{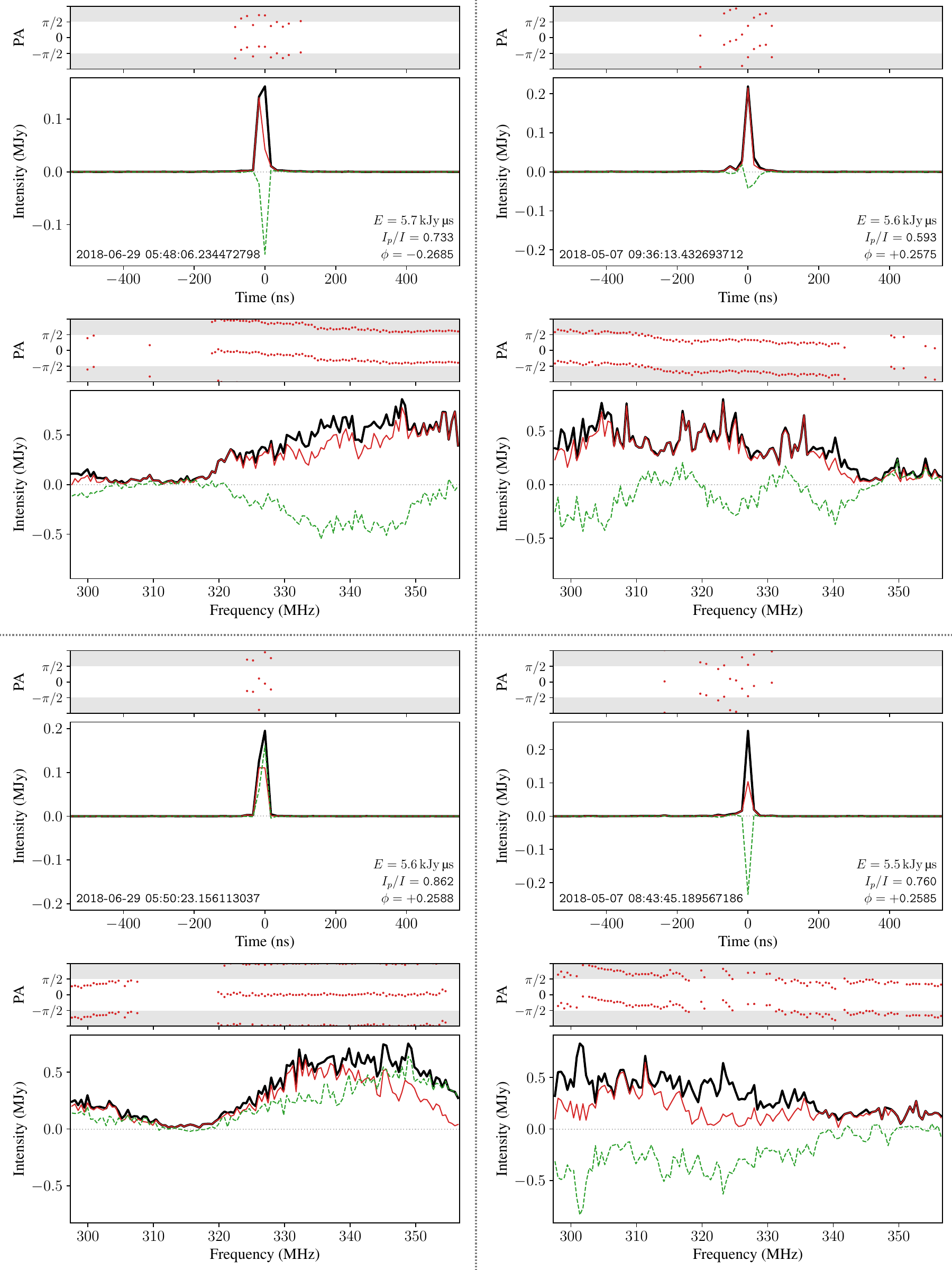}
  \caption{A selection of the most energetic giant pulses. See Figure~\ref{fig:gp_brights} for a full description.}
\end{figure*}

\begin{figure*}
  \centering
  \includegraphics[width=0.9\textwidth,trim=0 0 0 0,clip]{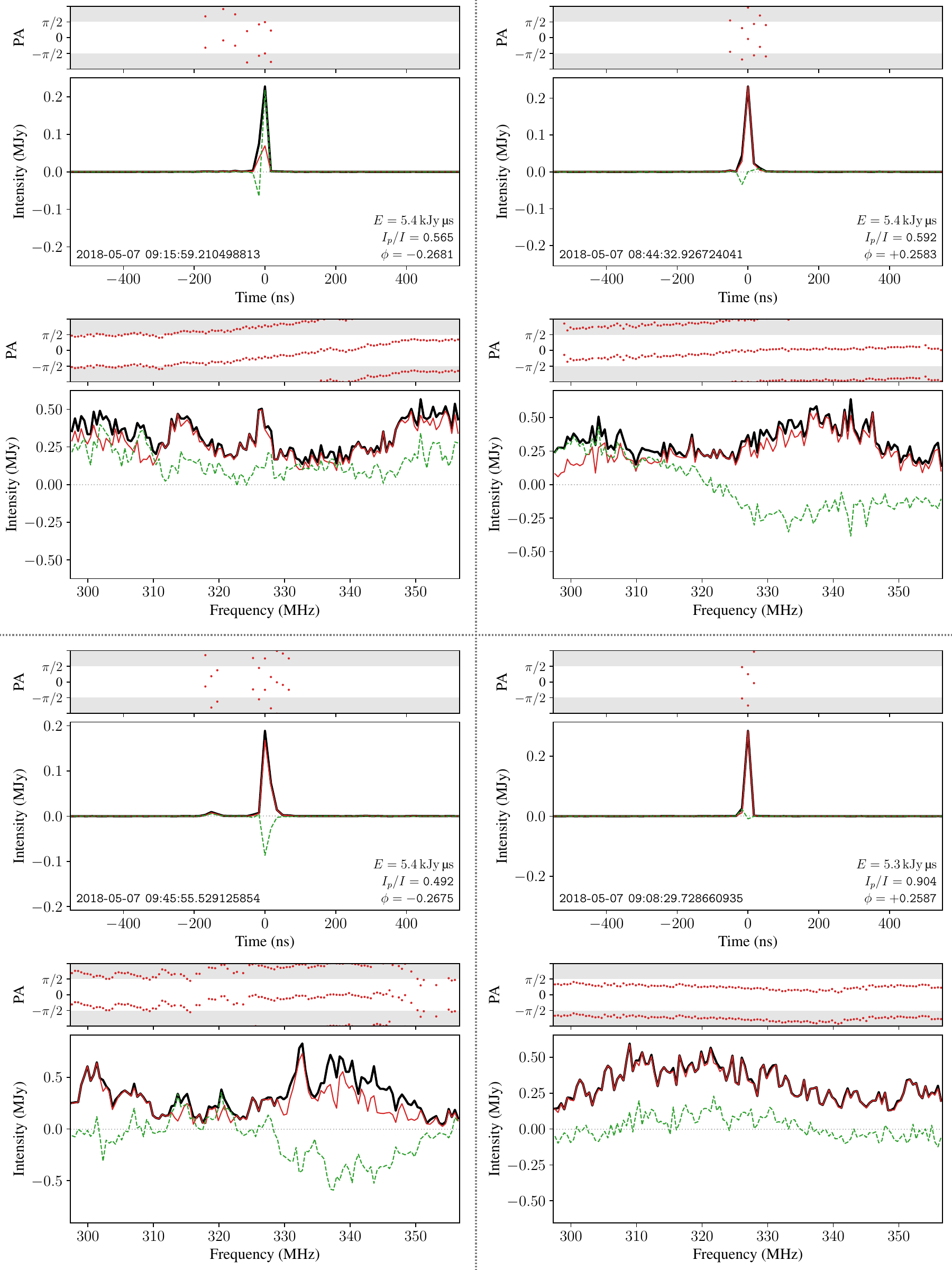}
  \caption{A selection of the most energetic giant pulses. See Figure~\ref{fig:gp_brights} for a full description.}
\end{figure*}

\begin{figure*}
  \centering
  \includegraphics[width=0.9\textwidth,trim=0 0 0 0,clip]{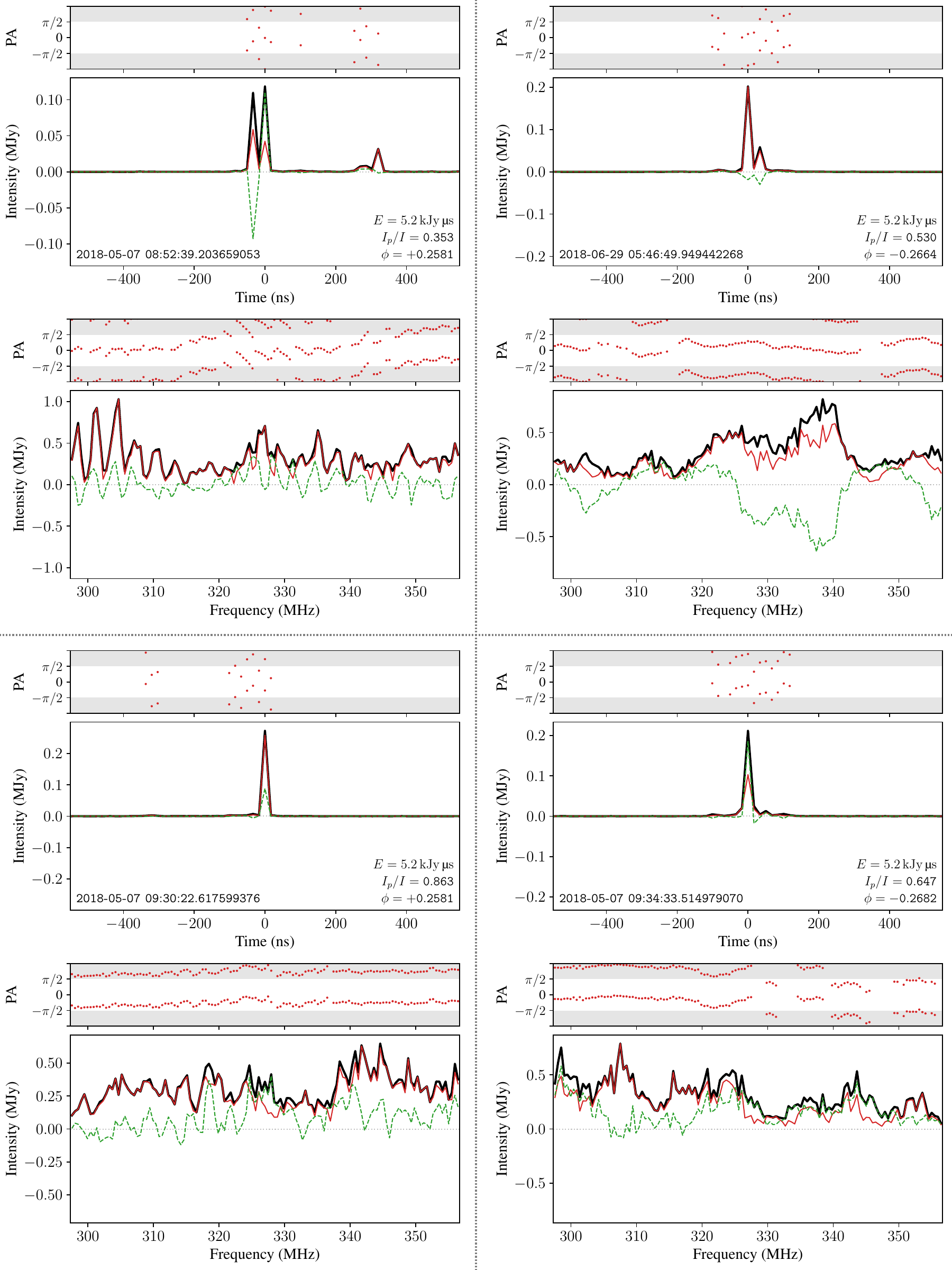}
  \caption{A selection of the most energetic giant pulses. See Figure~\ref{fig:gp_brights} for a full description.}
\end{figure*}

\begin{figure*}
  \centering
  \includegraphics[width=0.9\textwidth,trim=0 0 0 0,clip]{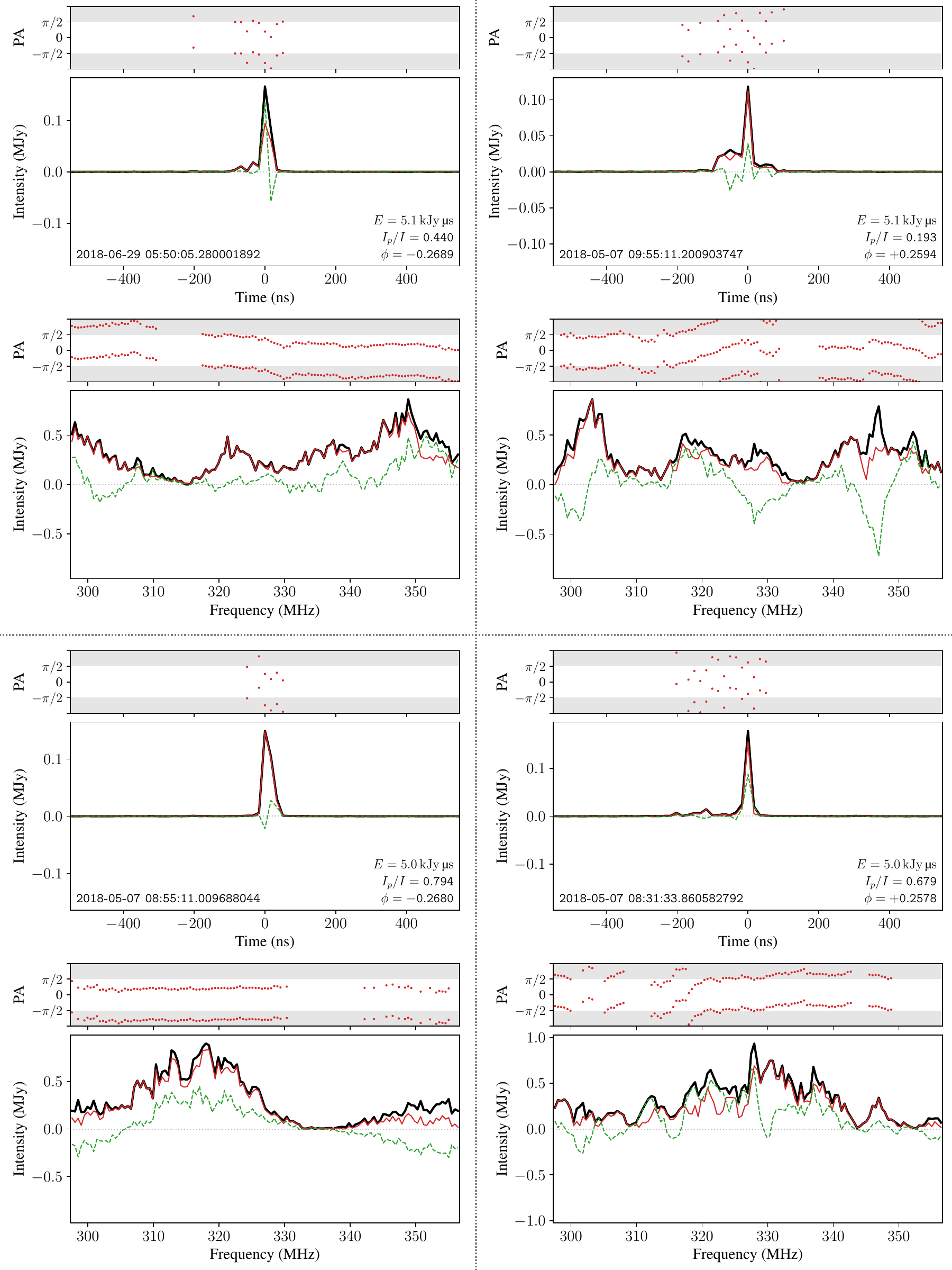}
  \caption{A selection of the most energetic giant pulses. See Figure~\ref{fig:gp_brights} for a full description.}
\end{figure*}

\begin{figure*}
  \centering
  \includegraphics[width=0.9\textwidth,trim=0 0 0 0,clip]{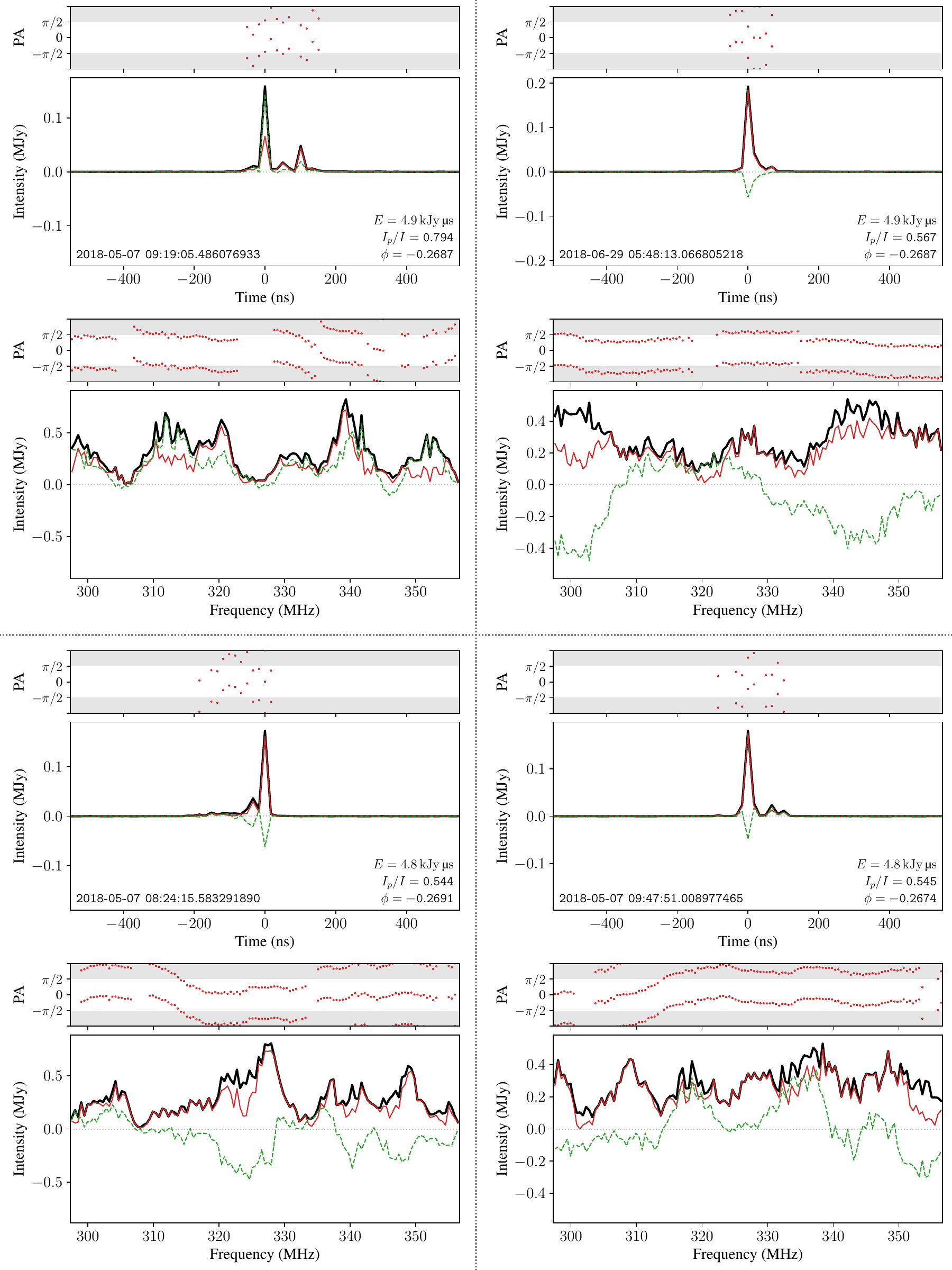}
  \caption{A selection of the most energetic giant pulses. See Figure~\ref{fig:gp_brights} for a full description.}
\end{figure*}

\begin{figure*}
  \centering
  \includegraphics[width=0.9\textwidth,trim=0 0 0 0,clip]{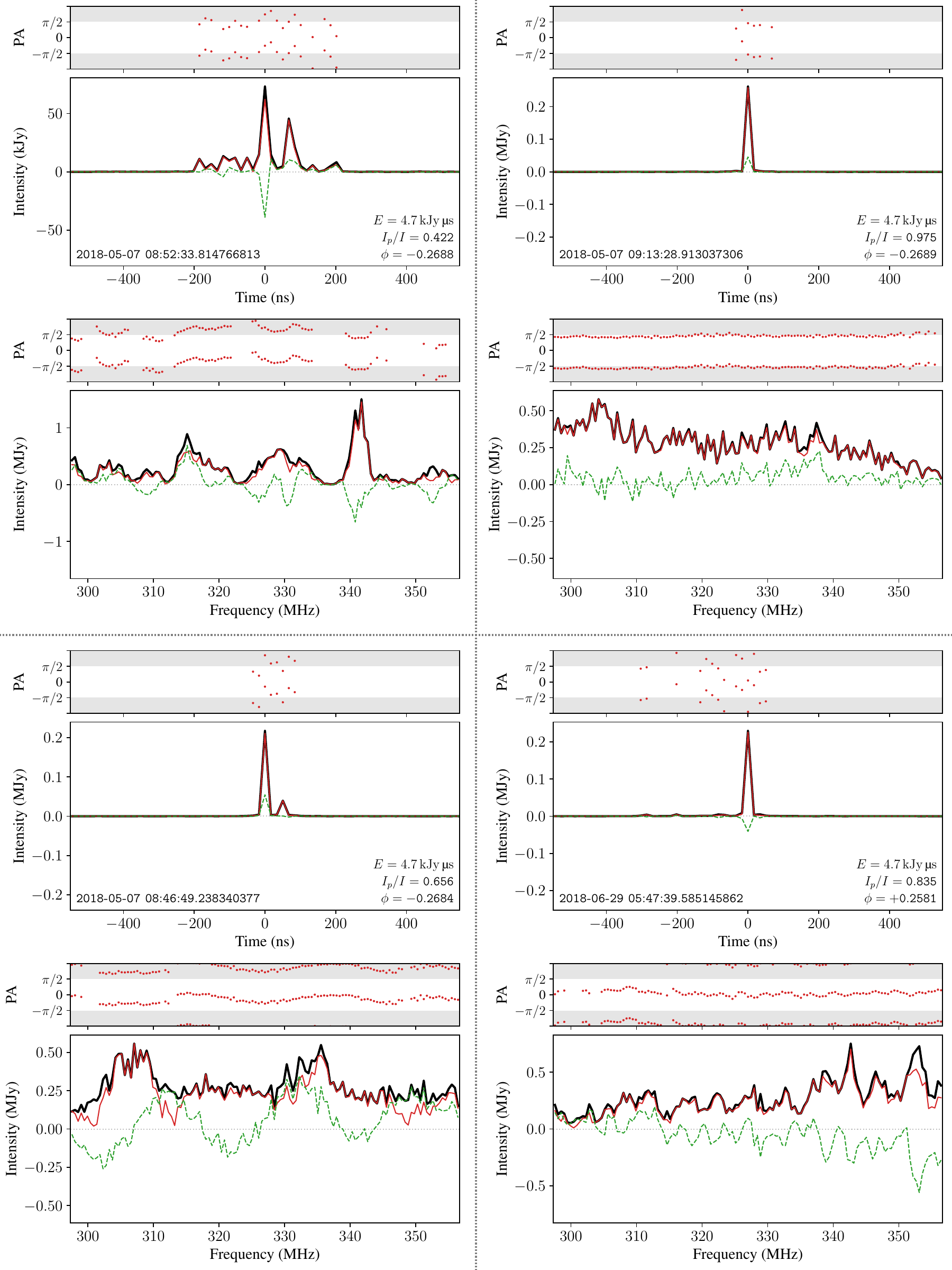}
  \caption{A selection of the most energetic giant pulses. See Figure~\ref{fig:gp_brights} for a full description.}
\end{figure*}

\begin{figure*}
  \centering
  \includegraphics[width=0.9\textwidth,trim=0 0 0 0,clip]{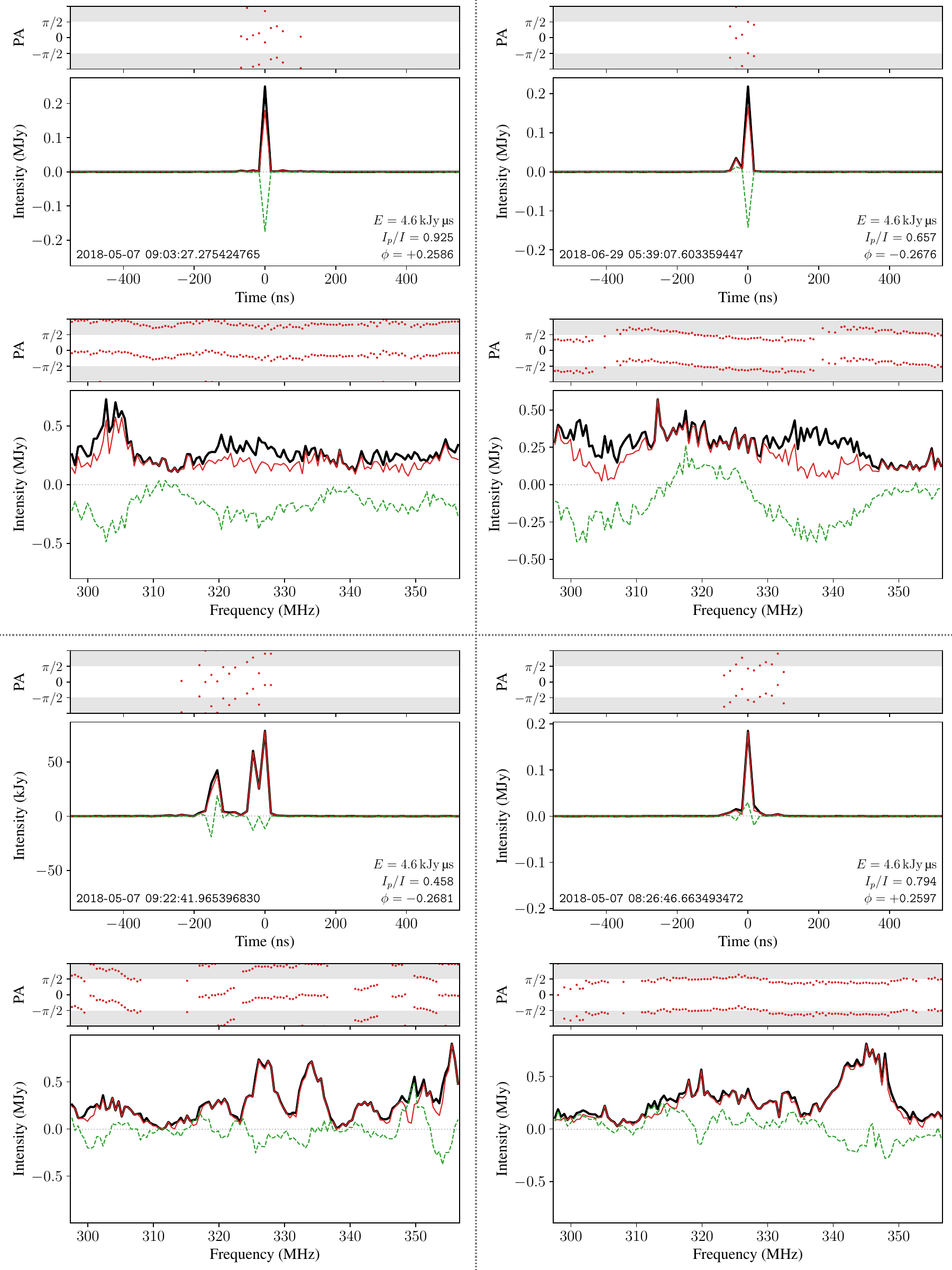}
  \caption{A selection of the most energetic giant pulses. See Figure~\ref{fig:gp_brights} for a full description.}
\end{figure*}

\begin{figure*}
  \centering
  \includegraphics[width=0.9\textwidth,trim=0 0 0 0,clip]{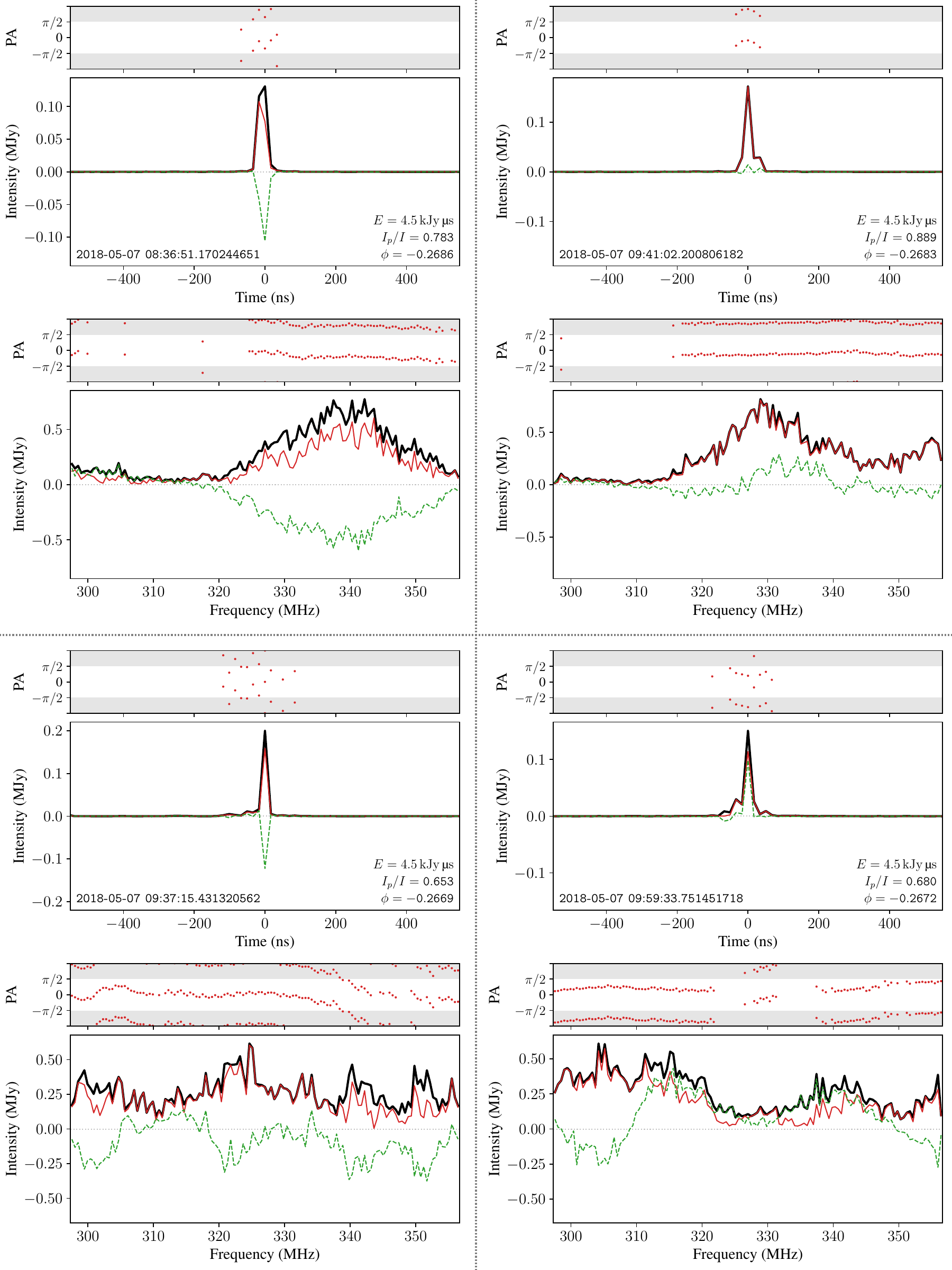}
  \caption{A selection of the most energetic giant pulses. See Figure~\ref{fig:gp_brights} for a full description.}
\end{figure*}

\begin{figure*}
  \centering
  \includegraphics[width=0.9\textwidth,trim=0 0 0 0,clip]{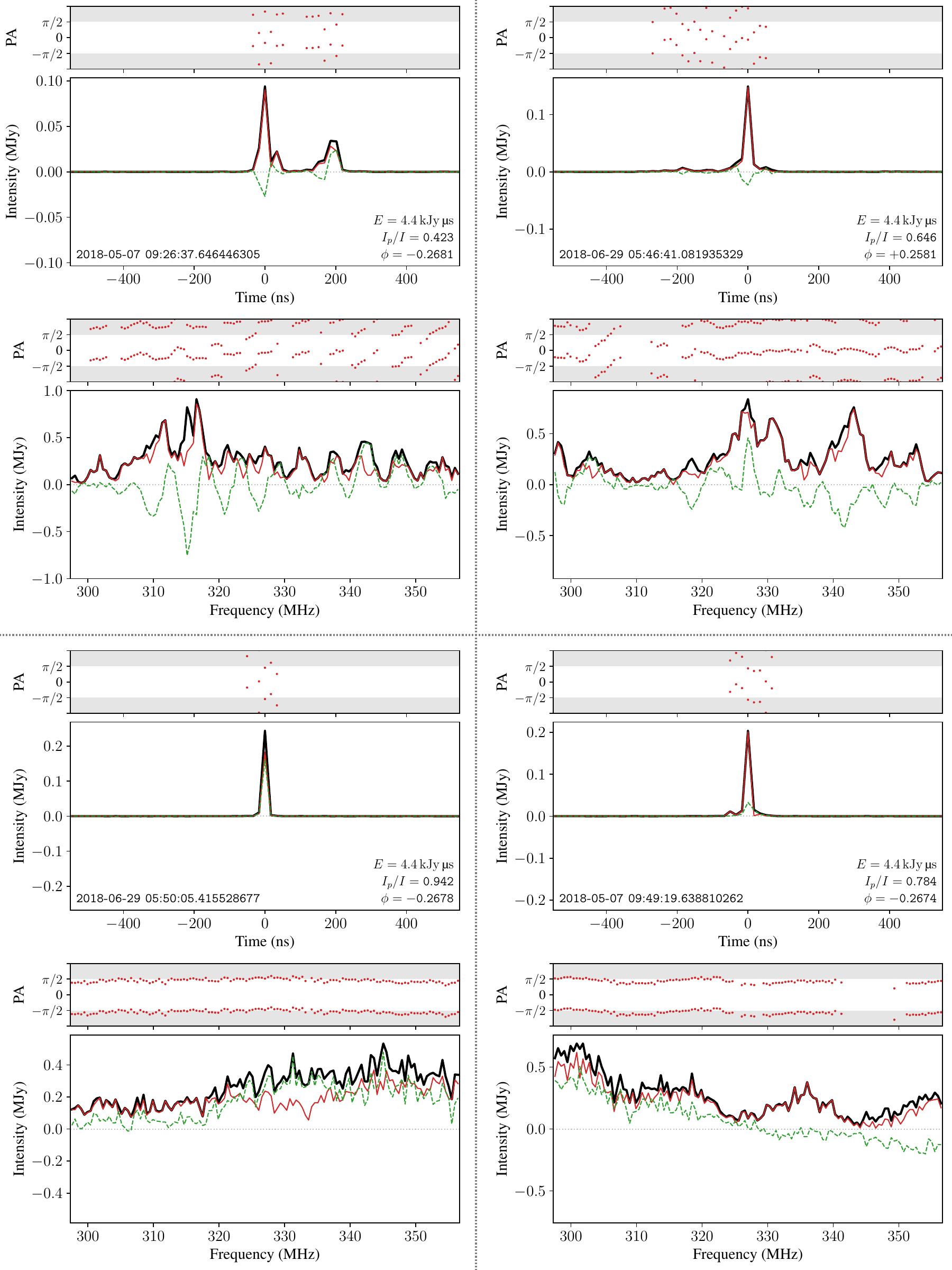}
  \caption{A selection of the most energetic giant pulses. See Figure~\ref{fig:gp_brights} for a full description.}
\end{figure*}

\begin{figure*}
  \centering
  \includegraphics[width=0.9\textwidth,trim=0 0 0 0,clip]{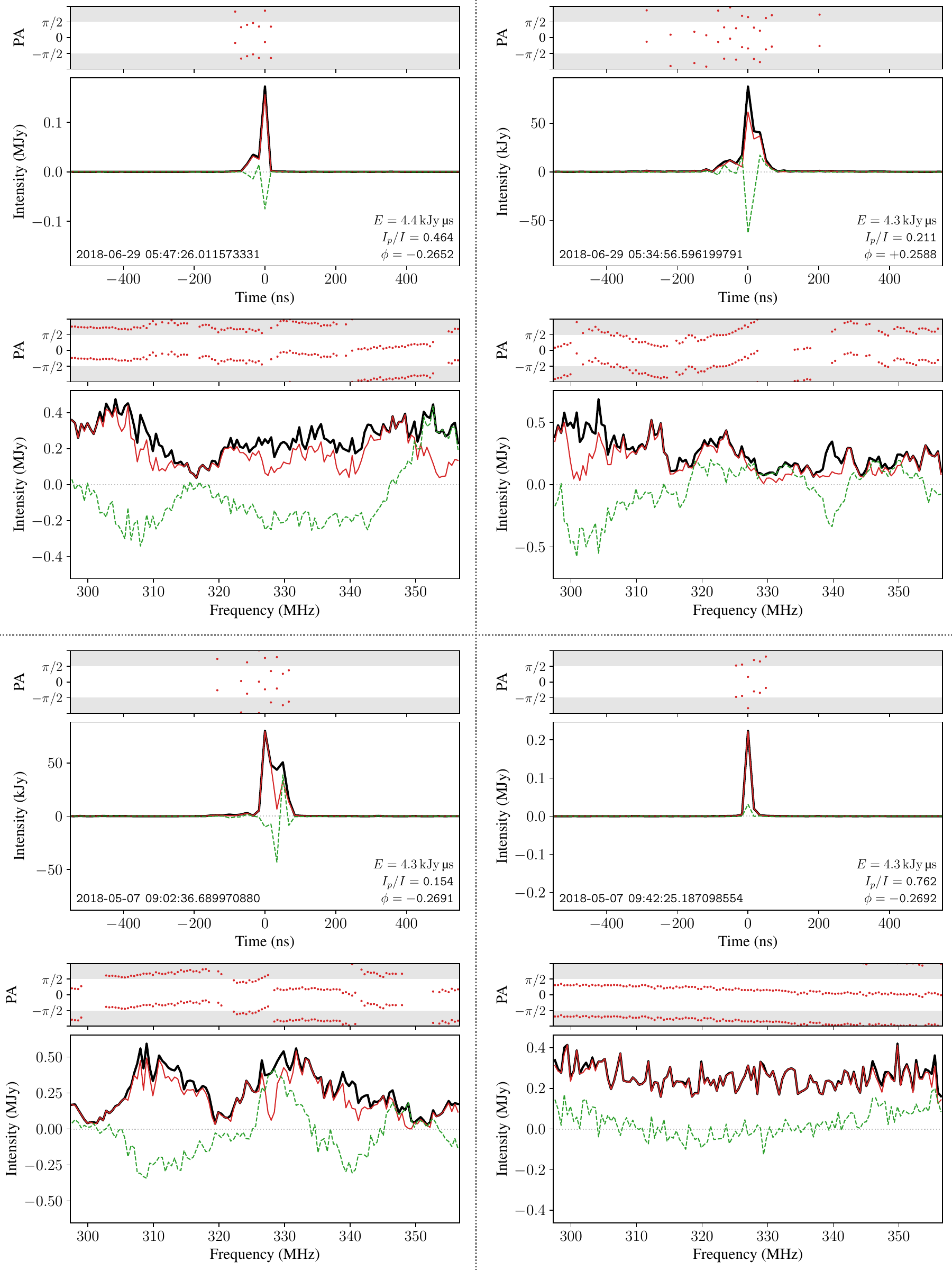}
  \caption{A selection of the most energetic giant pulses. See Figure~\ref{fig:gp_brights} for a full description.}
\end{figure*}

\begin{figure*}
  \centering
  \includegraphics[width=0.9\textwidth,trim=0 0 0 0,clip]{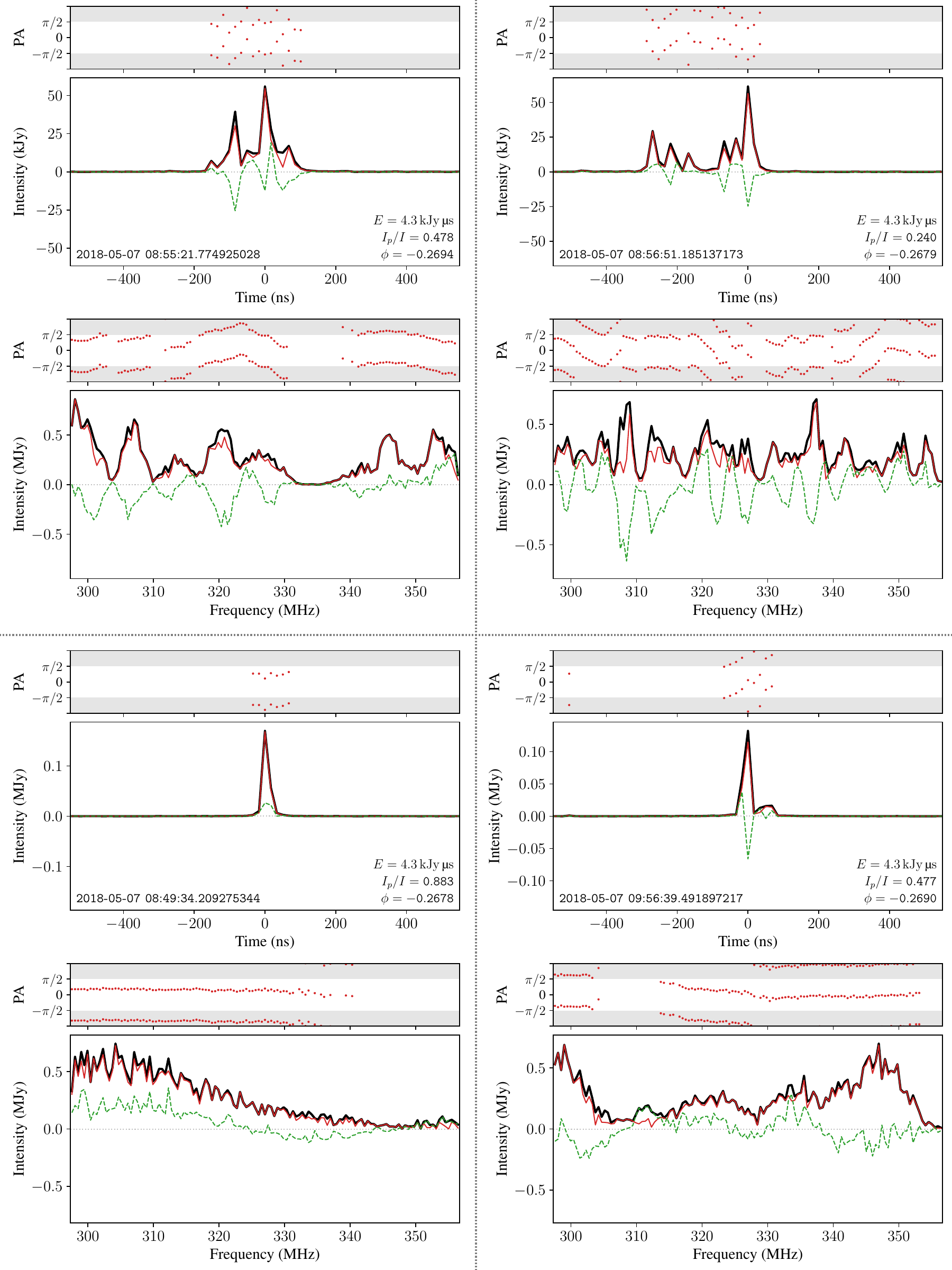}
  \caption{A selection of the most energetic giant pulses. See Figure~\ref{fig:gp_brights} for a full description.}
\end{figure*}

\begin{figure*}
  \centering
  \includegraphics[width=0.9\textwidth,trim=0 0 0 0,clip]{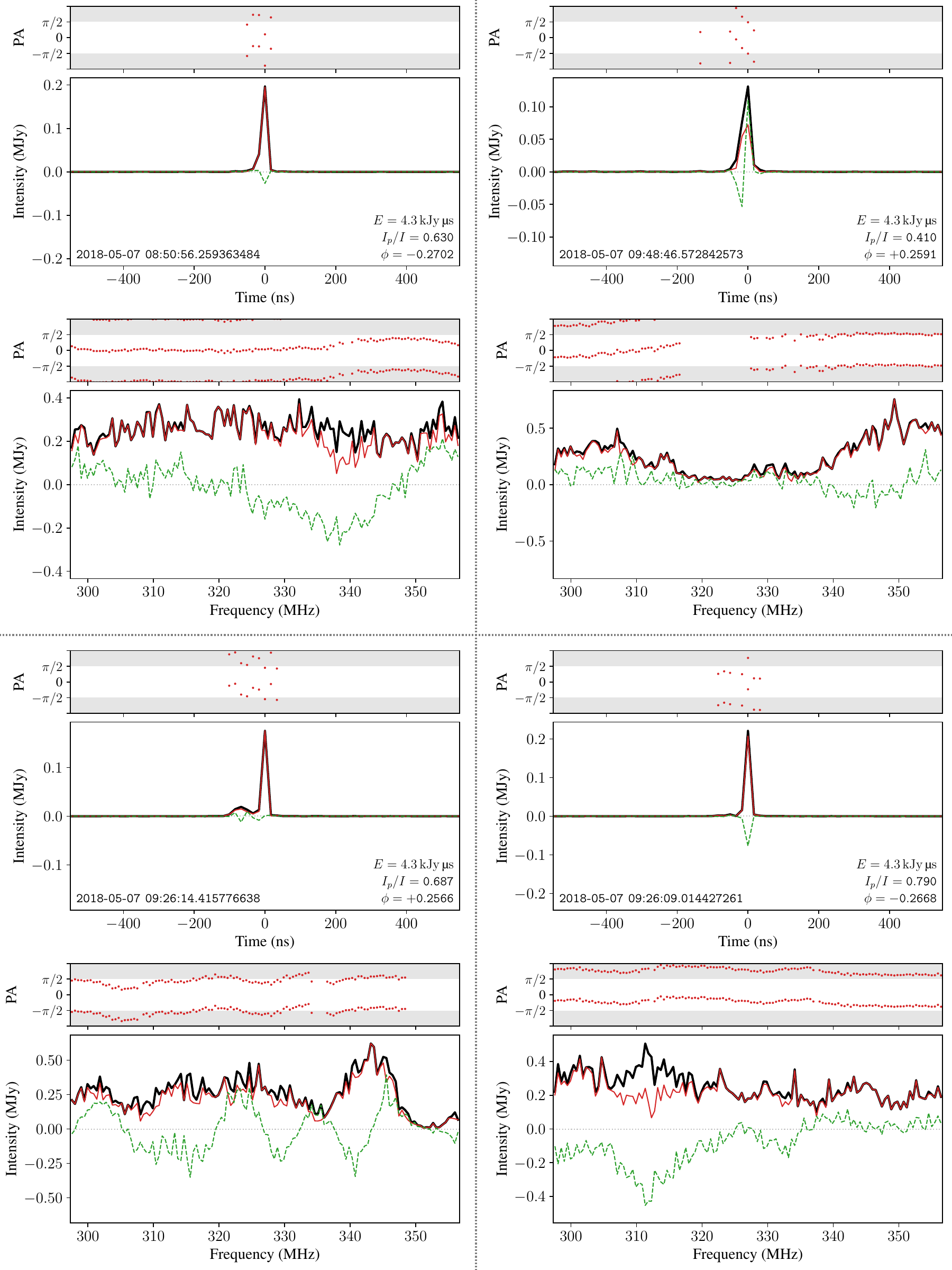}
  \caption{A selection of the most energetic giant pulses. See Figure~\ref{fig:gp_brights} for a full description.}
\end{figure*}

\begin{figure*}
  \centering
  \includegraphics[width=0.9\textwidth,trim=0 0 0 0,clip]{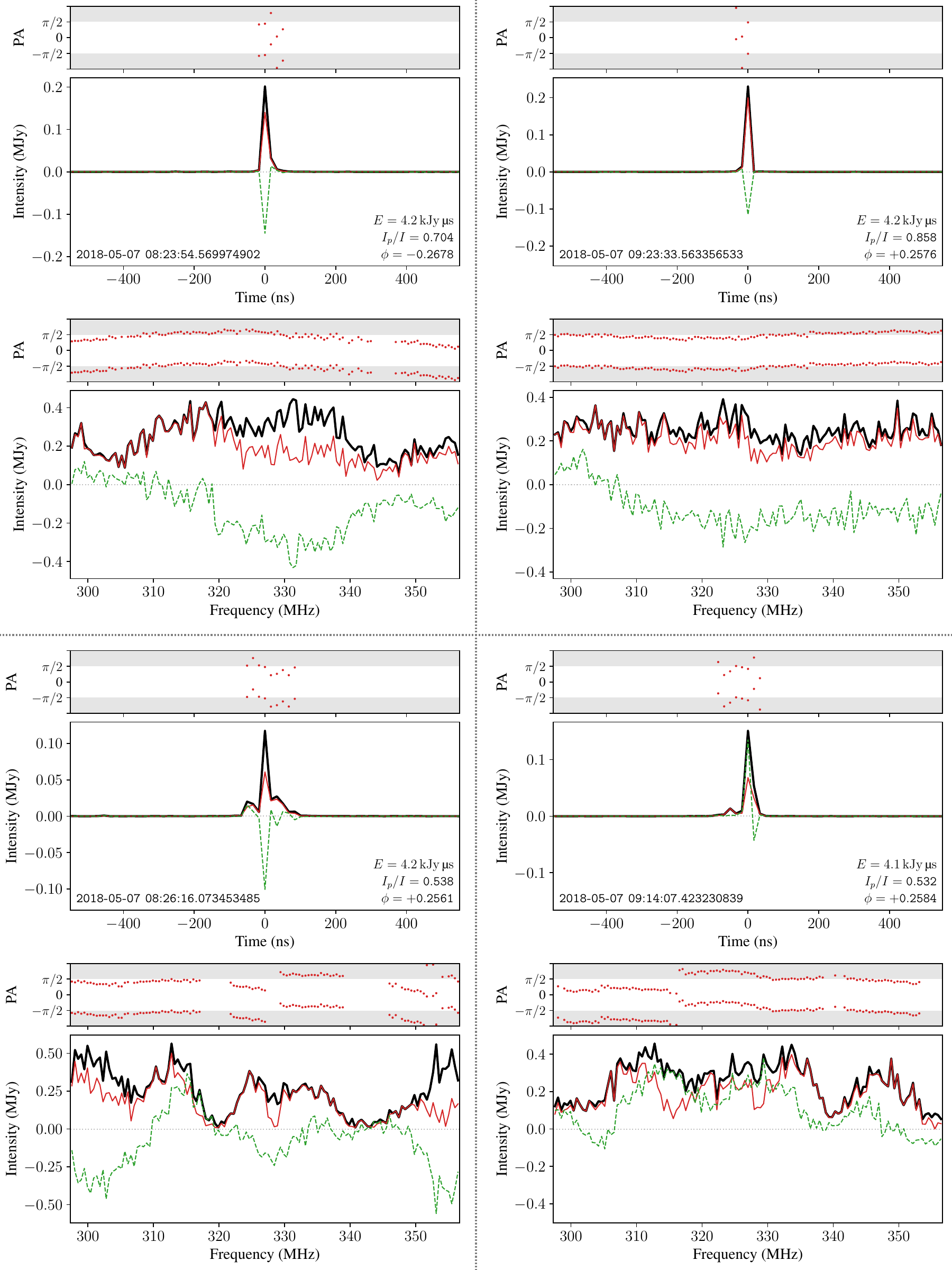}
  \caption{A selection of the most energetic giant pulses. See Figure~\ref{fig:gp_brights} for a full description.}
\end{figure*}

\begin{figure*}
  \centering
  \includegraphics[width=0.9\textwidth,trim=0 0 0 0,clip]{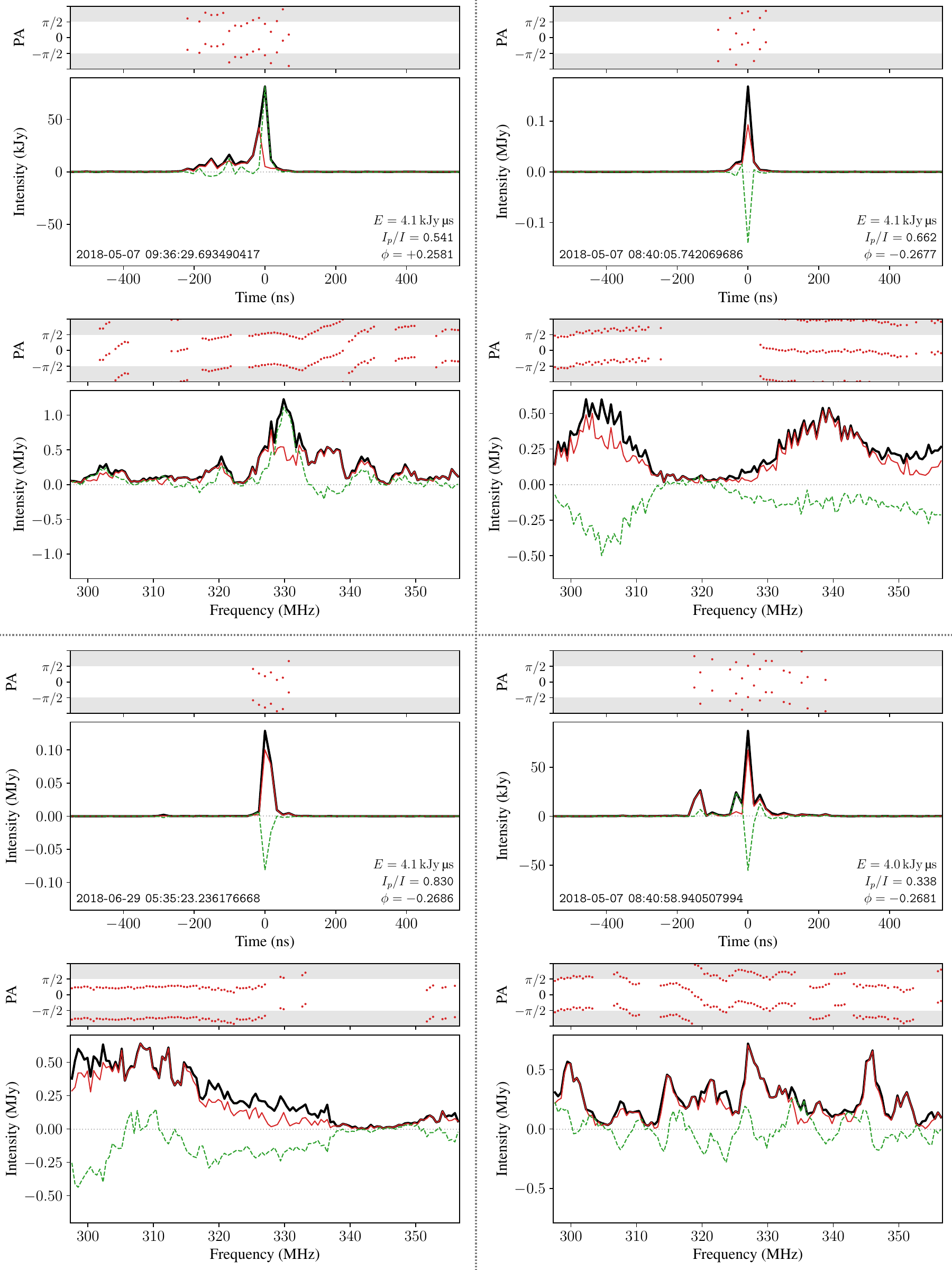}
  \caption{A selection of the most energetic giant pulses. See Figure~\ref{fig:gp_brights} for a full description.}
\end{figure*}

\begin{figure*}
  \centering
  \includegraphics[width=0.9\textwidth,trim=0 0 0 0,clip]{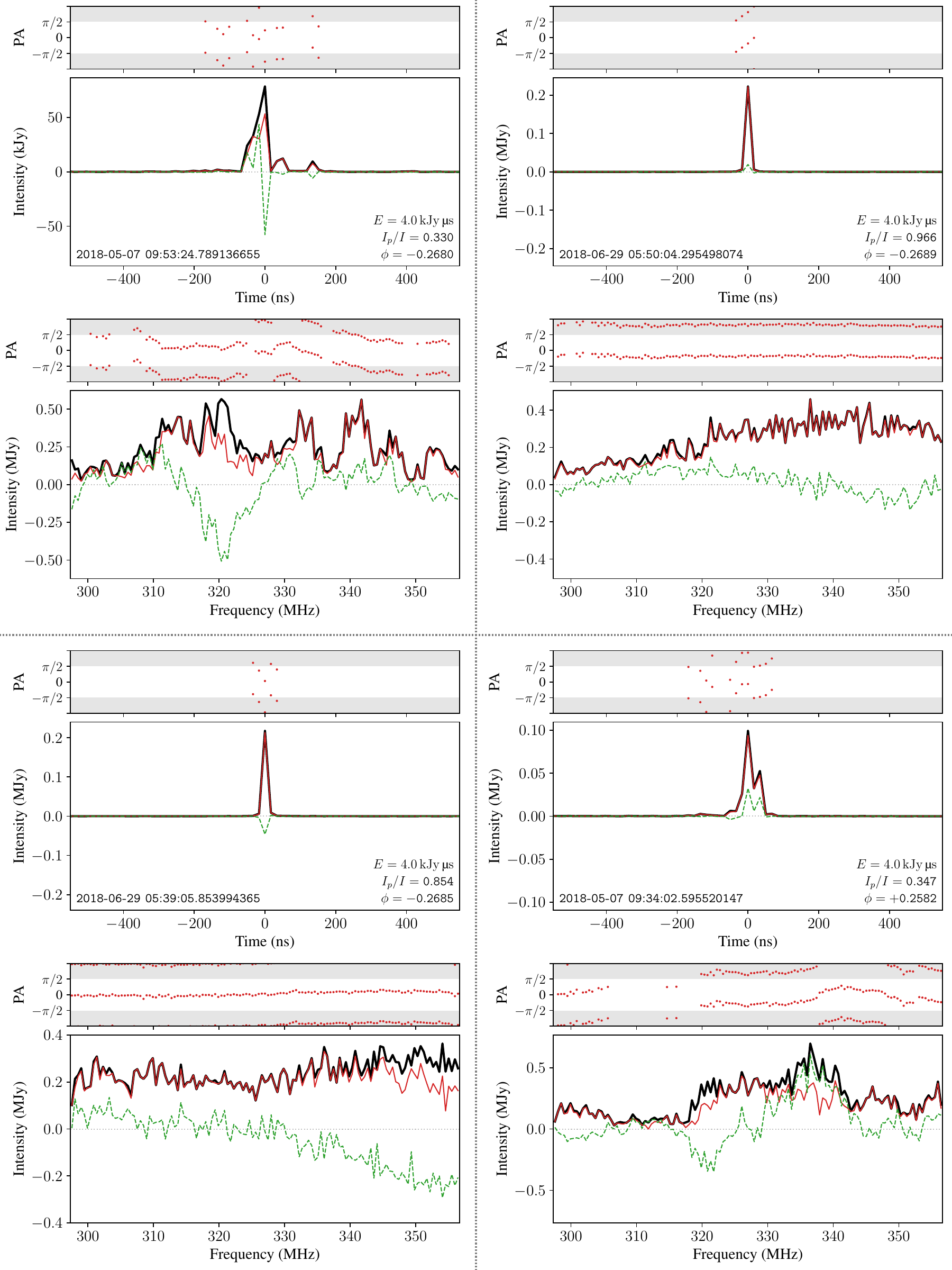}
  \caption{A selection of the most energetic giant pulses. See Figure~\ref{fig:gp_brights} for a full description.}
\end{figure*}

\begin{figure*}
  \centering
  \includegraphics[width=0.9\textwidth,trim=0 0 0 0,clip]{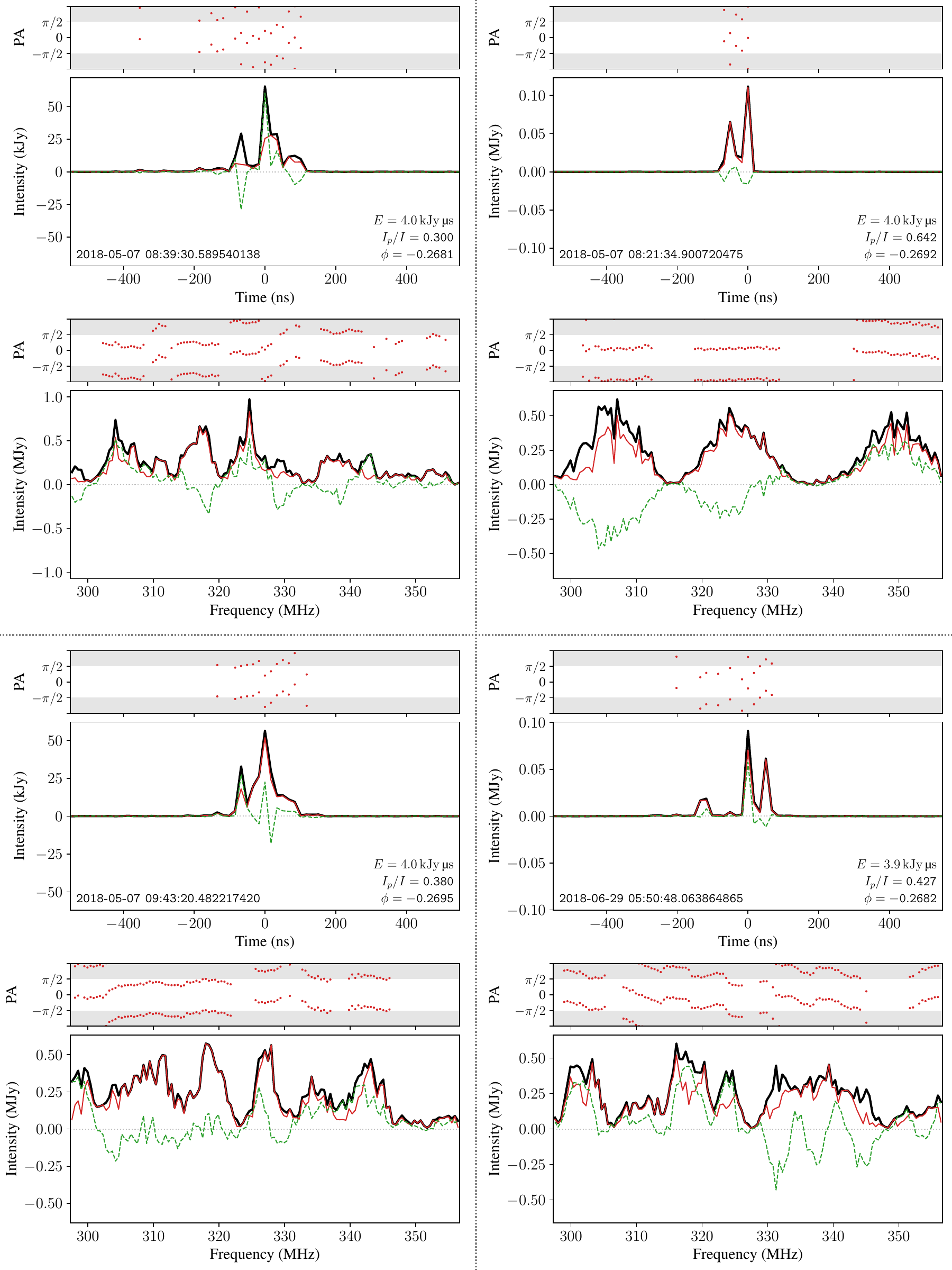}
  \caption{A selection of the most energetic giant pulses. See Figure~\ref{fig:gp_brights} for a full description.}
\end{figure*}

\begin{figure*}
  \centering
  \includegraphics[width=0.9\textwidth,trim=0 0 0 0,clip]{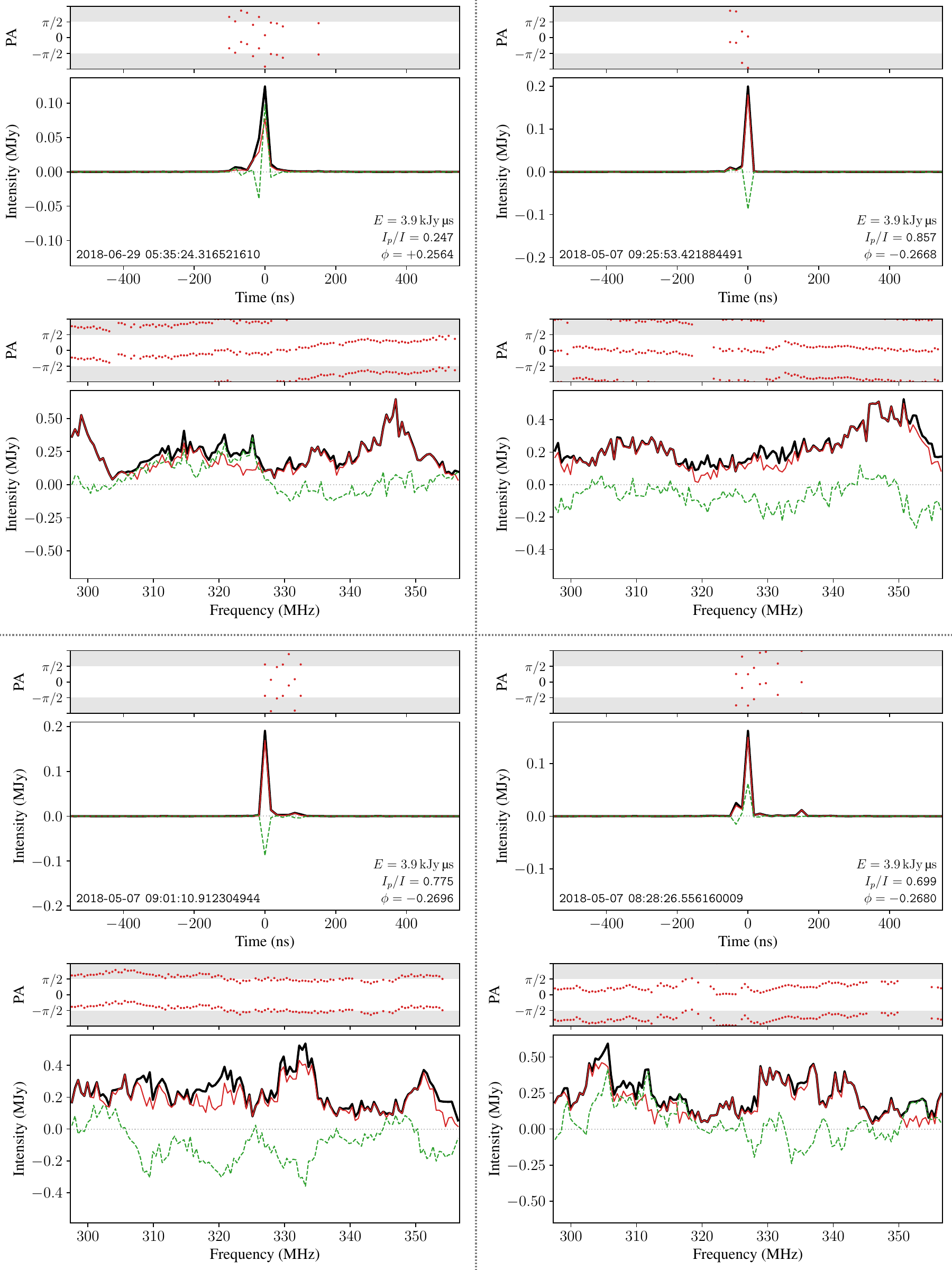}
  \caption{A selection of the most energetic giant pulses. See Figure~\ref{fig:gp_brights} for a full description.}
\end{figure*}

\begin{figure*}
  \centering
  \includegraphics[width=0.9\textwidth,trim=0 0 0 0,clip]{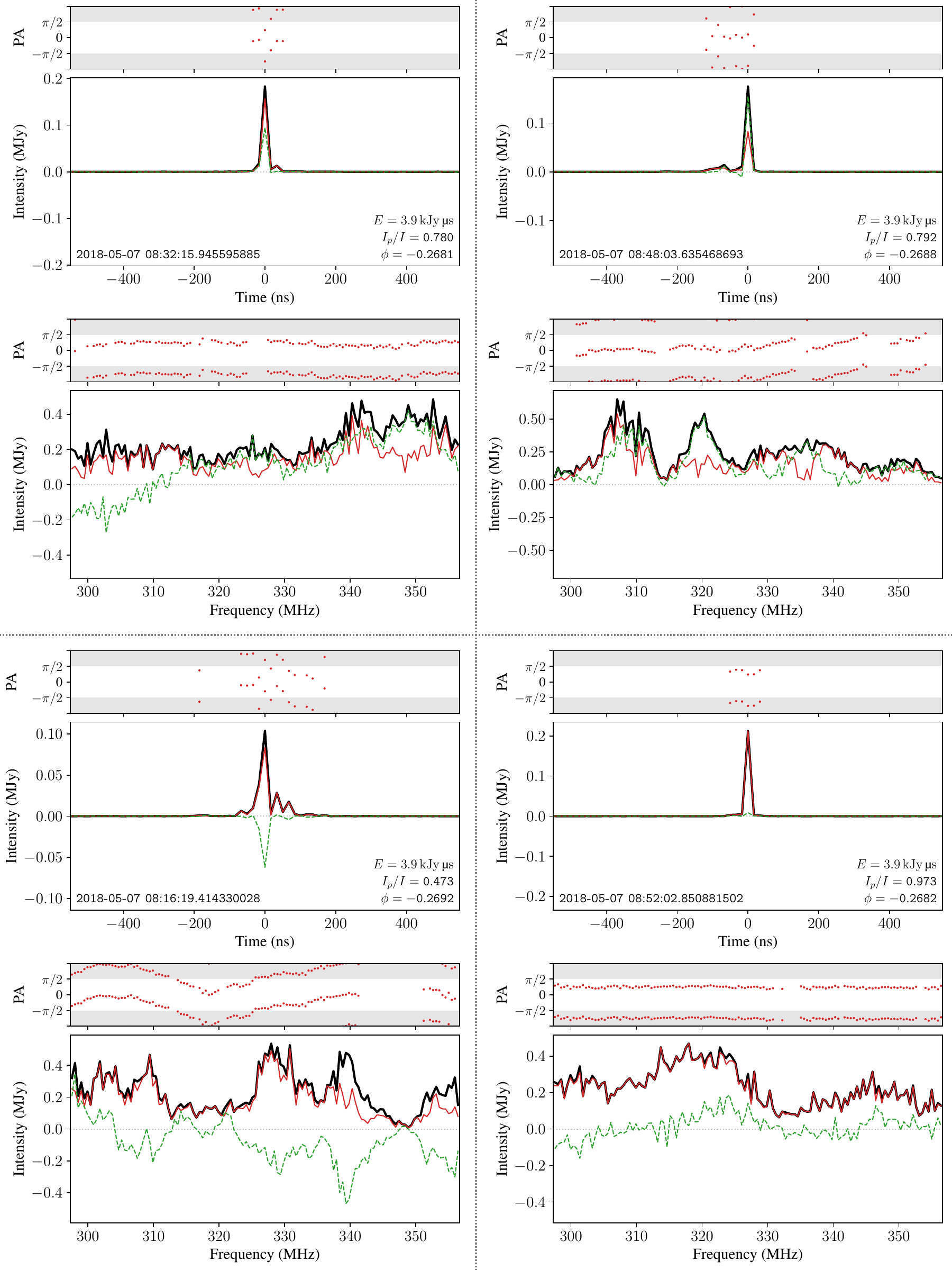}
  \caption{A selection of the most energetic giant pulses. See Figure~\ref{fig:gp_brights} for a full description.}
\end{figure*}

\begin{figure*}
  \centering
  \includegraphics[width=0.9\textwidth,trim=0 0 0 0,clip]{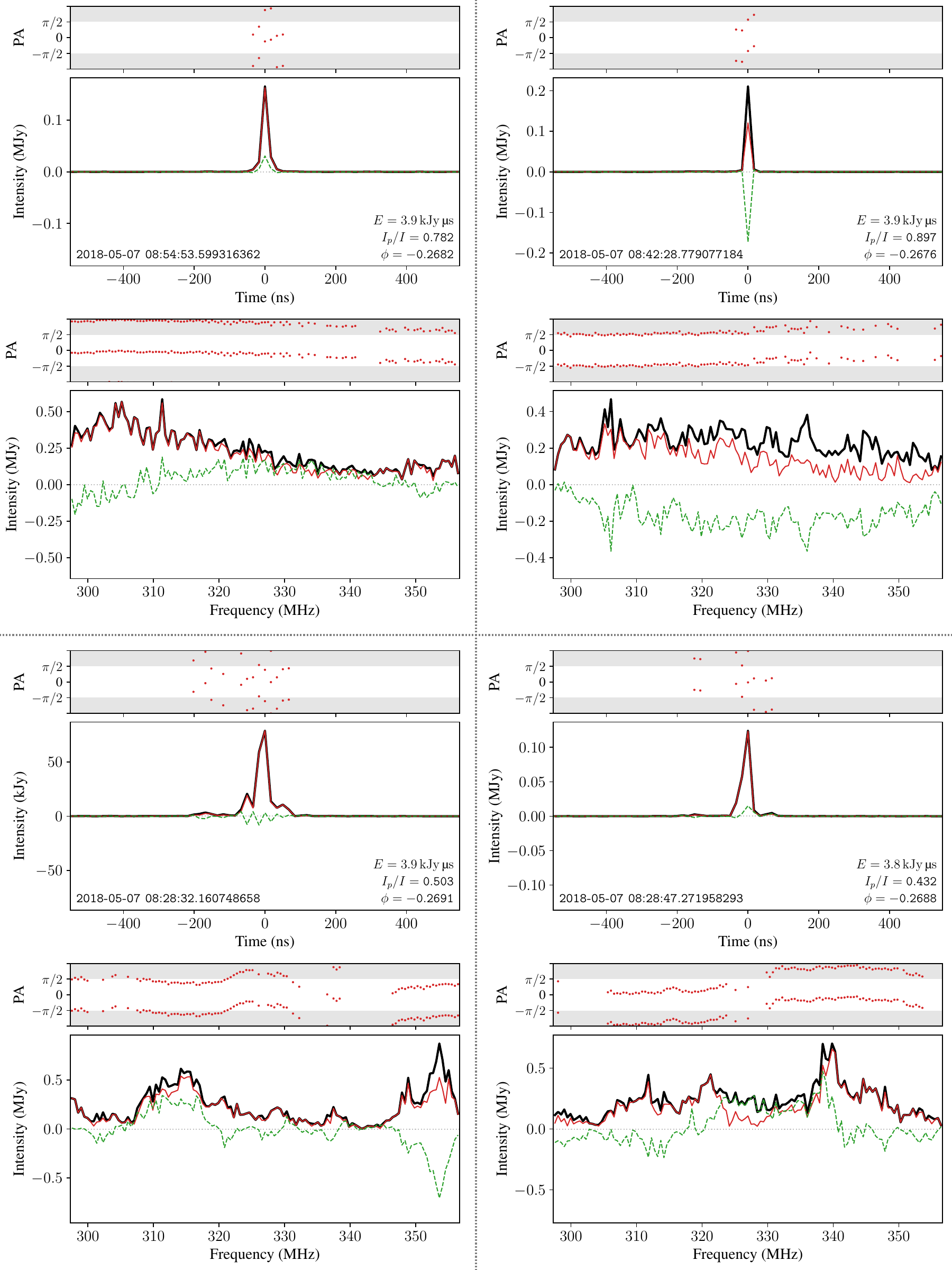}
  \caption{A selection of the most energetic giant pulses. See Figure~\ref{fig:gp_brights} for a full description.}
\end{figure*}

\begin{figure*}
  \centering
  \includegraphics[width=0.9\textwidth,trim=0 0 0 0,clip]{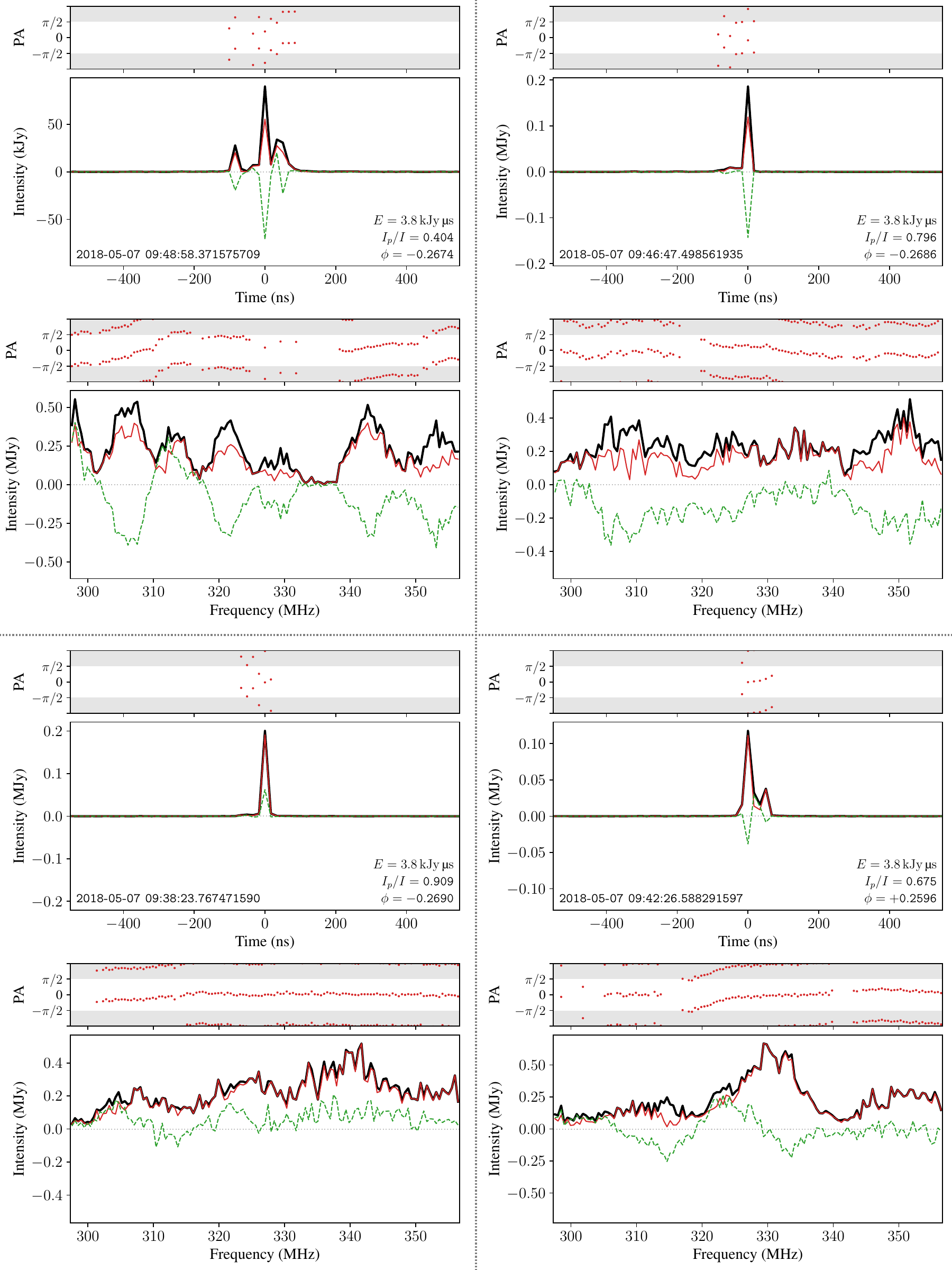}
  \caption{A selection of the most energetic giant pulses. See Figure~\ref{fig:gp_brights} for a full description.}
\end{figure*}

\begin{figure*}
  \centering
  \includegraphics[width=0.9\textwidth,trim=0 0 0 0,clip]{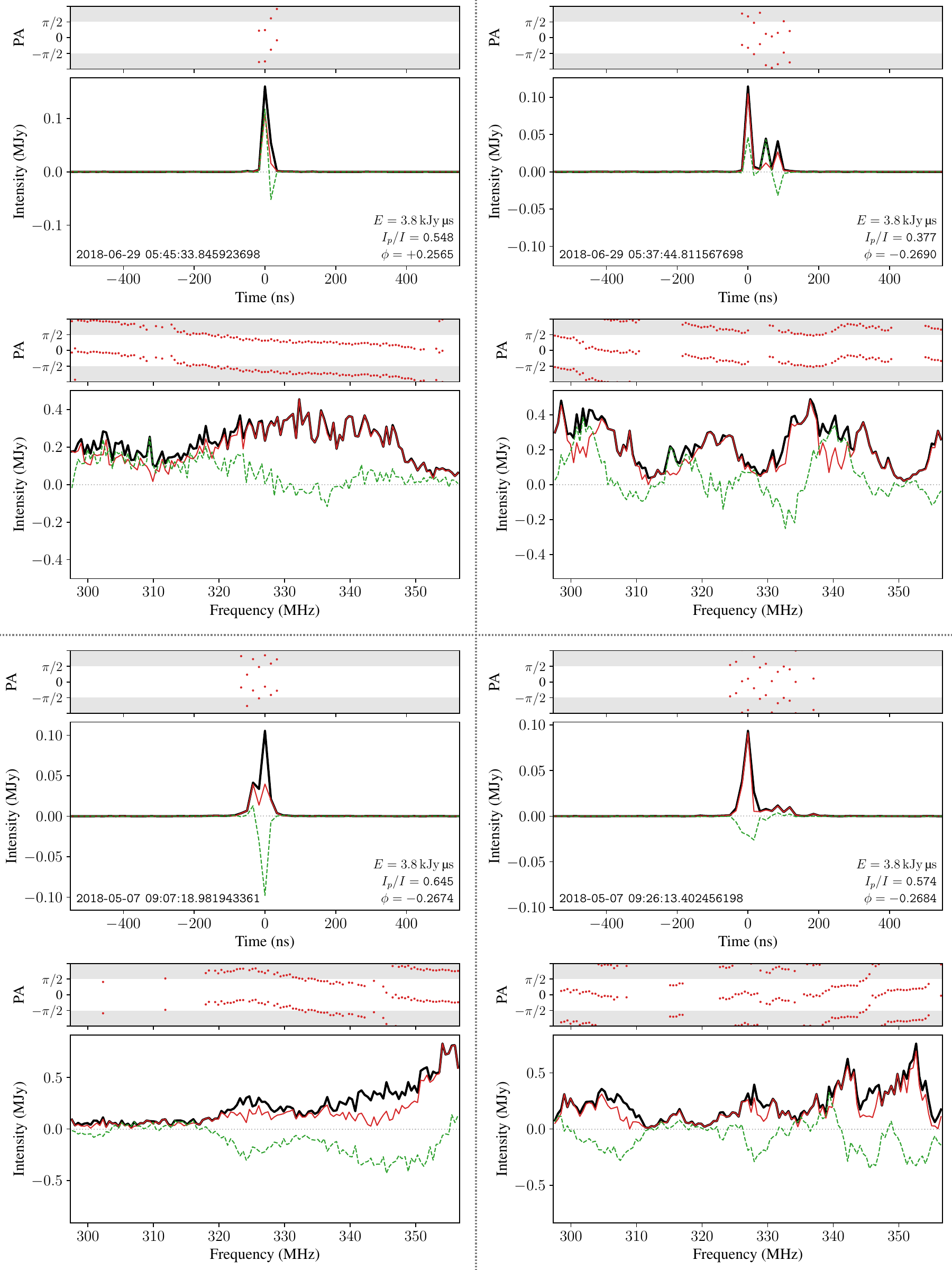}
  \caption{A selection of the most energetic giant pulses. See Figure~\ref{fig:gp_brights} for a full description.}
\end{figure*}

\begin{figure*}
  \centering
  \includegraphics[width=0.9\textwidth,trim=0 0 0 0,clip]{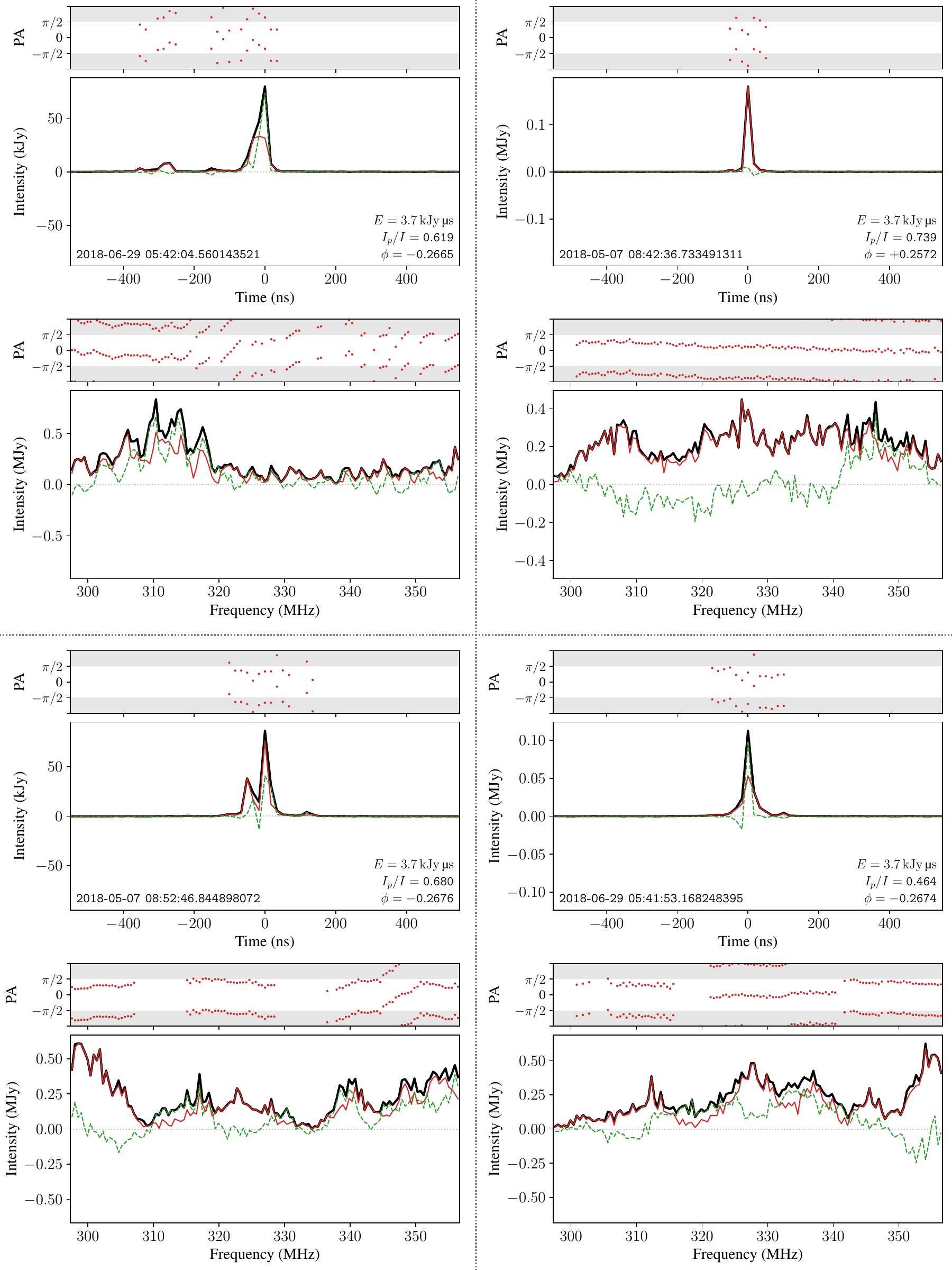}
  \caption{A selection of the most energetic giant pulses. See Figure~\ref{fig:gp_brights} for a full description.}
  \label{fig:figset_last}
\end{figure*}

\end{document}